\documentclass[aps,prd,nofootinbib,twocolumn,superscriptaddress,preprintnumbers,balancelastpage,longbibliography]{revtex4-1}
\usepackage{aas_macros}
\usepackage[utf8]{inputenc}
\usepackage{amsmath,amssymb,mathtools,bm}
\usepackage[section]{placeins}
\usepackage{graphicx, color, hepunits}
\usepackage[dvipsnames]{xcolor}
\usepackage{float}
\usepackage{multirow}
 \usepackage{hyperref} %Automatically links \label and \ref commands; Always load last
\hypersetup{
    colorlinks=true,       % false: boxed links; true: colored links
    linkcolor=blue,        % color of internal links
    citecolor=blue,        % color of links to bibliography
    filecolor=magenta,     % color of file links
    urlcolor=blue          % color of external links
}
\usepackage[utf8]{inputenc}
\usepackage[english]{babel}
\usepackage{lipsum}

\usepackage{tensor}
\usepackage{slashed}
\usepackage{lineno}

\newcommand{\pvec}{{\bm p}}
\newcommand{\kvec}{{\bm k}}

\newcommand{\Mcal}{\mathcal{M}}

\newcommand{\qvec}{{\bm q}}

\newcommand{\gagg}{{g_{a\gamma\gamma}}}
\newcommand{\aagg}{{\alpha_{a\gamma\gamma}}}
\newcommand{\gaee}{{g_{aee}}}
\newcommand{\aaee}{{\alpha_{aee}}}
\newcommand{\Mol}{\text{M{\o}l}}

\newcommand{\ba}[1]{\begin{align} #1 \end{align}}
\newcommand{\bes}[1]{\begin{equation}\begin{split} #1 \end{split}\end{equation}}

\newcommand{\es}[2] {\begin{equation} \label{#1} \begin{split} #2 \end{split} \end{equation}}
\newcommand{\farcs}{\mbox{\ensuremath{.\!\!^{\prime\prime}}}}%  % fractional arcsecond symbol: 0.''0
\begin{document}

\title{No evidence for axions from {\it Chandra} observation of magnetic white dwarf}

\author{Christopher Dessert}
%\email{dessert@umich.edu}
\affiliation{Leinweber Center for Theoretical Physics, Department of Physics, University of Michigan, Ann Arbor, MI 48109 U.S.A.}
\affiliation{Berkeley Center for Theoretical Physics, University of California, Berkeley, CA 94720, U.S.A.}
\affiliation{Theoretical Physics Group, Lawrence Berkeley National Laboratory, Berkeley, CA 94720, U.S.A.}

\author{Andrew J. Long}
%\email{andrewjlong@rice.edu}
\affiliation{Department of Physics and Astronomy, Rice University, Houston TX 77005, U.S.A.}

\author{Benjamin R. Safdi}
%\email{brsafdi@lbl.gov}
\affiliation{Berkeley Center for Theoretical Physics, University of California, Berkeley, CA 94720, U.S.A.}
\affiliation{Theoretical Physics Group, Lawrence Berkeley National Laboratory, Berkeley, CA 94720, U.S.A.}

\date{\today}

\begin{abstract}
Ultralight axions with axion-photon couplings $g_{a\gamma\gamma} \sim {\rm few} \times 10^{-11}$ GeV$^{-1}$ may resolve a number of astrophysical anomalies, such as unexpected $\sim$TeV transparency, anomalous stellar 
cooling, and $X$-ray excesses from nearby neutron stars. 
We show, however, that such axions are severely constrained by the non-observation of $X$-rays from the magnetic white dwarf (MWD) RE J0317-853 using $\sim$40 ks of data acquired from a dedicated observation with the {\it Chandra} $X$-ray Observatory.  Axions may be produced in the core of the MWD through electron bremsstrahlung and then convert to $X$-rays in the
magnetosphere.  The non-observation of $X$-rays constrains the axion-photon coupling to $g_{a\gamma\gamma} \lesssim 5.5 \times 10^{-13} \sqrt{C_{a\gamma\gamma}/C_{aee}}$ GeV$^{-1}$ at 95\% confidence for axion masses $m_a \lesssim 5 \times 10^{-6}$ eV, with $C_{aee}$ and $C_{a\gamma\gamma}$ the dimensionless coupling constants to electrons and photons.  Considering that $C_{aee}$ is generated from the renormalization group, our results robustly disfavor $g_{a\gamma\gamma} \gtrsim 4.4 \times 10^{-11}$ GeV$^{-1}$ even for models with no ultraviolet contribution to $C_{aee}$.        
\end{abstract}
\maketitle

Axions are hypothetical ultralight pseudoscalar particles that couple through dimension-5 operators to the Standard Model. 
In particular the quantum chromodynamics (QCD) axion couples to QCD, which allows it to solve the strong-{\it CP} problem~\cite{Peccei:1977hh,Peccei:1977ur,Weinberg:1977ma,Wilczek:1977pj}; this coupling also generates a mass $m_a^{\rm QCD} \sim \Lambda_{\rm QCD}^2 / f_a$ for the particle, with $f_a$ the axion decay constant and $\Lambda_{\rm QCD}$ the QCD confinement scale.  In this work we probe axions with masses $m_a \lesssim 10^{-2}$ eV that do not couple to QCD (but see~\cite{Farina:2016tgd,DiLuzio:2017pfr,Darme:2020gyx})
though they couple to electromagnetism and matter.  Such ultralight axions, often referred to as axion-like particles, are especially motivated theoretically in the context of the String Axiverse~\cite{Svrcek:2006yi,Arvanitaki:2009fg,Acharya:2010zx,Ringwald:2012cu,Stott:2017hvl,Halverson:2019cmy}.
In the Axiverse it is natural to expect a large number $N$ of ultralight axions, with $m_a \ll m_a^{\rm QCD}$.  One linear combination couples to QCD and receives a mass from QCD, becoming the QCD axion, while the rest of the $N-1$ states remain ultralight and retain their non-QCD couplings to the Standard Model.  It is well established that axions may be produced within stars including white dwarfs (WDs) (see {\it e.g.}~\cite{Raffelt:1985nj,Raffelt:1990yz,Giannotti:2017hny}) and escape the stars due to their weak interaction strengths with matter.  Recently it has been pointed out that 
such axions could produce $X$-ray signatures through axion-photon conversion in magnetic WD (MWD) magnetospheres~\cite{Dessert:2019sgw} (see~\cite{Morris:1984iz,Raffelt:1987im,Fortin:2018ehg,Fortin:2018aom,Buschmann:2019pfp,Fortin:2021sst} for related discussions in neutron star (NS) magnetospheres).  In this work we collect and analyze data from the MWD RE J0317-853 to look for evidence of this process.

\begin{figure}[htb]
\begin{center}
\includegraphics[width=0.475\textwidth]{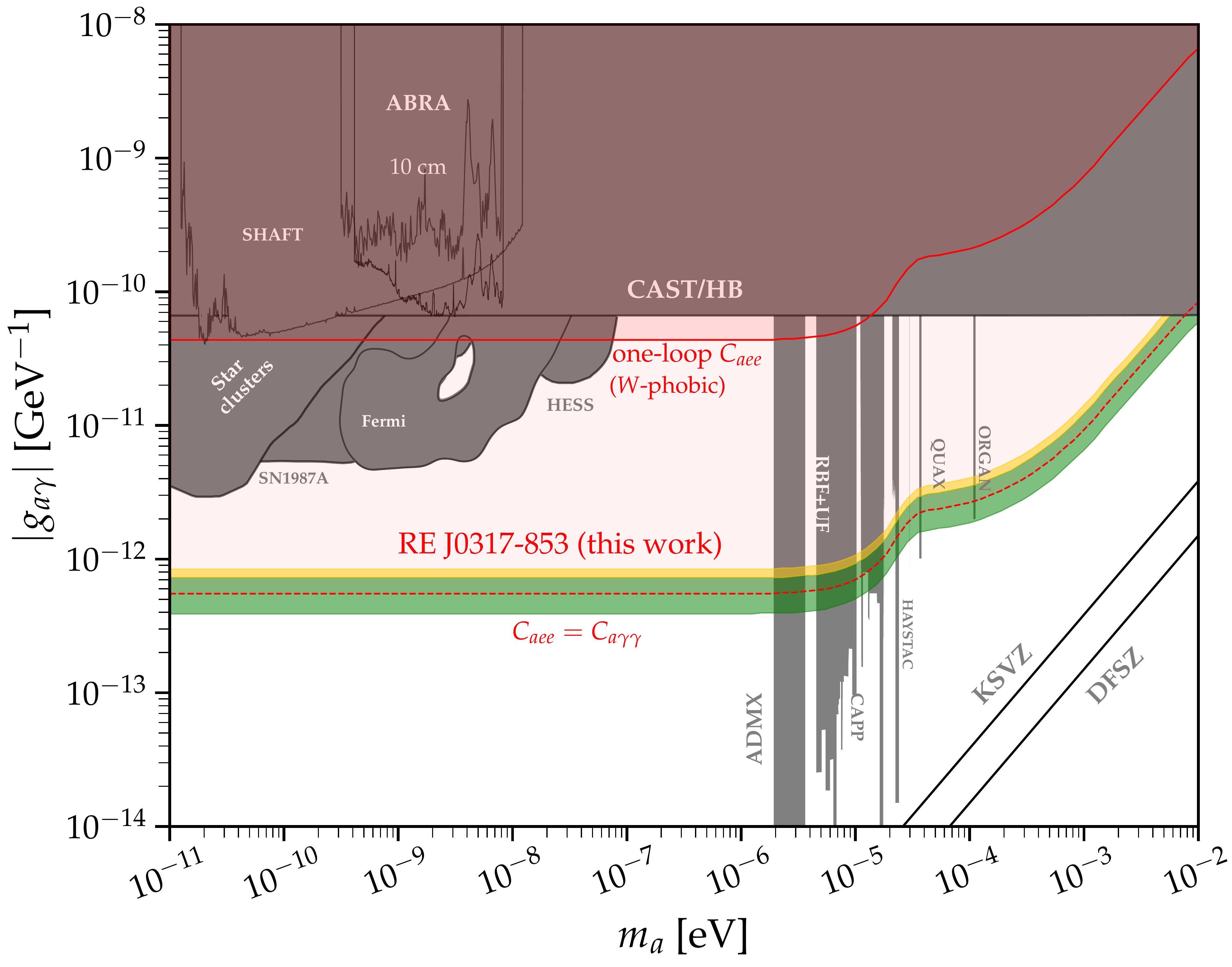} 
\caption{ We constrain $g_{a\gamma\gamma} g_{aee} \lesssim 1.3 \times 10^{-25}$ GeV$^{-1}$ at 95\% confidence for low $m_a$ from the non-observation of $X$-rays from the MWD RE J0317-853.  We translate this result to constraints on $g_{a\gamma\gamma}$ assuming: (i) a tree-level axion-electron coupling with $C_{aee} = C_{a\gamma\gamma}$, and (ii) the loop-induced $C_{aee} \approx 1.5 \cdot 10^{-4} C_{a\gamma\gamma}$ that represents  a conservative $W$-phobic axion (the loop-induced $C_{aee}$ is generically larger). The expected 68\% (95\%) containment region for the power-constrained 95\% upper limit  is shaded in green (gold) for the $C_{aee} = C_{a\gamma\gamma}$ scenario.  Previous constraints are shaded in grey~\cite{ciaran_o_hare_2020_3932430}.
}
\label{fig:gagg_limits}
\end{center}
\end{figure}

The couplings of the axion $a$ with mass $m_a$ to electromagnetism and electronic matter are described through the Lagrangian terms
\es{eq:Lag}{
\mathcal{L}_{\rm int} \supset  - {1 \over 4} g_{a\gamma\gamma} a F_{\mu \nu} \tilde F^{\mu \nu} + {g_{aee} \over 2 m_e} (\partial_\mu a) \bar e\gamma^\mu \gamma_5 e \,, 
}
with $F$ ($\tilde F$) the (dual) quantum electrodynamics (QED) field strength, $e$ the electron field, and $m_e$ the electron mass.  It is convenient to parameterize the coupling constants by $g_{a\gamma\gamma} = C_{a\gamma\gamma} \alpha_{\rm EM} / (2 \pi f_a)$ and $g_{aee} = C_{aee} m_e / f_a$, where the $C$'s are dimensionless.  Most laboratory and astrophysical searches for axions focus on the axion-photon coupling, with current constraints illustrated in Fig.~\ref{fig:gagg_limits}.  Low-mass constraints arise from the non-observation of photons from super star clusters (SSCs)~\cite{Dessert:2020lil} (see also~\cite{Xiao:2020pra}) and SN1987A~\cite{Payez:2014xsa} and searches for spectral modulations with {\it Fermi}~\cite{TheFermi-LAT:2016zue}, H.E.S.S.~\cite{Abramowski:2013oea}, and {\it Chandra}~\cite{Reynolds:2019uqt} (but see~\cite{Libanov:2019fzq}).  The constraints from the solar axion search with the CAST experiment~\cite{Anastassopoulos:2017ftl} and from Horizontal Branch (HB) star cooling~\cite{Ayala:2014pea} are comparable and extend over the whole mass range in Fig.~\ref{fig:gagg_limits}, which also shows the predicted coupling-mass relations in the DFSZ~\cite{Dine:1981rt,Zhitnitsky:1980tq} and KSVZ~\cite{Kim:1979if,Shifman:1979if} QCD axion models.
The additional constraints shown in Fig.~\ref{fig:gagg_limits} require the axion to be dark matter~\cite{Gramolin:2020ict,Ouellet:2018beu,Salemi:2021gck,Du:2018uak,Braine:2019fqb,Zhong:2018rsr,Backes:2020ajv,Jeong:2020cwz,Alesini:2020vny,McAllister:2017lkb} (see~\cite{ciaran_o_hare_2020_3932430} for a summary).   

As described in~\cite{Dessert:2019sgw} axions may be produced within the cores of MWD stars through electron bremsstrahlung off of ions, using the $g_{aee}$ coupling, and converted to $X$-rays in the stellar magnetospheres with the $g_{a\gamma\gamma}$ term in~\eqref{eq:Lag}.  Ref.~\cite{Dessert:2019sgw} identified RE J0317-853 as being the most promising currently-known MWD because of a combination of (i) the close distance $d = 29.38 \pm 0.02$ pc, as measured by {\it Gaia}~\cite{2020arXiv201201533G}, (ii) the large magnetic field $B_{\rm pole} \sim 500$ MG, and (iii) the high core temperature $T_{\rm core} \sim 1.5$ keV.  The predicted axion-induced $X$-ray signal is expected to be roughly thermal at the core temperature, meaning that it should peak at a few keV where {\it Chandra} is the most sensitive currently-operating $X$-ray telescope.    

\begin{figure}[ht]
\begin{center}
\includegraphics[width=0.48\textwidth]{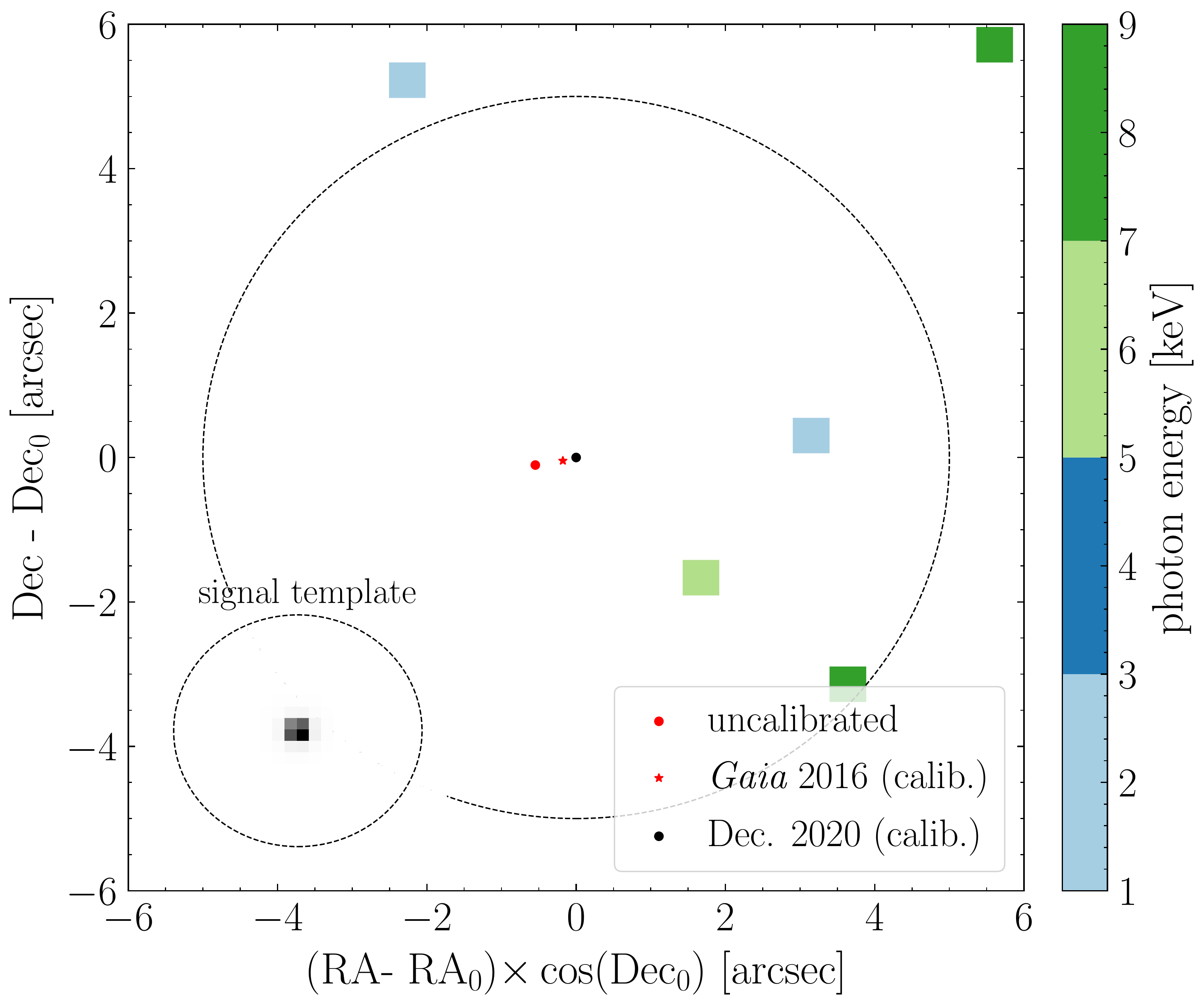} 
\caption{ The binned counts over 1--9 keV from our $\sim$40 ks {\it Chandra} observation of the MWD RE J0317-853.  No counts are observed within the vicinity of the source, whose location is indicated along with intermediate locations at various stages in the astrometric calibration process (see text for details), and also no more than one count is observed in any pixel.  
The dashed circle indicates the extent of the ROI used in our analysis.  The inset panel illustrates the signal template in grey scale, for the first energy bin, over the analysis ROI.
}
\label{fig:cts_map}
\end{center}
\end{figure}

We observed the MWD RE J0317-853 on 2020-12-18 using the {\it Chandra} ACIS-I instrument with no grating for a total of 37.42 ks (PI Safdi, observation ID \texttt{22326}).  After data reduction -- see the Supplementary Material (SM) -- we produce pixelated counts maps in four energy bins from 1 to 9 keV of width 2 keV each.  Each square pixel in right ascension (RA) and declination (DEC) has physical length of $\sim$$0\farcs492$ (note in the RA direction this is the width in ${\rm RA} \times \cos({\rm Dec})$).  In Fig.~\ref{fig:cts_map} we show the binned counts over 1--9 keV in the vicinity of the MWD; note that in this region no pixel has more than one count.  The figure is centered at the current location of the MWD, labeled `Dec. 2020 (calib.)': ${\rm RA}_0 \approx 49^\circ \, 18' \, 37\farcs77$, ${\rm DEC}_0 \approx -85^\circ \, 32' 25\farcs81$.  Fig.~\ref{fig:cts_map} also shows intermediate source locations determined during the astrometric calibration process (see the SM).
The 68\% energy containment radius at 1 keV (9 keV) is approximately $0\farcs5$ ($0\farcs6$).  
The inset illustrates the expected template for emission associated with the MWD at 1 keV.
No photon counts are observed near the MWD.  The circle in Fig.~\ref{fig:cts_map} has radius $5''$ and is the extent of our region of interest (ROI); that is, we exclude pixels whose centers are beyond this radius in our analysis.

We analyze the pixelated data ${\bf d} = \{n_{i,j} \}$, with $n_{i,j}$ the number of counts in energy bin $i$ and pixel $j$, in the context of the axion model, which is discussed more shortly, using the joint Poisson likelihood
\es{}{
p({\bf d}| \mathcal{M}, {\bm \theta}) = \prod_{i=1}^4 \prod_{j=1}^{\rm N_{\rm pix}} {\mu_{i,j}({\bm \theta})^{n_{i,j}} e^{-\mu_{i,j}({\bm \theta})} \over n_{i,j}!} \,,
}
with ${\mathcal M}$ denoting the joint signal and background model, with model parameters ${\bm \theta} = \{ {\bf A_{\rm bkg}}, g_{aee} g_{a\gamma\gamma}, m_a\}$, and $N_{\rm pix}$ the number of spatial pixels.  The model predicts $\mu_{i,j}({\bm \theta})$ counts in energy and spatial pixel $i,j$.  The background parameter vector ${\bf A}_{\rm bkg}$ consists of a single normalization parameter in each of the four energy bins that re-scales the background counts spatial template.  For our background template, which we profile over, we use the exposure map, which is flat to less than $0.5$\% over our ROI.  The signal model has the two parameters $\{g_{aee} g_{a\gamma\gamma}, m_a\}$, which predict the counts in each of the four energy bins.
The signal template is centered on the MWD and accounts for the point spread function (PSF), as illustrated in the inset of Fig.~\ref{fig:cts_map}. 

At a fixed $m_a$ we construct the profile likelihood for $g_{a\gamma\gamma} g_{aee}$ by maximizing the log-likelihood over ${\bf A}_{\rm bkg}$ at each $g_{a\gamma\gamma} g_{aee}$.  Our 95\% upper limit on $g_{a\gamma\gamma} g_{aee}$  is constructed directly by Monte Carlo simulations of the signal and null hypotheses instead of relying on Wilks' theorem, since we are in the low-counts limit (see {\it e.g.}~\cite{Cowan:2010js} for details).  A priori we decided to power constrain~\cite{Cowan:2011an} our limits to account for the possibility of under fluctuations, though this was not necessary in practice.

\begin{figure}[t]
\begin{center}
\includegraphics[width=0.48\textwidth]{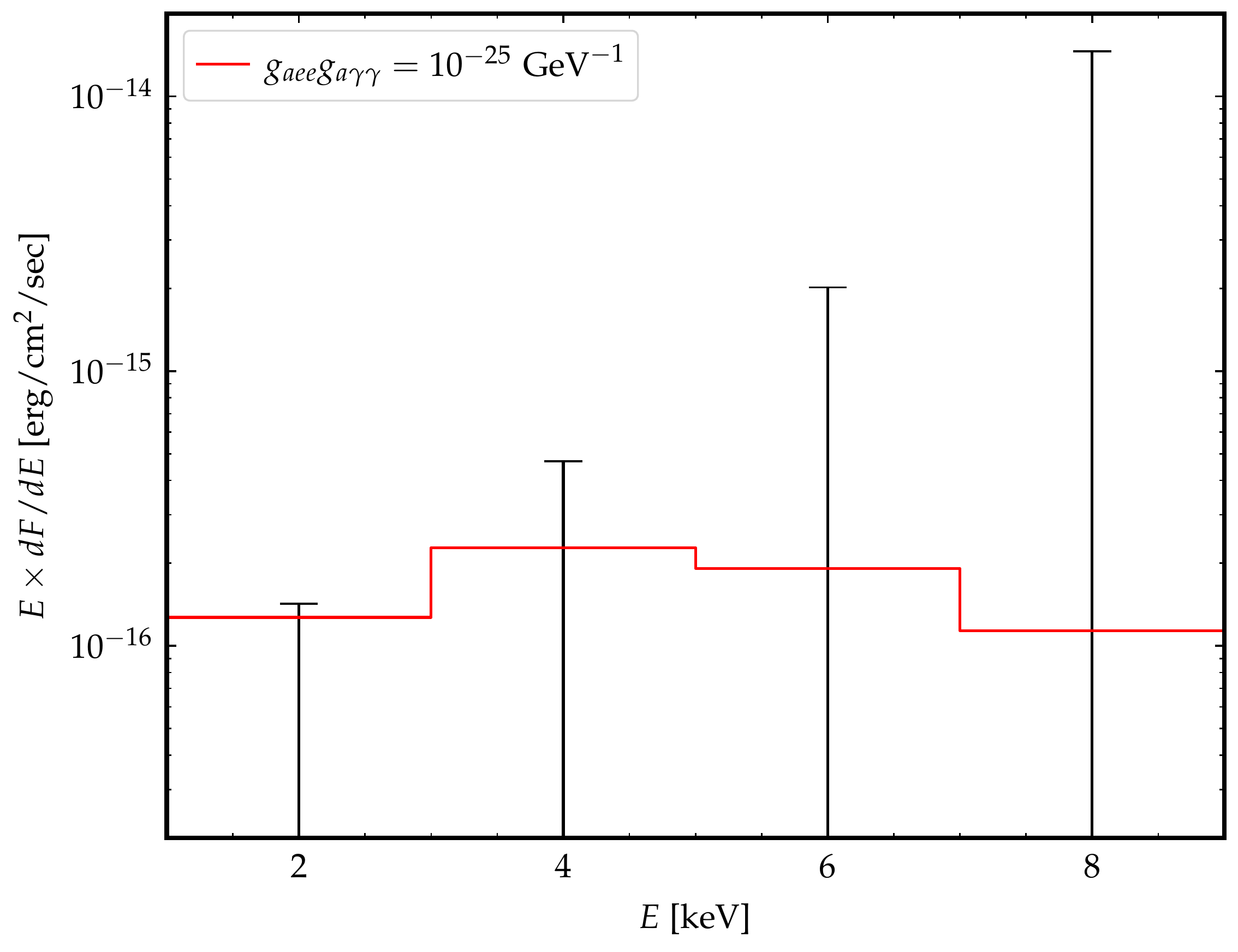} 
\caption{ The energy spectrum found from our analysis of the {\it Chandra} data from the MWD RE J0317-853.  In each of the four energy bins the best-fit fluxes are consistent with zero (the 68\% containment intervals are shown).  We also illustrate the predicted axion-induced signal that would be seen from an axion with the indicated couplings and $m_a \ll 10^{-5}$ eV. }
\label{fig:spectrum}
\end{center}
\end{figure}

We also analyze the data using the Poisson likelihood in the individual energy bins to extract the spectrum $dF/dE$, which is illustrated in Fig.~\ref{fig:spectrum}.  In that figure we overlay the axion model prediction, which we now detail.
For production via axion bremsstrahlung from electron-ion scattering~\cite{Nakagawa:1987pga,Raffelt:1990yz}, we broadly follow the formalism developed in~\cite{Dessert:2019sgw}, though we make improvements thanks to updated WD models and luminosity data from {\it Gaia}. 
Firstly, we improve our modeling of the density profile and composition of RE J0317-853 using \texttt{MESA}~\cite{Paxton2010} version \texttt{12778}. We simulate a WD of RE J0317-853's mass from stellar birth until it has cooled below RE J0317-853's observed luminosity. These simulations account for core electrostatic effects including ionic correlations and crystallization in the core that modify the profiles from that of a fully degenerate ideal electron gas, which were neglected in~\cite{Dessert:2019sgw}. We find RE J0317-853 has a predominantly oxygen-neon core because it completed carbon-burning while ascending the asymptotic giant branch, typical for a WD of its mass undergoing single-star evolution. We take as our fiducial profiles those density and composition profiles from the model for which the luminosity matches the observed luminosity of RE J0317-853 (see Sec.~\ref{app:model} of the SM for further details). 

The second improvement we make is in estimating the core temperature of RE J0317-853. Ref.~\cite{Dessert:2019sgw} estimated the core temperature from an empirical core temperature-luminosity relation using an assumed luminosity from~\cite{Kulebi:2010pd}. Ref.~\cite{Kulebi:2010pd} used Hubble parallax and photometric data along with WD cooling sequences to estimate the luminosity of RE J0317-853. Here, we estimate the core temperature from WD cooling sequences~\cite{2019AA...625A..87C} which predict {\it Gaia} DR2 band magnitudes. These cooling sequences are improved over those of~\cite{Kulebi:2010pd} because they better account for ionic correlation effects than previous sequences, and our use of {\it Gaia} data rather than Hubble represents an improvement because of smaller uncertainties on the magnitudes, partly due to improved parallax measurements. In particular, we fit the models in~\cite{2019AA...625A..87C} over cooling age and mass to the measured RE J0317-853 {\it Gaia} DR2 data~\cite{GaiaDR2}. Although previous measurements indicated a mass for RE J0317-853 of $\gtrsim 1.26$ $M_\odot$, we find that the $1.22$ $M_\odot$ model provides the best fit to the data. In the context of that model, we find that the {\it Gaia} data prefers a core temperature $T_c = 1.388 \pm 0.005$ keV. Therefore we use this model and to be conservative assume a core temperature at the lower $1\sigma$ allowed value, $T_c = 1.383$ keV, since the emissivity increases with increasing $T_c$.

Axion emission from the stellar interior primarily results from the bremsstrahlung scattering $e + (A,Z) \to e + (A,Z) + a$ where an electron is incident on a nucleus with atomic number $Z$ and mass number $A$.  
The electrons in a WD core are strongly degenerate with a temperature $T \ll p_F$ that is much smaller than the Fermi momentum $p_F$.  
In this regime, the axion emissivity spectrum is thermal and given by~\cite{Nakagawa:1987pga,Raffelt:1990yz}
\es{eq:epsa_ebrem}{
    \frac{d\varepsilon_a}{d\omega} = \frac{\alpha_{\rm EM}^2 g_{aee}^2}{4\pi^3 \, m_e^2} \, \frac{\omega^3}{e^{\omega/T}-1} \, \sum_s \frac{Z_s^2 \rho_s F_s}{A_s u} 
    \;,
}
which includes a sum over the species $s$ of nuclei that are present in the plasma; $Z_s$ is the atomic number, $A_s$ is the mass number, $\rho_s$ is the mass density, and $u \simeq 931.5 \ \mathrm{MeV}$ is the atomic mass unit.  
The species-dependent, dimensionless factor $F_s$ accounts for medium effects, including screening of the electric field and interference between different scattering sites.  
For a strongly-coupled plasma~\cite{Ichimaru:1982zz} we use the empirical fitting functions provided by~\cite{Nakagawa:1988rhp}.  
Note that the axion luminosity is given by the integral of the emissivity over the WD core. 

Our fiducial WD model leads to the predicted axion luminosity \mbox{$L_a \approx 8 \cdot 10^{-4} L_\odot (g_{aee} / 10^{-13} )^2$}. 
Accounting for modeling uncertainties on RE J0317-853 we estimate the limit on $g_{a\gamma\gamma}$ may be $\sim$10\% stronger, as illustrated in SM Fig.~\ref{fig:gagg_limits_stellar}.  Axions may also be produced by the $g_{a\gamma\gamma}$ coupling from electro-Primakoff production, which we compute in the SM, though as we show in SM Figs.~\ref{fig:gagg_limits_zoom} and~\ref{fig:spectrum_ep} this process is subdominant compared to bremsstrahlung for RE J0317-853.

The axions then undergo conversion to $X$-rays in the MWD magnetic fields.  The conversion probability $p_{a\to\gamma}$ may be calculated numerically for arbitrary magnetic field configurations and axion masses $m_a$ by solving the axion-photon mixing equations in the presence of $g_{a\gamma\gamma}$, though it is important to incorporate the Euler-Heisenberg Lagrangian term which modifies the propagation of photons in strong magnetic fields and suppresses the mixing~\cite{Raffelt:1987im}.
The magnetic field of the MWD is found to vary over the rotation period between 200 MG and 800 MG~\cite{Burleigh:1998pqa}; we follow~\cite{Dessert:2019sgw} and assume a dipole field of strength 200 MG, to be conservative.  Note that at low axion masses and high $B$-field values the dependence of the conversion probability on magnetic field is mild: $p_{a\to\gamma} \propto B^{2/5}$~\cite{Dessert:2019sgw}.  Using the offset dipole model from~\cite{Burleigh:1998pqa} increases the conversion probabilities by up to $\sim$50\%~\cite{Dessert:2019sgw} at low masses, which may increase the limit by $\sim$10\% relative to our fiducial case. Numerically the conversion probabilities are $\mathcal{O}(10^{-4}) \times \big(g_{a\gamma\gamma} / 10^{-11} \, {\rm GeV}^{-1} \big)^2$ for $m_a \ll 10^{-5}$ eV and drop off for higher masses.  
The distance is fixed at the central value measured by {\it Gaia} $d = 29.38$ pc~\cite{2020arXiv201201533G} because the distance uncertainty only leads to a $\sim$0.1\% uncertainty on the flux. 
In Fig.~\ref{fig:spectrum} we illustrate the energy-binned spectrum prediction from axion-induced emission from the MWD for $m_a \ll 10^{-5}$ eV and $g_{aee} g_{a\gamma\gamma} = 10^{-25}$ GeV$^{-1}$.  

\begin{figure}[t]
\begin{center}
\includegraphics[width=0.48\textwidth]{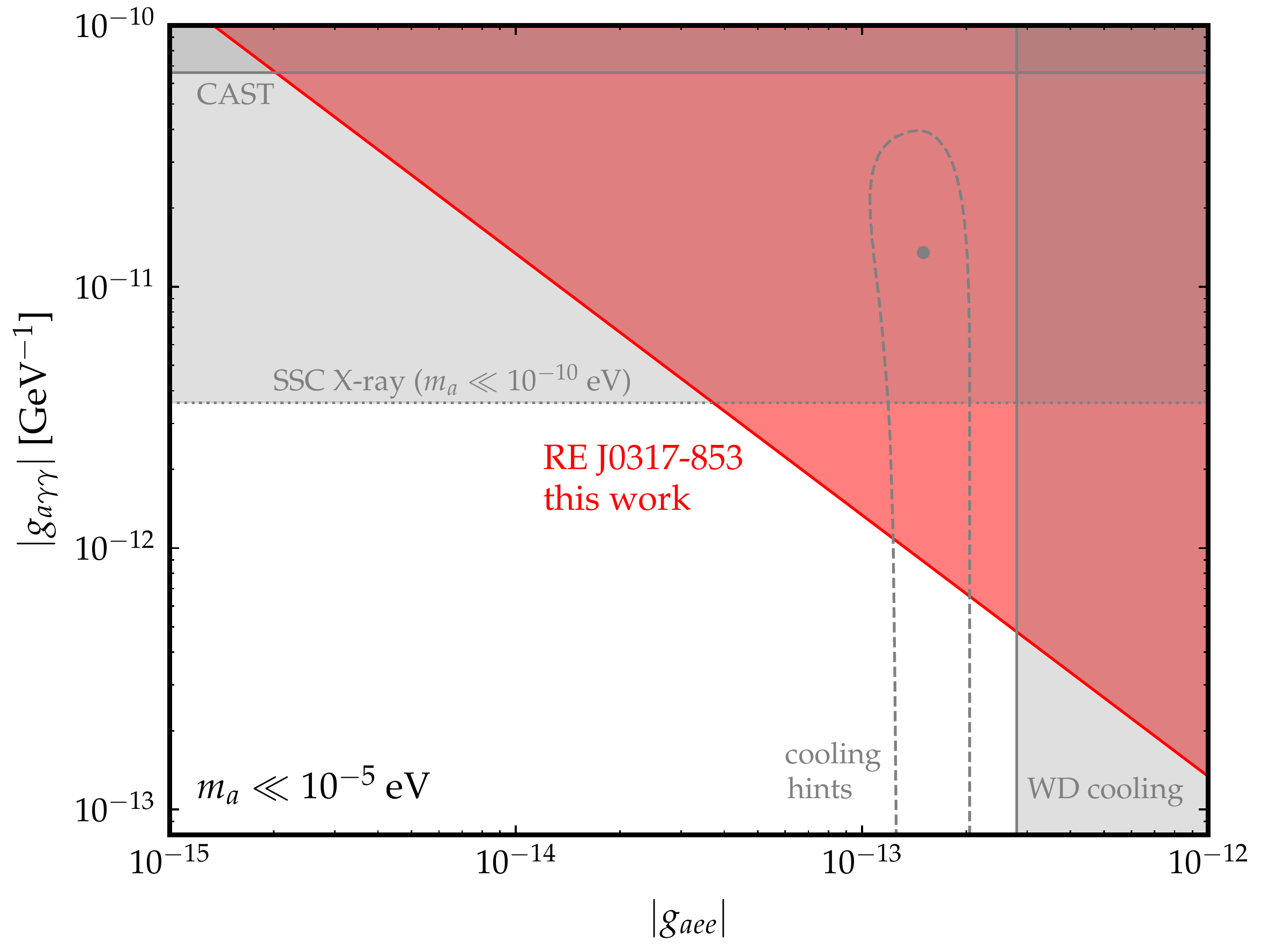} 
\caption{ The 95\% one-sided limit on the axion-photon and axion-electron coupling from this work $g_{aee}g_{a\gamma\gamma} < 1.3 \times 10^{-25}$ GeV$^{-1}$ assuming $m_a \ll 10^{-5}$ eV.  For $m_a \gtrsim 10^{-7}$ eV the leading constraint on $g_{a\gamma\gamma}$ is from the CAST experiment~\cite{Anastassopoulos:2017ftl} and HB star cooling~\cite{Ayala:2014pea}, while for $m_a \lesssim 10^{-10}$ eV it is from $X$-ray observations of SSCs~\cite{Dessert:2020lil}.  The leading limit on $g_{aee}$ is from WD cooling~\cite{Bertolami:2014wua}, while the 68\% containment region for explaining stellar cooling anomalies~\cite{Giannotti:2017hny}, along with the best-fit coupling, is also indicated and in tension with our null results. }
\label{fig:limit_gaee_gagg}
\end{center}
\end{figure}

We find no evidence for the axion model, with the best-fit coupling combination being zero for all masses.
We thus set 95\% one-sided upper limits on the coupling combination $g_{aee} g_{a\gamma\gamma}$ at fixed axion masses $m_a$ using the profile likelihood procedure.  For low masses $m_a \ll 10^{-5}$ eV the limit is $g_{aee} g_{a\gamma\gamma} \lesssim 1.3 \times 10^{-25}$ GeV$^{-1}$.  This limit is around three orders of magnitude stronger than that set by the CAST experiment on this coupling combination~\cite{Anastassopoulos:2017ftl}.  Our limit also severely constrains the low-mass axion explanation of stellar cooling anomalies~\cite{Giannotti:2017hny}, which prefer $g_{a\gamma\gamma} g_{aee} \sim 2 \times 10^{-24}$ GeV$^{-1}$ as illustrated in Fig.~\ref{fig:limit_gaee_gagg}, where we show our low-mass limit in the $g_{a\gamma\gamma}-g_{aee}$ plane, along with current constraints.

It is instructive to translate our limit to one on $g_{a\gamma\gamma}$ alone by assuming a relation between the dimensionless coupling constants $C_{aee}$ and $C_{a\gamma\gamma}$.  Note that in the DFSZ QCD axion model there is a tree-level coupling between the axion and electron, such that $C_{aee} \sim C_{a\gamma\gamma}$, while in the KSVZ model no ordinary matter is charged under the Peccei-Quinn (PQ) symmetry and so $C_{aee} = 0$ at tree level, though it is generated at one loop~\cite{Srednicki:1985xd}.
The loop-induced value of $C_{a\gamma\gamma}$ depends on the relative coupling of the axion to $SU(2)_L$ versus hypercharge $U(1)_Y$.  If the axion couples only to $SU(2)_L$ ($U(1)_Y$) then we expect, at one loop, \mbox{$C_{aee} \sim 4.8 \times 10^{-4} C_{a\gamma\gamma}$} (\mbox{$C_{aee} \sim 1.6 \times 10^{-4} C_{a\gamma\gamma}$}) for $f_a \approx 10^9$ GeV$^{-1}$ (see~\cite{Srednicki:1985xd,Chang:1993gm,Dessert:2019sgw} and the SM).  
To be conservative we assume in Fig.~\ref{fig:gagg_limits} the $W$-phobic axion scenario, where the axion only couples to $U(1)_Y$ (but see SM Fig.~\ref{fig:gagg_limits_zoom}).  
We also show the limit on $g_{a\gamma\gamma}$ for axion models with $C_{aee} = C_{a\gamma\gamma}$, which is nearly two orders of magnitude stronger than the loop-induced limit.

Our results have strong implications for a number of astrophysical anomalies and planned laboratory experiments. For example, the WD cooling anomaly prefers $g_{aee} \sim 1.6 \times 10^{-13}$~\cite{Giannotti:2017hny}.  In order for a low mass axion to explain this result and be compatible with our upper limit, one would need $C_{a\gamma\gamma} \lesssim 2.2 C_{aee}$ ($g_{a\gamma\gamma} \lesssim 8.1 \times 10^{-13}$ GeV$^{-1}$), which would not be able to also explain the  axion-photon coupling $g_{a\gamma\gamma} \sim 10^{-11}$ GeV$^{-1}$ suggested by the global fit to stellar cooling data~\cite{Giannotti:2017hny} (see Fig.~\ref{fig:limit_gaee_gagg}) or the TeV transparency anomalies, which prefer $g_{a\gamma\gamma} \gtrsim 2 \times 10^{-11}$ GeV$^{-1}$ for $m_a \ll 10^{-8}$ eV~\cite{Meyer:2013pny}.  
Anomalous $X$-ray emission from nearby isolated Magnificent Seven NSs may be interpreted as low-mass ($m_a \ll 10^{-5}$ eV) axion production from nucleon bremsstrahlung in the NS cores and conversion to $X$-rays in the NS magnetospheres~\cite{Dessert:2019dos,Buschmann:2019pfp}.  The required coupling combination to explain the $X$-ray excesses is $g_{a\gamma\gamma} g_{aNN} \gtrsim 10^{-21}$ GeV$^{-1}$, with $g_{aNN} = C_{aNN} m_N / f_a$ the axion-nucleon coupling, with $m_N$ the nucleon mass and $C_{aNN}$ the dimensionless coupling.  The non-observation of $X$-rays in this work from the MWD implies that if axions explain the Magnificent Seven excess they must be electro-phobic, with $C_{aee} \lesssim 4 \, C_{aNN}$.  
Lastly, we note that our results are especially relevant for the upcoming ALPS II light-shining-through-walls experiment~\cite{Bahre:2013ywa}.  The last stage of the experiment will have sensitivity to $g_{a\gamma\gamma} \gtrsim 2 \cdot 10^{-11}$ GeV$^{-1}$ for $m_a \lesssim 10^{-4}$ eV, meaning that much of the axion parameter space to be probed is constrained by the current analysis (see SM Fig.~\ref{fig:gagg_limits_zoom}).

As evident in {\it e.g.} Fig.~\ref{fig:cts_map} with $\sim$40 ks of {\it Chandra} data we are able to perform a nearly zero-background search; an additional order of magnitude in exposure time would allow us to improve the sensitivity to $g_{a\gamma\gamma}$ by a factor $\sim$1.5.  The proposed {\it Lynx} X-ray Observatory~\cite{LynxTeam:2018usc} aims to improve the point source sensitivity by roughly two orders of magnitude compared to {\it Chandra}.  A $\sim$400 ks observation with {\it Lynx} or a similar future telescope of RE J0317-853 (see SM Fig.~\ref{fig:gagg_limits_prog}) may be sensitive to axions with $g_{a\gamma\gamma} \sim   10^{-13}$ GeV$^{-1}$ for $C_{aee} \sim C_{a\gamma\gamma}$, which may probe photo-philic QCD axion models in addition to vast regions of uncharted parameter space for the hypothetical Axiverse. 

\begin{acknowledgments}
{\it 
We thank Josh Foster and Anson Hook for useful conversations. 
C.D. and B.R.S. were supported  in  part  by  the  DOE  Early Career  Grant  DESC0019225. This research used resources from the National Energy Research Scientific Computing Center (NERSC) and the Lawrencium computational cluster provided by the IT Division at the Lawrence Berkeley National Laboratory, supported by the Director, Office of Science, and Office of Basic Energy Sciences, of the U.S. Department of Energy under Contract No.  DE-AC02-05CH11231. Support for this work was provided by the National Aeronautics and Space Administration through Chandra Award Number GO0-21013X issued by the {\it Chandra} X-ray Center (CXC), which is operated by the Smithsonian Astrophysical Observatory for and on behalf of the National Aeronautics Space Administration under contract NAS8-03060. The scientific results reported in this article are based to a significant degree on observations made by the {\it Chandra} X-ray Observatory. This research has made use of software provided by the CXC in the application package CIAO.
}

\end{acknowledgments}

\bibliography{axion}

\begin{thebibliography}{69}%
\makeatletter
\providecommand \@ifxundefined [1]{%
 \@ifx{#1\undefined}
}%
\providecommand \@ifnum [1]{%
 \ifnum #1\expandafter \@firstoftwo
 \else \expandafter \@secondoftwo
 \fi
}%
\providecommand \@ifx [1]{%
 \ifx #1\expandafter \@firstoftwo
 \else \expandafter \@secondoftwo
 \fi
}%
\providecommand \natexlab [1]{#1}%
\providecommand \enquote  [1]{``#1''}%
\providecommand \bibnamefont  [1]{#1}%
\providecommand \bibfnamefont [1]{#1}%
\providecommand \citenamefont [1]{#1}%
\providecommand \href@noop [0]{\@secondoftwo}%
\providecommand \href [0]{\begingroup \@sanitize@url \@href}%
\providecommand \@href[1]{\@@startlink{#1}\@@href}%
\providecommand \@@href[1]{\endgroup#1\@@endlink}%
\providecommand \@sanitize@url [0]{\catcode `\\12\catcode `\$12\catcode
  `\&12\catcode `\#12\catcode `\^12\catcode `\_12\catcode `\%12\relax}%
\providecommand \@@startlink[1]{}%
\providecommand \@@endlink[0]{}%
\providecommand \url  [0]{\begingroup\@sanitize@url \@url }%
\providecommand \@url [1]{\endgroup\@href {#1}{\urlprefix }}%
\providecommand \urlprefix  [0]{URL }%
\providecommand \Eprint [0]{\href }%
\providecommand \doibase [0]{http://dx.doi.org/}%
\providecommand \selectlanguage [0]{\@gobble}%
\providecommand \bibinfo  [0]{\@secondoftwo}%
\providecommand \bibfield  [0]{\@secondoftwo}%
\providecommand \translation [1]{[#1]}%
\providecommand \BibitemOpen [0]{}%
\providecommand \bibitemStop [0]{}%
\providecommand \bibitemNoStop [0]{.\EOS\space}%
\providecommand \EOS [0]{\spacefactor3000\relax}%
\providecommand \BibitemShut  [1]{\csname bibitem#1\endcsname}%
\let\auto@bib@innerbib\@empty
%</preamble>
\bibitem [{\citenamefont {Peccei}\ and\ \citenamefont
  {Quinn}(1977{\natexlab{a}})}]{Peccei:1977hh}%
  \BibitemOpen
  \bibfield  {author} {\bibinfo {author} {\bibfnamefont {R.~D.}\ \bibnamefont
  {Peccei}}\ and\ \bibinfo {author} {\bibfnamefont {Helen~R.}\ \bibnamefont
  {Quinn}},\ }\bibfield  {title} {\enquote {\bibinfo {title} {{CP Conservation
  in the Presence of Instantons}},}\ }\href {\doibase
  10.1103/PhysRevLett.38.1440} {\bibfield  {journal} {\bibinfo  {journal}
  {Phys. Rev. Lett.}\ }\textbf {\bibinfo {volume} {38}},\ \bibinfo {pages}
  {1440--1443} (\bibinfo {year} {1977}{\natexlab{a}})}\BibitemShut {NoStop}%
%%CITATION = PRLTA,38,1440;%%
\bibitem [{\citenamefont {Peccei}\ and\ \citenamefont
  {Quinn}(1977{\natexlab{b}})}]{Peccei:1977ur}%
  \BibitemOpen
  \bibfield  {author} {\bibinfo {author} {\bibfnamefont {R.~D.}\ \bibnamefont
  {Peccei}}\ and\ \bibinfo {author} {\bibfnamefont {Helen~R.}\ \bibnamefont
  {Quinn}},\ }\bibfield  {title} {\enquote {\bibinfo {title} {{Constraints
  Imposed by CP Conservation in the Presence of Instantons}},}\ }\href
  {\doibase 10.1103/PhysRevD.16.1791} {\bibfield  {journal} {\bibinfo
  {journal} {Phys. Rev.}\ }\textbf {\bibinfo {volume} {D16}},\ \bibinfo {pages}
  {1791--1797} (\bibinfo {year} {1977}{\natexlab{b}})}\BibitemShut {NoStop}%
%%CITATION = PHRVA,D16,1791;%%
\bibitem [{\citenamefont {Weinberg}(1978)}]{Weinberg:1977ma}%
  \BibitemOpen
  \bibfield  {author} {\bibinfo {author} {\bibfnamefont {Steven}\ \bibnamefont
  {Weinberg}},\ }\bibfield  {title} {\enquote {\bibinfo {title} {{A New Light
  Boson?}}}\ }\href {\doibase 10.1103/PhysRevLett.40.223} {\bibfield  {journal}
  {\bibinfo  {journal} {Phys. Rev. Lett.}\ }\textbf {\bibinfo {volume} {40}},\
  \bibinfo {pages} {223--226} (\bibinfo {year} {1978})}\BibitemShut {NoStop}%
%%CITATION = PRLTA,40,223;%%
\bibitem [{\citenamefont {Wilczek}(1978)}]{Wilczek:1977pj}%
  \BibitemOpen
  \bibfield  {author} {\bibinfo {author} {\bibfnamefont {Frank}\ \bibnamefont
  {Wilczek}},\ }\bibfield  {title} {\enquote {\bibinfo {title} {{Problem of
  Strong p and t Invariance in the Presence of Instantons}},}\ }\href {\doibase
  10.1103/PhysRevLett.40.279} {\bibfield  {journal} {\bibinfo  {journal} {Phys.
  Rev. Lett.}\ }\textbf {\bibinfo {volume} {40}},\ \bibinfo {pages} {279--282}
  (\bibinfo {year} {1978})}\BibitemShut {NoStop}%
%%CITATION = PRLTA,40,279;%%
\bibitem [{\citenamefont {Farina}\ \emph {et~al.}(2017)\citenamefont {Farina},
  \citenamefont {Pappadopulo}, \citenamefont {Rompineve},\ and\ \citenamefont
  {Tesi}}]{Farina:2016tgd}%
  \BibitemOpen
  \bibfield  {author} {\bibinfo {author} {\bibfnamefont {Marco}\ \bibnamefont
  {Farina}}, \bibinfo {author} {\bibfnamefont {Duccio}\ \bibnamefont
  {Pappadopulo}}, \bibinfo {author} {\bibfnamefont {Fabrizio}\ \bibnamefont
  {Rompineve}}, \ and\ \bibinfo {author} {\bibfnamefont {Andrea}\ \bibnamefont
  {Tesi}},\ }\bibfield  {title} {\enquote {\bibinfo {title} {{The photo-philic
  QCD axion}},}\ }\href {\doibase 10.1007/JHEP01(2017)095} {\bibfield
  {journal} {\bibinfo  {journal} {JHEP}\ }\textbf {\bibinfo {volume} {01}},\
  \bibinfo {pages} {095} (\bibinfo {year} {2017})},\ \Eprint
  {http://arxiv.org/abs/1611.09855} {arXiv:1611.09855 [hep-ph]} \BibitemShut
  {NoStop}%
\bibitem [{\citenamefont {Di~Luzio}\ \emph {et~al.}(2017)\citenamefont
  {Di~Luzio}, \citenamefont {Mescia},\ and\ \citenamefont
  {Nardi}}]{DiLuzio:2017pfr}%
  \BibitemOpen
  \bibfield  {author} {\bibinfo {author} {\bibfnamefont {Luca}\ \bibnamefont
  {Di~Luzio}}, \bibinfo {author} {\bibfnamefont {Federico}\ \bibnamefont
  {Mescia}}, \ and\ \bibinfo {author} {\bibfnamefont {Enrico}\ \bibnamefont
  {Nardi}},\ }\bibfield  {title} {\enquote {\bibinfo {title} {{Window for
  preferred axion models}},}\ }\href {\doibase 10.1103/PhysRevD.96.075003}
  {\bibfield  {journal} {\bibinfo  {journal} {Phys. Rev. D}\ }\textbf {\bibinfo
  {volume} {96}},\ \bibinfo {pages} {075003} (\bibinfo {year} {2017})},\
  \Eprint {http://arxiv.org/abs/1705.05370} {arXiv:1705.05370 [hep-ph]}
  \BibitemShut {NoStop}%
\bibitem [{\citenamefont {Darm\'e}\ \emph {et~al.}(2021)\citenamefont
  {Darm\'e}, \citenamefont {Di~Luzio}, \citenamefont {Giannotti},\ and\
  \citenamefont {Nardi}}]{Darme:2020gyx}%
  \BibitemOpen
  \bibfield  {author} {\bibinfo {author} {\bibfnamefont {Luc}\ \bibnamefont
  {Darm\'e}}, \bibinfo {author} {\bibfnamefont {Luca}\ \bibnamefont
  {Di~Luzio}}, \bibinfo {author} {\bibfnamefont {Maurizio}\ \bibnamefont
  {Giannotti}}, \ and\ \bibinfo {author} {\bibfnamefont {Enrico}\ \bibnamefont
  {Nardi}},\ }\bibfield  {title} {\enquote {\bibinfo {title} {{Selective
  enhancement of the QCD axion couplings}},}\ }\href {\doibase
  10.1103/PhysRevD.103.015034} {\bibfield  {journal} {\bibinfo  {journal}
  {Phys. Rev. D}\ }\textbf {\bibinfo {volume} {103}},\ \bibinfo {pages}
  {015034} (\bibinfo {year} {2021})},\ \Eprint
  {http://arxiv.org/abs/2010.15846} {arXiv:2010.15846 [hep-ph]} \BibitemShut
  {NoStop}%
\bibitem [{\citenamefont {Svrcek}\ and\ \citenamefont
  {Witten}(2006)}]{Svrcek:2006yi}%
  \BibitemOpen
  \bibfield  {author} {\bibinfo {author} {\bibfnamefont {Peter}\ \bibnamefont
  {Svrcek}}\ and\ \bibinfo {author} {\bibfnamefont {Edward}\ \bibnamefont
  {Witten}},\ }\bibfield  {title} {\enquote {\bibinfo {title} {{Axions In
  String Theory}},}\ }\href {\doibase 10.1088/1126-6708/2006/06/051} {\bibfield
   {journal} {\bibinfo  {journal} {JHEP}\ }\textbf {\bibinfo {volume} {06}},\
  \bibinfo {pages} {051} (\bibinfo {year} {2006})},\ \Eprint
  {http://arxiv.org/abs/hep-th/0605206} {arXiv:hep-th/0605206 [hep-th]}
  \BibitemShut {NoStop}%
%%CITATION = HEP-TH/0605206;%%
\bibitem [{\citenamefont {Arvanitaki}\ \emph {et~al.}(2010)\citenamefont
  {Arvanitaki}, \citenamefont {Dimopoulos}, \citenamefont {Dubovsky},
  \citenamefont {Kaloper},\ and\ \citenamefont
  {March-Russell}}]{Arvanitaki:2009fg}%
  \BibitemOpen
  \bibfield  {author} {\bibinfo {author} {\bibfnamefont {Asimina}\ \bibnamefont
  {Arvanitaki}}, \bibinfo {author} {\bibfnamefont {Savas}\ \bibnamefont
  {Dimopoulos}}, \bibinfo {author} {\bibfnamefont {Sergei}\ \bibnamefont
  {Dubovsky}}, \bibinfo {author} {\bibfnamefont {Nemanja}\ \bibnamefont
  {Kaloper}}, \ and\ \bibinfo {author} {\bibfnamefont {John}\ \bibnamefont
  {March-Russell}},\ }\bibfield  {title} {\enquote {\bibinfo {title} {{String
  Axiverse}},}\ }\href {\doibase 10.1103/PhysRevD.81.123530} {\bibfield
  {journal} {\bibinfo  {journal} {Phys. Rev.}\ }\textbf {\bibinfo {volume}
  {D81}},\ \bibinfo {pages} {123530} (\bibinfo {year} {2010})},\ \Eprint
  {http://arxiv.org/abs/0905.4720} {arXiv:0905.4720 [hep-th]} \BibitemShut
  {NoStop}%
%%CITATION = ARXIV:0905.4720;%%
\bibitem [{\citenamefont {Acharya}\ \emph {et~al.}(2010)\citenamefont
  {Acharya}, \citenamefont {Bobkov},\ and\ \citenamefont
  {Kumar}}]{Acharya:2010zx}%
  \BibitemOpen
  \bibfield  {author} {\bibinfo {author} {\bibfnamefont {Bobby~Samir}\
  \bibnamefont {Acharya}}, \bibinfo {author} {\bibfnamefont {Konstantin}\
  \bibnamefont {Bobkov}}, \ and\ \bibinfo {author} {\bibfnamefont {Piyush}\
  \bibnamefont {Kumar}},\ }\bibfield  {title} {\enquote {\bibinfo {title} {{An
  M Theory Solution to the Strong CP Problem and Constraints on the
  Axiverse}},}\ }\href {\doibase 10.1007/JHEP11(2010)105} {\bibfield  {journal}
  {\bibinfo  {journal} {JHEP}\ }\textbf {\bibinfo {volume} {11}},\ \bibinfo
  {pages} {105} (\bibinfo {year} {2010})},\ \Eprint
  {http://arxiv.org/abs/1004.5138} {arXiv:1004.5138 [hep-th]} \BibitemShut
  {NoStop}%
\bibitem [{\citenamefont {Ringwald}(2014)}]{Ringwald:2012cu}%
  \BibitemOpen
  \bibfield  {author} {\bibinfo {author} {\bibfnamefont {Andreas}\ \bibnamefont
  {Ringwald}},\ }\bibfield  {title} {\enquote {\bibinfo {title} {{Searching for
  axions and ALPs from string theory}},}\ }\href {\doibase
  10.1088/1742-6596/485/1/012013} {\bibfield  {journal} {\bibinfo  {journal}
  {J. Phys. Conf. Ser.}\ }\textbf {\bibinfo {volume} {485}},\ \bibinfo {pages}
  {012013} (\bibinfo {year} {2014})},\ \Eprint {http://arxiv.org/abs/1209.2299}
  {arXiv:1209.2299 [hep-ph]} \BibitemShut {NoStop}%
\bibitem [{\citenamefont {Stott}\ \emph {et~al.}(2017)\citenamefont {Stott},
  \citenamefont {Marsh}, \citenamefont {Pongkitivanichkul}, \citenamefont
  {Price},\ and\ \citenamefont {Acharya}}]{Stott:2017hvl}%
  \BibitemOpen
  \bibfield  {author} {\bibinfo {author} {\bibfnamefont {Matthew~J.}\
  \bibnamefont {Stott}}, \bibinfo {author} {\bibfnamefont {David J.~E.}\
  \bibnamefont {Marsh}}, \bibinfo {author} {\bibfnamefont {Chakrit}\
  \bibnamefont {Pongkitivanichkul}}, \bibinfo {author} {\bibfnamefont
  {Layne~C.}\ \bibnamefont {Price}}, \ and\ \bibinfo {author} {\bibfnamefont
  {Bobby~S.}\ \bibnamefont {Acharya}},\ }\bibfield  {title} {\enquote {\bibinfo
  {title} {{Spectrum of the axion dark sector}},}\ }\href {\doibase
  10.1103/PhysRevD.96.083510} {\bibfield  {journal} {\bibinfo  {journal} {Phys.
  Rev. D}\ }\textbf {\bibinfo {volume} {96}},\ \bibinfo {pages} {083510}
  (\bibinfo {year} {2017})},\ \Eprint {http://arxiv.org/abs/1706.03236}
  {arXiv:1706.03236 [astro-ph.CO]} \BibitemShut {NoStop}%
\bibitem [{\citenamefont {Halverson}\ \emph {et~al.}(2019)\citenamefont
  {Halverson}, \citenamefont {Long}, \citenamefont {Nelson},\ and\
  \citenamefont {Salinas}}]{Halverson:2019cmy}%
  \BibitemOpen
  \bibfield  {author} {\bibinfo {author} {\bibfnamefont {James}\ \bibnamefont
  {Halverson}}, \bibinfo {author} {\bibfnamefont {Cody}\ \bibnamefont {Long}},
  \bibinfo {author} {\bibfnamefont {Brent}\ \bibnamefont {Nelson}}, \ and\
  \bibinfo {author} {\bibfnamefont {Gustavo}\ \bibnamefont {Salinas}},\
  }\bibfield  {title} {\enquote {\bibinfo {title} {{Towards string theory
  expectations for photon couplings to axionlike particles}},}\ }\href
  {\doibase 10.1103/PhysRevD.100.106010} {\bibfield  {journal} {\bibinfo
  {journal} {Phys. Rev. D}\ }\textbf {\bibinfo {volume} {100}},\ \bibinfo
  {pages} {106010} (\bibinfo {year} {2019})},\ \Eprint
  {http://arxiv.org/abs/1909.05257} {arXiv:1909.05257 [hep-th]} \BibitemShut
  {NoStop}%
\bibitem [{\citenamefont {Raffelt}(1986)}]{Raffelt:1985nj}%
  \BibitemOpen
  \bibfield  {author} {\bibinfo {author} {\bibfnamefont {Georg~G.}\
  \bibnamefont {Raffelt}},\ }\bibfield  {title} {\enquote {\bibinfo {title}
  {{Axion Constraints From White Dwarf Cooling Times}},}\ }\href {\doibase
  10.1016/0370-2693(86)91588-1} {\bibfield  {journal} {\bibinfo  {journal}
  {Phys. Lett.}\ }\textbf {\bibinfo {volume} {166B}},\ \bibinfo {pages}
  {402--406} (\bibinfo {year} {1986})}\BibitemShut {NoStop}%
%%CITATION = PHLTA,166B,402;%%
\bibitem [{\citenamefont {Raffelt}(1990)}]{Raffelt:1990yz}%
  \BibitemOpen
  \bibfield  {author} {\bibinfo {author} {\bibfnamefont {Georg~G.}\
  \bibnamefont {Raffelt}},\ }\bibfield  {title} {\enquote {\bibinfo {title}
  {{Astrophysical methods to constrain axions and other novel particle
  phenomena}},}\ }\href {\doibase 10.1016/0370-1573(90)90054-6} {\bibfield
  {journal} {\bibinfo  {journal} {Phys. Rept.}\ }\textbf {\bibinfo {volume}
  {198}},\ \bibinfo {pages} {1--113} (\bibinfo {year} {1990})}\BibitemShut
  {NoStop}%
\bibitem [{\citenamefont {Giannotti}\ \emph {et~al.}(2017)\citenamefont
  {Giannotti}, \citenamefont {Irastorza}, \citenamefont {Redondo},
  \citenamefont {Ringwald},\ and\ \citenamefont {Saikawa}}]{Giannotti:2017hny}%
  \BibitemOpen
  \bibfield  {author} {\bibinfo {author} {\bibfnamefont {Maurizio}\
  \bibnamefont {Giannotti}}, \bibinfo {author} {\bibfnamefont {Igor~G.}\
  \bibnamefont {Irastorza}}, \bibinfo {author} {\bibfnamefont {Javier}\
  \bibnamefont {Redondo}}, \bibinfo {author} {\bibfnamefont {Andreas}\
  \bibnamefont {Ringwald}}, \ and\ \bibinfo {author} {\bibfnamefont {Ken'ichi}\
  \bibnamefont {Saikawa}},\ }\bibfield  {title} {\enquote {\bibinfo {title}
  {{Stellar Recipes for Axion Hunters}},}\ }\href {\doibase
  10.1088/1475-7516/2017/10/010} {\bibfield  {journal} {\bibinfo  {journal}
  {JCAP}\ }\textbf {\bibinfo {volume} {10}},\ \bibinfo {pages} {010} (\bibinfo
  {year} {2017})},\ \Eprint {http://arxiv.org/abs/1708.02111} {arXiv:1708.02111
  [hep-ph]} \BibitemShut {NoStop}%
\bibitem [{\citenamefont {Dessert}\ \emph {et~al.}(2019)\citenamefont
  {Dessert}, \citenamefont {Long},\ and\ \citenamefont
  {Safdi}}]{Dessert:2019sgw}%
  \BibitemOpen
  \bibfield  {author} {\bibinfo {author} {\bibfnamefont {Christopher}\
  \bibnamefont {Dessert}}, \bibinfo {author} {\bibfnamefont {Andrew~J.}\
  \bibnamefont {Long}}, \ and\ \bibinfo {author} {\bibfnamefont {Benjamin~R.}\
  \bibnamefont {Safdi}},\ }\bibfield  {title} {\enquote {\bibinfo {title}
  {{X-ray signatures of axion conversion in magnetic white dwarf stars}},}\
  }\href {\doibase 10.1103/PhysRevLett.123.061104} {\bibfield  {journal}
  {\bibinfo  {journal} {Phys. Rev. Lett.}\ }\textbf {\bibinfo {volume} {123}},\
  \bibinfo {pages} {061104} (\bibinfo {year} {2019})},\ \Eprint
  {http://arxiv.org/abs/1903.05088} {arXiv:1903.05088 [hep-ph]} \BibitemShut
  {NoStop}%
\bibitem [{\citenamefont {Morris}(1986)}]{Morris:1984iz}%
  \BibitemOpen
  \bibfield  {author} {\bibinfo {author} {\bibfnamefont {Donald~E.}\
  \bibnamefont {Morris}},\ }\bibfield  {title} {\enquote {\bibinfo {title}
  {{Axion Mass Limits From Pulsar X-rays}},}\ }\href {\doibase
  10.1103/PhysRevD.34.843} {\bibfield  {journal} {\bibinfo  {journal} {Phys.
  Rev.}\ }\textbf {\bibinfo {volume} {D34}},\ \bibinfo {pages} {843} (\bibinfo
  {year} {1986})}\BibitemShut {NoStop}%
%%CITATION = PHRVA,D34,843;%%
\bibitem [{\citenamefont {Raffelt}\ and\ \citenamefont
  {Stodolsky}(1988)}]{Raffelt:1987im}%
  \BibitemOpen
  \bibfield  {author} {\bibinfo {author} {\bibfnamefont {Georg}\ \bibnamefont
  {Raffelt}}\ and\ \bibinfo {author} {\bibfnamefont {Leo}\ \bibnamefont
  {Stodolsky}},\ }\bibfield  {title} {\enquote {\bibinfo {title} {{Mixing of
  the Photon with Low Mass Particles}},}\ }\href {\doibase
  10.1103/PhysRevD.37.1237} {\bibfield  {journal} {\bibinfo  {journal} {Phys.
  Rev.}\ }\textbf {\bibinfo {volume} {D37}},\ \bibinfo {pages} {1237} (\bibinfo
  {year} {1988})}\BibitemShut {NoStop}%
%%CITATION = PHRVA,D37,1237;%%
\bibitem [{\citenamefont {Fortin}\ and\ \citenamefont
  {Sinha}(2018)}]{Fortin:2018ehg}%
  \BibitemOpen
  \bibfield  {author} {\bibinfo {author} {\bibfnamefont {Jean-Fran{\c c}ois}\
  \bibnamefont {Fortin}}\ and\ \bibinfo {author} {\bibfnamefont {Kuver}\
  \bibnamefont {Sinha}},\ }\bibfield  {title} {\enquote {\bibinfo {title}
  {{Constraining Axion-Like-Particles with Hard X-ray Emission from
  Magnetars}},}\ }\href {\doibase 10.1007/JHEP06(2018)048} {\bibfield
  {journal} {\bibinfo  {journal} {JHEP}\ }\textbf {\bibinfo {volume} {06}},\
  \bibinfo {pages} {048} (\bibinfo {year} {2018})},\ \Eprint
  {http://arxiv.org/abs/1804.01992} {arXiv:1804.01992 [hep-ph]} \BibitemShut
  {NoStop}%
%%CITATION = ARXIV:1804.01992;%%
\bibitem [{\citenamefont {Fortin}\ and\ \citenamefont
  {Sinha}(2019)}]{Fortin:2018aom}%
  \BibitemOpen
  \bibfield  {author} {\bibinfo {author} {\bibfnamefont {Jean-Fran\c{c}ois}\
  \bibnamefont {Fortin}}\ and\ \bibinfo {author} {\bibfnamefont {Kuver}\
  \bibnamefont {Sinha}},\ }\bibfield  {title} {\enquote {\bibinfo {title}
  {{X-Ray Polarization Signals from Magnetars with Axion-Like-Particles}},}\
  }\href {\doibase 10.1007/JHEP01(2019)163} {\bibfield  {journal} {\bibinfo
  {journal} {JHEP}\ }\textbf {\bibinfo {volume} {01}},\ \bibinfo {pages} {163}
  (\bibinfo {year} {2019})},\ \Eprint {http://arxiv.org/abs/1807.10773}
  {arXiv:1807.10773 [hep-ph]} \BibitemShut {NoStop}%
\bibitem [{\citenamefont {Buschmann}\ \emph {et~al.}(2021)\citenamefont
  {Buschmann}, \citenamefont {Co}, \citenamefont {Dessert},\ and\ \citenamefont
  {Safdi}}]{Buschmann:2019pfp}%
  \BibitemOpen
  \bibfield  {author} {\bibinfo {author} {\bibfnamefont {Malte}\ \bibnamefont
  {Buschmann}}, \bibinfo {author} {\bibfnamefont {Raymond~T.}\ \bibnamefont
  {Co}}, \bibinfo {author} {\bibfnamefont {Christopher}\ \bibnamefont
  {Dessert}}, \ and\ \bibinfo {author} {\bibfnamefont {Benjamin~R.}\
  \bibnamefont {Safdi}},\ }\bibfield  {title} {\enquote {\bibinfo {title}
  {{Axion Emission Can Explain a New Hard X-Ray Excess from Nearby Isolated
  Neutron Stars}},}\ }\href {\doibase 10.1103/PhysRevLett.126.021102}
  {\bibfield  {journal} {\bibinfo  {journal} {Phys. Rev. Lett.}\ }\textbf
  {\bibinfo {volume} {126}},\ \bibinfo {pages} {021102} (\bibinfo {year}
  {2021})},\ \Eprint {http://arxiv.org/abs/1910.04164} {arXiv:1910.04164
  [hep-ph]} \BibitemShut {NoStop}%
\bibitem [{\citenamefont {Fortin}\ \emph {et~al.}(2021)\citenamefont {Fortin},
  \citenamefont {Guo}, \citenamefont {Harris}, \citenamefont {Sheridan},\ and\
  \citenamefont {Sinha}}]{Fortin:2021sst}%
  \BibitemOpen
  \bibfield  {author} {\bibinfo {author} {\bibfnamefont {Jean-Fran\c{c}ois}\
  \bibnamefont {Fortin}}, \bibinfo {author} {\bibfnamefont {Huai-Ke}\
  \bibnamefont {Guo}}, \bibinfo {author} {\bibfnamefont {Steven~P.}\
  \bibnamefont {Harris}}, \bibinfo {author} {\bibfnamefont {Elijah}\
  \bibnamefont {Sheridan}}, \ and\ \bibinfo {author} {\bibfnamefont {Kuver}\
  \bibnamefont {Sinha}},\ }\bibfield  {title} {\enquote {\bibinfo {title}
  {{Magnetars and Axion-like Particles: Probes with the Hard X-ray
  Spectrum}},}\ }\href@noop {} {\  (\bibinfo {year} {2021})},\ \Eprint
  {http://arxiv.org/abs/2101.05302} {arXiv:2101.05302 [hep-ph]} \BibitemShut
  {NoStop}%
\bibitem [{\citenamefont {O'HARE}(2020)}]{ciaran_o_hare_2020_3932430}%
  \BibitemOpen
  \bibfield  {author} {\bibinfo {author} {\bibfnamefont {Ciaran}\ \bibnamefont
  {O'HARE}},\ }\href {\doibase 10.5281/zenodo.3932430} {\enquote {\bibinfo
  {title} {cajohare/axionlimits: Axionlimits},}\ } (\bibinfo {year}
  {2020})\BibitemShut {NoStop}%
\bibitem [{\citenamefont {Dessert}\ \emph
  {et~al.}(2020{\natexlab{a}})\citenamefont {Dessert}, \citenamefont {Foster},\
  and\ \citenamefont {Safdi}}]{Dessert:2020lil}%
  \BibitemOpen
  \bibfield  {author} {\bibinfo {author} {\bibfnamefont {Christopher}\
  \bibnamefont {Dessert}}, \bibinfo {author} {\bibfnamefont {Joshua~W.}\
  \bibnamefont {Foster}}, \ and\ \bibinfo {author} {\bibfnamefont
  {Benjamin~R.}\ \bibnamefont {Safdi}},\ }\bibfield  {title} {\enquote
  {\bibinfo {title} {{X-ray Searches for Axions from Super Star Clusters}},}\
  }\href {\doibase 10.1103/PhysRevLett.125.261102} {\bibfield  {journal}
  {\bibinfo  {journal} {Phys. Rev. Lett.}\ }\textbf {\bibinfo {volume} {125}},\
  \bibinfo {pages} {261102} (\bibinfo {year} {2020}{\natexlab{a}})},\ \Eprint
  {http://arxiv.org/abs/2008.03305} {arXiv:2008.03305 [hep-ph]} \BibitemShut
  {NoStop}%
\bibitem [{\citenamefont {Xiao}\ \emph {et~al.}(2021)\citenamefont {Xiao},
  \citenamefont {Perez}, \citenamefont {Giannotti}, \citenamefont {Straniero},
  \citenamefont {Mirizzi}, \citenamefont {Grefenstette}, \citenamefont
  {Roach},\ and\ \citenamefont {Nynka}}]{Xiao:2020pra}%
  \BibitemOpen
  \bibfield  {author} {\bibinfo {author} {\bibfnamefont {Mengjiao}\
  \bibnamefont {Xiao}}, \bibinfo {author} {\bibfnamefont {Kerstin~M.}\
  \bibnamefont {Perez}}, \bibinfo {author} {\bibfnamefont {Maurizio}\
  \bibnamefont {Giannotti}}, \bibinfo {author} {\bibfnamefont {Oscar}\
  \bibnamefont {Straniero}}, \bibinfo {author} {\bibfnamefont {Alessandro}\
  \bibnamefont {Mirizzi}}, \bibinfo {author} {\bibfnamefont {Brian~W.}\
  \bibnamefont {Grefenstette}}, \bibinfo {author} {\bibfnamefont {Brandon~M.}\
  \bibnamefont {Roach}}, \ and\ \bibinfo {author} {\bibfnamefont {Melania}\
  \bibnamefont {Nynka}},\ }\bibfield  {title} {\enquote {\bibinfo {title}
  {{Constraints on Axionlike Particles from a Hard X-Ray Observation of
  Betelgeuse}},}\ }\href {\doibase 10.1103/PhysRevLett.126.031101} {\bibfield
  {journal} {\bibinfo  {journal} {Phys. Rev. Lett.}\ }\textbf {\bibinfo
  {volume} {126}},\ \bibinfo {pages} {031101} (\bibinfo {year} {2021})},\
  \Eprint {http://arxiv.org/abs/2009.09059} {arXiv:2009.09059 [astro-ph.HE]}
  \BibitemShut {NoStop}%
\bibitem [{\citenamefont {Payez}\ \emph {et~al.}(2015)\citenamefont {Payez},
  \citenamefont {Evoli}, \citenamefont {Fischer}, \citenamefont {Giannotti},
  \citenamefont {Mirizzi},\ and\ \citenamefont {Ringwald}}]{Payez:2014xsa}%
  \BibitemOpen
  \bibfield  {author} {\bibinfo {author} {\bibfnamefont {Alexandre}\
  \bibnamefont {Payez}}, \bibinfo {author} {\bibfnamefont {Carmelo}\
  \bibnamefont {Evoli}}, \bibinfo {author} {\bibfnamefont {Tobias}\
  \bibnamefont {Fischer}}, \bibinfo {author} {\bibfnamefont {Maurizio}\
  \bibnamefont {Giannotti}}, \bibinfo {author} {\bibfnamefont {Alessandro}\
  \bibnamefont {Mirizzi}}, \ and\ \bibinfo {author} {\bibfnamefont {Andreas}\
  \bibnamefont {Ringwald}},\ }\bibfield  {title} {\enquote {\bibinfo {title}
  {{Revisiting the SN1987A gamma-ray limit on ultralight axion-like
  particles}},}\ }\href {\doibase 10.1088/1475-7516/2015/02/006} {\bibfield
  {journal} {\bibinfo  {journal} {JCAP}\ }\textbf {\bibinfo {volume} {02}},\
  \bibinfo {pages} {006} (\bibinfo {year} {2015})},\ \Eprint
  {http://arxiv.org/abs/1410.3747} {arXiv:1410.3747 [astro-ph.HE]} \BibitemShut
  {NoStop}%
\bibitem [{\citenamefont {Ajello}\ \emph {et~al.}(2016)\citenamefont {Ajello}
  \emph {et~al.}}]{TheFermi-LAT:2016zue}%
  \BibitemOpen
  \bibfield  {author} {\bibinfo {author} {\bibfnamefont {M.}~\bibnamefont
  {Ajello}} \emph {et~al.} (\bibinfo {collaboration} {Fermi-LAT}),\ }\bibfield
  {title} {\enquote {\bibinfo {title} {{Search for Spectral Irregularities due
  to Photon\textendash{}Axionlike-Particle Oscillations with the Fermi Large
  Area Telescope}},}\ }\href {\doibase 10.1103/PhysRevLett.116.161101}
  {\bibfield  {journal} {\bibinfo  {journal} {Phys. Rev. Lett.}\ }\textbf
  {\bibinfo {volume} {116}},\ \bibinfo {pages} {161101} (\bibinfo {year}
  {2016})},\ \Eprint {http://arxiv.org/abs/1603.06978} {arXiv:1603.06978
  [astro-ph.HE]} \BibitemShut {NoStop}%
\bibitem [{\citenamefont {Abramowski}\ \emph {et~al.}(2013)\citenamefont
  {Abramowski} \emph {et~al.}}]{Abramowski:2013oea}%
  \BibitemOpen
  \bibfield  {author} {\bibinfo {author} {\bibfnamefont {A.}~\bibnamefont
  {Abramowski}} \emph {et~al.} (\bibinfo {collaboration} {H.E.S.S.}),\
  }\bibfield  {title} {\enquote {\bibinfo {title} {{Constraints on axionlike
  particles with H.E.S.S. from the irregularity of the PKS 2155-304 energy
  spectrum}},}\ }\href {\doibase 10.1103/PhysRevD.88.102003} {\bibfield
  {journal} {\bibinfo  {journal} {Phys. Rev. D}\ }\textbf {\bibinfo {volume}
  {88}},\ \bibinfo {pages} {102003} (\bibinfo {year} {2013})},\ \Eprint
  {http://arxiv.org/abs/1311.3148} {arXiv:1311.3148 [astro-ph.HE]} \BibitemShut
  {NoStop}%
\bibitem [{\citenamefont {Reynolds}\ \emph {et~al.}(2019)\citenamefont
  {Reynolds}, \citenamefont {Marsh}, \citenamefont {Russell}, \citenamefont
  {Fabian}, \citenamefont {Smith}, \citenamefont {Tombesi},\ and\ \citenamefont
  {Veilleux}}]{Reynolds:2019uqt}%
  \BibitemOpen
  \bibfield  {author} {\bibinfo {author} {\bibfnamefont {Christopher~S.}\
  \bibnamefont {Reynolds}}, \bibinfo {author} {\bibfnamefont {M.~C.~David}\
  \bibnamefont {Marsh}}, \bibinfo {author} {\bibfnamefont {Helen~R.}\
  \bibnamefont {Russell}}, \bibinfo {author} {\bibfnamefont {Andrew~C.}\
  \bibnamefont {Fabian}}, \bibinfo {author} {\bibfnamefont {Robyn}\
  \bibnamefont {Smith}}, \bibinfo {author} {\bibfnamefont {Francesco}\
  \bibnamefont {Tombesi}}, \ and\ \bibinfo {author} {\bibfnamefont {Sylvain}\
  \bibnamefont {Veilleux}},\ }\bibfield  {title} {\enquote {\bibinfo {title}
  {{Astrophysical limits on very light axion-like particles from Chandra
  grating spectroscopy of NGC 1275}},}\ }\href {\doibase
  10.3847/1538-4357/ab6a0c} {\  (\bibinfo {year} {2019}),\
  10.3847/1538-4357/ab6a0c},\ \Eprint {http://arxiv.org/abs/1907.05475}
  {arXiv:1907.05475 [hep-ph]} \BibitemShut {NoStop}%
\bibitem [{\citenamefont {Libanov}\ and\ \citenamefont
  {Troitsky}(2020)}]{Libanov:2019fzq}%
  \BibitemOpen
  \bibfield  {author} {\bibinfo {author} {\bibfnamefont {Maxim}\ \bibnamefont
  {Libanov}}\ and\ \bibinfo {author} {\bibfnamefont {Sergey}\ \bibnamefont
  {Troitsky}},\ }\bibfield  {title} {\enquote {\bibinfo {title} {{On the impact
  of magnetic-field models in galaxy clusters on constraints on axion-like
  particles from the lack of irregularities in high-energy spectra of
  astrophysical sources}},}\ }\href {\doibase 10.1016/j.physletb.2020.135252}
  {\bibfield  {journal} {\bibinfo  {journal} {Phys. Lett. B}\ }\textbf
  {\bibinfo {volume} {802}},\ \bibinfo {pages} {135252} (\bibinfo {year}
  {2020})},\ \Eprint {http://arxiv.org/abs/1908.03084} {arXiv:1908.03084
  [astro-ph.HE]} \BibitemShut {NoStop}%
\bibitem [{\citenamefont {Anastassopoulos}\ \emph {et~al.}(2017)\citenamefont
  {Anastassopoulos} \emph {et~al.}}]{Anastassopoulos:2017ftl}%
  \BibitemOpen
  \bibfield  {author} {\bibinfo {author} {\bibfnamefont {V.}~\bibnamefont
  {Anastassopoulos}} \emph {et~al.} (\bibinfo {collaboration} {CAST}),\
  }\bibfield  {title} {\enquote {\bibinfo {title} {{New CAST Limit on the
  Axion-Photon Interaction}},}\ }\href {\doibase 10.1038/nphys4109} {\bibfield
  {journal} {\bibinfo  {journal} {Nature Phys.}\ }\textbf {\bibinfo {volume}
  {13}},\ \bibinfo {pages} {584--590} (\bibinfo {year} {2017})},\ \Eprint
  {http://arxiv.org/abs/1705.02290} {arXiv:1705.02290 [hep-ex]} \BibitemShut
  {NoStop}%
\bibitem [{\citenamefont {Ayala}\ \emph {et~al.}(2014)\citenamefont {Ayala},
  \citenamefont {Dom\'\i{}nguez}, \citenamefont {Giannotti}, \citenamefont
  {Mirizzi},\ and\ \citenamefont {Straniero}}]{Ayala:2014pea}%
  \BibitemOpen
  \bibfield  {author} {\bibinfo {author} {\bibfnamefont {Adrian}\ \bibnamefont
  {Ayala}}, \bibinfo {author} {\bibfnamefont {Inma}\ \bibnamefont
  {Dom\'\i{}nguez}}, \bibinfo {author} {\bibfnamefont {Maurizio}\ \bibnamefont
  {Giannotti}}, \bibinfo {author} {\bibfnamefont {Alessandro}\ \bibnamefont
  {Mirizzi}}, \ and\ \bibinfo {author} {\bibfnamefont {Oscar}\ \bibnamefont
  {Straniero}},\ }\bibfield  {title} {\enquote {\bibinfo {title} {{Revisiting
  the bound on axion-photon coupling from Globular Clusters}},}\ }\href
  {\doibase 10.1103/PhysRevLett.113.191302} {\bibfield  {journal} {\bibinfo
  {journal} {Phys. Rev. Lett.}\ }\textbf {\bibinfo {volume} {113}},\ \bibinfo
  {pages} {191302} (\bibinfo {year} {2014})},\ \Eprint
  {http://arxiv.org/abs/1406.6053} {arXiv:1406.6053 [astro-ph.SR]} \BibitemShut
  {NoStop}%
\bibitem [{\citenamefont {Dine}\ \emph {et~al.}(1981)\citenamefont {Dine},
  \citenamefont {Fischler},\ and\ \citenamefont {Srednicki}}]{Dine:1981rt}%
  \BibitemOpen
  \bibfield  {author} {\bibinfo {author} {\bibfnamefont {Michael}\ \bibnamefont
  {Dine}}, \bibinfo {author} {\bibfnamefont {Willy}\ \bibnamefont {Fischler}},
  \ and\ \bibinfo {author} {\bibfnamefont {Mark}\ \bibnamefont {Srednicki}},\
  }\bibfield  {title} {\enquote {\bibinfo {title} {{A Simple Solution to the
  Strong CP Problem with a Harmless Axion}},}\ }\href {\doibase
  10.1016/0370-2693(81)90590-6} {\bibfield  {journal} {\bibinfo  {journal}
  {Phys. Lett. B}\ }\textbf {\bibinfo {volume} {104}},\ \bibinfo {pages}
  {199--202} (\bibinfo {year} {1981})}\BibitemShut {NoStop}%
\bibitem [{\citenamefont {Zhitnitsky}(1980)}]{Zhitnitsky:1980tq}%
  \BibitemOpen
  \bibfield  {author} {\bibinfo {author} {\bibfnamefont {A.~R.}\ \bibnamefont
  {Zhitnitsky}},\ }\bibfield  {title} {\enquote {\bibinfo {title} {{On Possible
  Suppression of the Axion Hadron Interactions. (In Russian)}},}\ }\href@noop
  {} {\bibfield  {journal} {\bibinfo  {journal} {Sov. J. Nucl. Phys.}\ }\textbf
  {\bibinfo {volume} {31}},\ \bibinfo {pages} {260} (\bibinfo {year}
  {1980})}\BibitemShut {NoStop}%
\bibitem [{\citenamefont {Kim}(1979)}]{Kim:1979if}%
  \BibitemOpen
  \bibfield  {author} {\bibinfo {author} {\bibfnamefont {Jihn~E.}\ \bibnamefont
  {Kim}},\ }\bibfield  {title} {\enquote {\bibinfo {title} {{Weak Interaction
  Singlet and Strong CP Invariance}},}\ }\href {\doibase
  10.1103/PhysRevLett.43.103} {\bibfield  {journal} {\bibinfo  {journal} {Phys.
  Rev. Lett.}\ }\textbf {\bibinfo {volume} {43}},\ \bibinfo {pages} {103}
  (\bibinfo {year} {1979})}\BibitemShut {NoStop}%
\bibitem [{\citenamefont {Shifman}\ \emph {et~al.}(1980)\citenamefont
  {Shifman}, \citenamefont {Vainshtein},\ and\ \citenamefont
  {Zakharov}}]{Shifman:1979if}%
  \BibitemOpen
  \bibfield  {author} {\bibinfo {author} {\bibfnamefont {Mikhail~A.}\
  \bibnamefont {Shifman}}, \bibinfo {author} {\bibfnamefont {A.~I.}\
  \bibnamefont {Vainshtein}}, \ and\ \bibinfo {author} {\bibfnamefont
  {Valentin~I.}\ \bibnamefont {Zakharov}},\ }\bibfield  {title} {\enquote
  {\bibinfo {title} {{Can Confinement Ensure Natural CP Invariance of Strong
  Interactions?}}}\ }\href {\doibase 10.1016/0550-3213(80)90209-6} {\bibfield
  {journal} {\bibinfo  {journal} {Nucl. Phys. B}\ }\textbf {\bibinfo {volume}
  {166}},\ \bibinfo {pages} {493--506} (\bibinfo {year} {1980})}\BibitemShut
  {NoStop}%
\bibitem [{\citenamefont {Gramolin}\ \emph {et~al.}(2021)\citenamefont
  {Gramolin}, \citenamefont {Aybas}, \citenamefont {Johnson}, \citenamefont
  {Adam},\ and\ \citenamefont {Sushkov}}]{Gramolin:2020ict}%
  \BibitemOpen
  \bibfield  {author} {\bibinfo {author} {\bibfnamefont {Alexander~V.}\
  \bibnamefont {Gramolin}}, \bibinfo {author} {\bibfnamefont {Deniz}\
  \bibnamefont {Aybas}}, \bibinfo {author} {\bibfnamefont {Dorian}\
  \bibnamefont {Johnson}}, \bibinfo {author} {\bibfnamefont {Janos}\
  \bibnamefont {Adam}}, \ and\ \bibinfo {author} {\bibfnamefont {Alexander~O.}\
  \bibnamefont {Sushkov}},\ }\bibfield  {title} {\enquote {\bibinfo {title}
  {{Search for axion-like dark matter with ferromagnets}},}\ }\href {\doibase
  10.1038/s41567-020-1006-6} {\bibfield  {journal} {\bibinfo  {journal} {Nature
  Phys.}\ }\textbf {\bibinfo {volume} {17}},\ \bibinfo {pages} {79--84}
  (\bibinfo {year} {2021})},\ \Eprint {http://arxiv.org/abs/2003.03348}
  {arXiv:2003.03348 [hep-ex]} \BibitemShut {NoStop}%
\bibitem [{\citenamefont {Ouellet}\ \emph {et~al.}(2019)\citenamefont {Ouellet}
  \emph {et~al.}}]{Ouellet:2018beu}%
  \BibitemOpen
  \bibfield  {author} {\bibinfo {author} {\bibfnamefont {Jonathan~L.}\
  \bibnamefont {Ouellet}} \emph {et~al.},\ }\bibfield  {title} {\enquote
  {\bibinfo {title} {{First Results from ABRACADABRA-10 cm: A Search for
  Sub-$\mu$eV Axion Dark Matter}},}\ }\href {\doibase
  10.1103/PhysRevLett.122.121802} {\bibfield  {journal} {\bibinfo  {journal}
  {Phys. Rev. Lett.}\ }\textbf {\bibinfo {volume} {122}},\ \bibinfo {pages}
  {121802} (\bibinfo {year} {2019})},\ \Eprint
  {http://arxiv.org/abs/1810.12257} {arXiv:1810.12257 [hep-ex]} \BibitemShut
  {NoStop}%
\bibitem [{\citenamefont {Salemi}\ \emph {et~al.}(2021)\citenamefont {Salemi}
  \emph {et~al.}}]{Salemi:2021gck}%
  \BibitemOpen
  \bibfield  {author} {\bibinfo {author} {\bibfnamefont {Chiara~P.}\
  \bibnamefont {Salemi}} \emph {et~al.},\ }\bibfield  {title} {\enquote
  {\bibinfo {title} {{The search for low-mass axion dark matter with
  ABRACADABRA-10cm}},}\ }\href@noop {} {\  (\bibinfo {year} {2021})},\ \Eprint
  {http://arxiv.org/abs/2102.06722} {arXiv:2102.06722 [hep-ex]} \BibitemShut
  {NoStop}%
\bibitem [{\citenamefont {Du}\ \emph {et~al.}(2018)\citenamefont {Du} \emph
  {et~al.}}]{Du:2018uak}%
  \BibitemOpen
  \bibfield  {author} {\bibinfo {author} {\bibfnamefont {N.}~\bibnamefont {Du}}
  \emph {et~al.} (\bibinfo {collaboration} {ADMX}),\ }\bibfield  {title}
  {\enquote {\bibinfo {title} {{A Search for Invisible Axion Dark Matter with
  the Axion Dark Matter Experiment}},}\ }\href {\doibase
  10.1103/PhysRevLett.120.151301} {\bibfield  {journal} {\bibinfo  {journal}
  {Phys. Rev. Lett.}\ }\textbf {\bibinfo {volume} {120}},\ \bibinfo {pages}
  {151301} (\bibinfo {year} {2018})},\ \Eprint
  {http://arxiv.org/abs/1804.05750} {arXiv:1804.05750 [hep-ex]} \BibitemShut
  {NoStop}%
\bibitem [{\citenamefont {Braine}\ \emph {et~al.}(2020)\citenamefont {Braine}
  \emph {et~al.}}]{Braine:2019fqb}%
  \BibitemOpen
  \bibfield  {author} {\bibinfo {author} {\bibfnamefont {T.}~\bibnamefont
  {Braine}} \emph {et~al.} (\bibinfo {collaboration} {ADMX}),\ }\bibfield
  {title} {\enquote {\bibinfo {title} {{Extended Search for the Invisible Axion
  with the Axion Dark Matter Experiment}},}\ }\href {\doibase
  10.1103/PhysRevLett.124.101303} {\bibfield  {journal} {\bibinfo  {journal}
  {Phys. Rev. Lett.}\ }\textbf {\bibinfo {volume} {124}},\ \bibinfo {pages}
  {101303} (\bibinfo {year} {2020})},\ \Eprint
  {http://arxiv.org/abs/1910.08638} {arXiv:1910.08638 [hep-ex]} \BibitemShut
  {NoStop}%
\bibitem [{\citenamefont {Zhong}\ \emph {et~al.}(2018)\citenamefont {Zhong}
  \emph {et~al.}}]{Zhong:2018rsr}%
  \BibitemOpen
  \bibfield  {author} {\bibinfo {author} {\bibfnamefont {L.}~\bibnamefont
  {Zhong}} \emph {et~al.} (\bibinfo {collaboration} {HAYSTAC}),\ }\bibfield
  {title} {\enquote {\bibinfo {title} {{Results from phase 1 of the HAYSTAC
  microwave cavity axion experiment}},}\ }\href {\doibase
  10.1103/PhysRevD.97.092001} {\bibfield  {journal} {\bibinfo  {journal} {Phys.
  Rev. D}\ }\textbf {\bibinfo {volume} {97}},\ \bibinfo {pages} {092001}
  (\bibinfo {year} {2018})},\ \Eprint {http://arxiv.org/abs/1803.03690}
  {arXiv:1803.03690 [hep-ex]} \BibitemShut {NoStop}%
\bibitem [{\citenamefont {Backes}\ \emph {et~al.}(2021)\citenamefont {Backes}
  \emph {et~al.}}]{Backes:2020ajv}%
  \BibitemOpen
  \bibfield  {author} {\bibinfo {author} {\bibfnamefont {K.~M.}\ \bibnamefont
  {Backes}} \emph {et~al.} (\bibinfo {collaboration} {HAYSTAC}),\ }\bibfield
  {title} {\enquote {\bibinfo {title} {{A quantum-enhanced search for dark
  matter axions}},}\ }\href {\doibase 10.1038/s41586-021-03226-7} {\bibfield
  {journal} {\bibinfo  {journal} {Nature}\ }\textbf {\bibinfo {volume} {590}},\
  \bibinfo {pages} {238--242} (\bibinfo {year} {2021})},\ \Eprint
  {http://arxiv.org/abs/2008.01853} {arXiv:2008.01853 [quant-ph]} \BibitemShut
  {NoStop}%
\bibitem [{\citenamefont {Jeong}\ \emph {et~al.}(2020)\citenamefont {Jeong},
  \citenamefont {Youn}, \citenamefont {Bae}, \citenamefont {Kim}, \citenamefont
  {Seong}, \citenamefont {Kim},\ and\ \citenamefont
  {Semertzidis}}]{Jeong:2020cwz}%
  \BibitemOpen
  \bibfield  {author} {\bibinfo {author} {\bibfnamefont {Junu}\ \bibnamefont
  {Jeong}}, \bibinfo {author} {\bibfnamefont {SungWoo}\ \bibnamefont {Youn}},
  \bibinfo {author} {\bibfnamefont {Sungjae}\ \bibnamefont {Bae}}, \bibinfo
  {author} {\bibfnamefont {Jihngeun}\ \bibnamefont {Kim}}, \bibinfo {author}
  {\bibfnamefont {Taehyeon}\ \bibnamefont {Seong}}, \bibinfo {author}
  {\bibfnamefont {Jihn~E.}\ \bibnamefont {Kim}}, \ and\ \bibinfo {author}
  {\bibfnamefont {Yannis~K.}\ \bibnamefont {Semertzidis}},\ }\bibfield  {title}
  {\enquote {\bibinfo {title} {{Search for Invisible Axion Dark Matter with a
  Multiple-Cell Haloscope}},}\ }\href {\doibase 10.1103/PhysRevLett.125.221302}
  {\bibfield  {journal} {\bibinfo  {journal} {Phys. Rev. Lett.}\ }\textbf
  {\bibinfo {volume} {125}},\ \bibinfo {pages} {221302} (\bibinfo {year}
  {2020})},\ \Eprint {http://arxiv.org/abs/2008.10141} {arXiv:2008.10141
  [hep-ex]} \BibitemShut {NoStop}%
\bibitem [{\citenamefont {Alesini}\ \emph {et~al.}(2020)\citenamefont {Alesini}
  \emph {et~al.}}]{Alesini:2020vny}%
  \BibitemOpen
  \bibfield  {author} {\bibinfo {author} {\bibfnamefont {D.}~\bibnamefont
  {Alesini}} \emph {et~al.},\ }\bibfield  {title} {\enquote {\bibinfo {title}
  {{Search for Invisible Axion Dark Matter of mass m$_a=43~\mu$eV with the
  QUAX--$a\gamma$ Experiment}},}\ }\href@noop {} {\  (\bibinfo {year}
  {2020})},\ \Eprint {http://arxiv.org/abs/2012.09498} {arXiv:2012.09498
  [hep-ex]} \BibitemShut {NoStop}%
\bibitem [{\citenamefont {McAllister}\ \emph {et~al.}(2017)\citenamefont
  {McAllister}, \citenamefont {Flower}, \citenamefont {Ivanov}, \citenamefont
  {Goryachev}, \citenamefont {Bourhill},\ and\ \citenamefont
  {Tobar}}]{McAllister:2017lkb}%
  \BibitemOpen
  \bibfield  {author} {\bibinfo {author} {\bibfnamefont {Ben~T.}\ \bibnamefont
  {McAllister}}, \bibinfo {author} {\bibfnamefont {Graeme}\ \bibnamefont
  {Flower}}, \bibinfo {author} {\bibfnamefont {Eugene~N.}\ \bibnamefont
  {Ivanov}}, \bibinfo {author} {\bibfnamefont {Maxim}\ \bibnamefont
  {Goryachev}}, \bibinfo {author} {\bibfnamefont {Jeremy}\ \bibnamefont
  {Bourhill}}, \ and\ \bibinfo {author} {\bibfnamefont {Michael~E.}\
  \bibnamefont {Tobar}},\ }\bibfield  {title} {\enquote {\bibinfo {title} {{The
  ORGAN Experiment: An axion haloscope above 15 GHz}},}\ }\href {\doibase
  10.1016/j.dark.2017.09.010} {\bibfield  {journal} {\bibinfo  {journal} {Phys.
  Dark Univ.}\ }\textbf {\bibinfo {volume} {18}},\ \bibinfo {pages} {67--72}
  (\bibinfo {year} {2017})},\ \Eprint {http://arxiv.org/abs/1706.00209}
  {arXiv:1706.00209 [physics.ins-det]} \BibitemShut {NoStop}%
\bibitem [{\citenamefont {{Gaia Collaboration}}\ \emph
  {et~al.}(2020)\citenamefont {{Gaia Collaboration}}, \citenamefont {{Brown}},
  \citenamefont {{Vallenari}}, \citenamefont {{Prusti}}, \citenamefont {{de
  Bruijne}}, \citenamefont {{Babusiaux}},\ and\ \citenamefont
  {{Biermann}}}]{2020arXiv201201533G}%
  \BibitemOpen
  \bibfield  {author} {\bibinfo {author} {\bibnamefont {{Gaia Collaboration}}},
  \bibinfo {author} {\bibfnamefont {A.~G.~A.}\ \bibnamefont {{Brown}}},
  \bibinfo {author} {\bibfnamefont {A.}~\bibnamefont {{Vallenari}}}, \bibinfo
  {author} {\bibfnamefont {T.}~\bibnamefont {{Prusti}}}, \bibinfo {author}
  {\bibfnamefont {J.~H.~J.}\ \bibnamefont {{de Bruijne}}}, \bibinfo {author}
  {\bibfnamefont {C.}~\bibnamefont {{Babusiaux}}}, \ and\ \bibinfo {author}
  {\bibfnamefont {M.}~\bibnamefont {{Biermann}}},\ }\bibfield  {title}
  {\enquote {\bibinfo {title} {{Gaia Early Data Release 3: Summary of the
  contents and survey properties}},}\ }\href@noop {} {\bibfield  {journal}
  {\bibinfo  {journal} {arXiv e-prints}\ ,\ \bibinfo {eid} {arXiv:2012.01533}}
  (\bibinfo {year} {2020})},\ \Eprint {http://arxiv.org/abs/2012.01533}
  {arXiv:2012.01533 [astro-ph.GA]} \BibitemShut {NoStop}%
\bibitem [{\citenamefont {Cowan}\ \emph
  {et~al.}(2011{\natexlab{a}})\citenamefont {Cowan}, \citenamefont {Cranmer},
  \citenamefont {Gross},\ and\ \citenamefont {Vitells}}]{Cowan:2010js}%
  \BibitemOpen
  \bibfield  {author} {\bibinfo {author} {\bibfnamefont {Glen}\ \bibnamefont
  {Cowan}}, \bibinfo {author} {\bibfnamefont {Kyle}\ \bibnamefont {Cranmer}},
  \bibinfo {author} {\bibfnamefont {Eilam}\ \bibnamefont {Gross}}, \ and\
  \bibinfo {author} {\bibfnamefont {Ofer}\ \bibnamefont {Vitells}},\ }\bibfield
   {title} {\enquote {\bibinfo {title} {{Asymptotic formulae for
  likelihood-based tests of new physics}},}\ }\href {\doibase
  10.1140/epjc/s10052-011-1554-0} {\bibfield  {journal} {\bibinfo  {journal}
  {Eur. Phys. J. C}\ }\textbf {\bibinfo {volume} {71}},\ \bibinfo {pages}
  {1554} (\bibinfo {year} {2011}{\natexlab{a}})},\ \bibinfo {note} {[Erratum:
  Eur.Phys.J.C 73, 2501 (2013)]},\ \Eprint {http://arxiv.org/abs/1007.1727}
  {arXiv:1007.1727 [physics.data-an]} \BibitemShut {NoStop}%
\bibitem [{\citenamefont {Cowan}\ \emph
  {et~al.}(2011{\natexlab{b}})\citenamefont {Cowan}, \citenamefont {Cranmer},
  \citenamefont {Gross},\ and\ \citenamefont {Vitells}}]{Cowan:2011an}%
  \BibitemOpen
  \bibfield  {author} {\bibinfo {author} {\bibfnamefont {Glen}\ \bibnamefont
  {Cowan}}, \bibinfo {author} {\bibfnamefont {Kyle}\ \bibnamefont {Cranmer}},
  \bibinfo {author} {\bibfnamefont {Eilam}\ \bibnamefont {Gross}}, \ and\
  \bibinfo {author} {\bibfnamefont {Ofer}\ \bibnamefont {Vitells}},\ }\bibfield
   {title} {\enquote {\bibinfo {title} {{Power-Constrained Limits}},}\
  }\href@noop {} {\  (\bibinfo {year} {2011}{\natexlab{b}})},\ \Eprint
  {http://arxiv.org/abs/1105.3166} {arXiv:1105.3166 [physics.data-an]}
  \BibitemShut {NoStop}%
\bibitem [{\citenamefont {Nakagawa}\ \emph {et~al.}(1987)\citenamefont
  {Nakagawa}, \citenamefont {Kohyama},\ and\ \citenamefont
  {Itoh}}]{Nakagawa:1987pga}%
  \BibitemOpen
  \bibfield  {author} {\bibinfo {author} {\bibfnamefont {Masayuki}\
  \bibnamefont {Nakagawa}}, \bibinfo {author} {\bibfnamefont {Yasuharu}\
  \bibnamefont {Kohyama}}, \ and\ \bibinfo {author} {\bibfnamefont {Naoki}\
  \bibnamefont {Itoh}},\ }\bibfield  {title} {\enquote {\bibinfo {title}
  {{Axion Bremsstrahlung in Dense Stars}},}\ }\href {\doibase 10.1086/165724}
  {\bibfield  {journal} {\bibinfo  {journal} {Astrophys. J.}\ }\textbf
  {\bibinfo {volume} {322}},\ \bibinfo {pages} {291} (\bibinfo {year}
  {1987})}\BibitemShut {NoStop}%
\bibitem [{\citenamefont {Paxton}\ \emph {et~al.}(2010)\citenamefont {Paxton},
  \citenamefont {Bildsten}, \citenamefont {Dotter}, \citenamefont {Herwig},
  \citenamefont {Lesaffre},\ and\ \citenamefont {Timmes}}]{Paxton2010}%
  \BibitemOpen
  \bibfield  {author} {\bibinfo {author} {\bibfnamefont {Bill}\ \bibnamefont
  {Paxton}}, \bibinfo {author} {\bibfnamefont {Lars}\ \bibnamefont {Bildsten}},
  \bibinfo {author} {\bibfnamefont {Aaron}\ \bibnamefont {Dotter}}, \bibinfo
  {author} {\bibfnamefont {Falk}\ \bibnamefont {Herwig}}, \bibinfo {author}
  {\bibfnamefont {Pierre}\ \bibnamefont {Lesaffre}}, \ and\ \bibinfo {author}
  {\bibfnamefont {Frank}\ \bibnamefont {Timmes}},\ }\bibfield  {title}
  {\enquote {\bibinfo {title} {Modules for experiments in stellar astrophysics
  (mesa)},}\ }\href {\doibase 10.1088/0067-0049/192/1/3} {\bibfield  {journal}
  {\bibinfo  {journal} {The Astrophysical Journal Supplement Series}\ }\textbf
  {\bibinfo {volume} {192}},\ \bibinfo {pages} {3} (\bibinfo {year}
  {2010})}\BibitemShut {NoStop}%
\bibitem [{\citenamefont {Kulebi}\ \emph {et~al.}(2010)\citenamefont {Kulebi},
  \citenamefont {Jordan}, \citenamefont {Nelan}, \citenamefont {Bastian},\ and\
  \citenamefont {Altmann}}]{Kulebi:2010pd}%
  \BibitemOpen
  \bibfield  {author} {\bibinfo {author} {\bibfnamefont {Baybars}\ \bibnamefont
  {Kulebi}}, \bibinfo {author} {\bibfnamefont {Stefan}\ \bibnamefont {Jordan}},
  \bibinfo {author} {\bibfnamefont {Edmund}\ \bibnamefont {Nelan}}, \bibinfo
  {author} {\bibfnamefont {Ulrich}\ \bibnamefont {Bastian}}, \ and\ \bibinfo
  {author} {\bibfnamefont {Martin}\ \bibnamefont {Altmann}},\ }\bibfield
  {title} {\enquote {\bibinfo {title} {{Constraints on the origin of the
  massive, hot, and rapidly rotating magnetic white dwarf RE J 0317-853 from an
  HST parallax measurement}},}\ }\href {\doibase 10.1051/0004-6361/201015237}
  {\bibfield  {journal} {\bibinfo  {journal} {Astron. Astrophys.}\ }\textbf
  {\bibinfo {volume} {524}},\ \bibinfo {pages} {A36} (\bibinfo {year}
  {2010})},\ \Eprint {http://arxiv.org/abs/1007.4978} {arXiv:1007.4978
  [astro-ph.SR]} \BibitemShut {NoStop}%
\bibitem [{\citenamefont {{Camisassa}}\ \emph {et~al.}(2019)\citenamefont
  {{Camisassa}}, \citenamefont {{Althaus}}, \citenamefont {{C{\'o}rsico}},
  \citenamefont {{De Ger{\'o}nimo}}, \citenamefont {{Miller Bertolami}},
  \citenamefont {{Novarino}}, \citenamefont {{Rohrmann}}, \citenamefont
  {{Wachlin}},\ and\ \citenamefont {{Garc{\'\i}a-Berro}}}]{2019AA...625A..87C}%
  \BibitemOpen
  \bibfield  {author} {\bibinfo {author} {\bibfnamefont {Mar{\'\i}a~E.}\
  \bibnamefont {{Camisassa}}}, \bibinfo {author} {\bibfnamefont {Leandro~G.}\
  \bibnamefont {{Althaus}}}, \bibinfo {author} {\bibfnamefont {Alejandro~H.}\
  \bibnamefont {{C{\'o}rsico}}}, \bibinfo {author} {\bibfnamefont
  {Francisco~C.}\ \bibnamefont {{De Ger{\'o}nimo}}}, \bibinfo {author}
  {\bibfnamefont {Marcelo~M.}\ \bibnamefont {{Miller Bertolami}}}, \bibinfo
  {author} {\bibfnamefont {Mar{\'\i}a~L.}\ \bibnamefont {{Novarino}}}, \bibinfo
  {author} {\bibfnamefont {Ren{\'e}~D.}\ \bibnamefont {{Rohrmann}}}, \bibinfo
  {author} {\bibfnamefont {Felipe~C.}\ \bibnamefont {{Wachlin}}}, \ and\
  \bibinfo {author} {\bibfnamefont {Enrique}\ \bibnamefont
  {{Garc{\'\i}a-Berro}}},\ }\bibfield  {title} {\enquote {\bibinfo {title}
  {{The evolution of ultra-massive white dwarfs}},}\ }\href {\doibase
  10.1051/0004-6361/201833822} {\bibfield  {journal} {\bibinfo  {journal}
  {\aap}\ }\textbf {\bibinfo {volume} {625}},\ \bibinfo {eid} {A87} (\bibinfo
  {year} {2019})},\ \Eprint {http://arxiv.org/abs/1807.03894} {arXiv:1807.03894
  [astro-ph.SR]} \BibitemShut {NoStop}%
\bibitem [{\citenamefont {Brown}\ \emph {et~al.}(2018)\citenamefont {Brown},
  \citenamefont {Vallenari}, \citenamefont {Prusti}, \citenamefont
  {de~Bruijne}, \citenamefont {Babusiaux}, \citenamefont {Bailer-Jones},
  \citenamefont {Biermann}, \citenamefont {Evans}, \citenamefont {Eyer},\ and\
  \citenamefont {et~al.}}]{GaiaDR2}%
  \BibitemOpen
  \bibfield  {author} {\bibinfo {author} {\bibfnamefont {A.~G.~A.}\
  \bibnamefont {Brown}}, \bibinfo {author} {\bibfnamefont {A.}~\bibnamefont
  {Vallenari}}, \bibinfo {author} {\bibfnamefont {T.}~\bibnamefont {Prusti}},
  \bibinfo {author} {\bibfnamefont {J.~H.~J.}\ \bibnamefont {de~Bruijne}},
  \bibinfo {author} {\bibfnamefont {C.}~\bibnamefont {Babusiaux}}, \bibinfo
  {author} {\bibfnamefont {C.~A.~L.}\ \bibnamefont {Bailer-Jones}}, \bibinfo
  {author} {\bibfnamefont {M.}~\bibnamefont {Biermann}}, \bibinfo {author}
  {\bibfnamefont {D.~W.}\ \bibnamefont {Evans}}, \bibinfo {author}
  {\bibfnamefont {L.}~\bibnamefont {Eyer}}, \ and\ \bibinfo {author}
  {\bibnamefont {et~al.}},\ }\bibfield  {title} {\enquote {\bibinfo {title}
  {Gaia data release 2},}\ }\href {\doibase 10.1051/0004-6361/201833051}
  {\bibfield  {journal} {\bibinfo  {journal} {Astronomy \& Astrophysics}\
  }\textbf {\bibinfo {volume} {616}},\ \bibinfo {pages} {A1} (\bibinfo {year}
  {2018})}\BibitemShut {NoStop}%
\bibitem [{\citenamefont {Ichimaru}(1982)}]{Ichimaru:1982zz}%
  \BibitemOpen
  \bibfield  {author} {\bibinfo {author} {\bibfnamefont {Setsuo}\ \bibnamefont
  {Ichimaru}},\ }\bibfield  {title} {\enquote {\bibinfo {title} {{Strongly
  coupled plasmas: high-density classical plasmas and degenerate electron
  liquids}},}\ }\href {\doibase 10.1103/RevModPhys.54.1017} {\bibfield
  {journal} {\bibinfo  {journal} {Rev. Mod. Phys.}\ }\textbf {\bibinfo {volume}
  {54}},\ \bibinfo {pages} {1017--1059} (\bibinfo {year} {1982})}\BibitemShut
  {NoStop}%
\bibitem [{\citenamefont {Nakagawa}\ \emph {et~al.}(1988)\citenamefont
  {Nakagawa}, \citenamefont {Adachi}, \citenamefont {Kohyama},\ and\
  \citenamefont {Itoh}}]{Nakagawa:1988rhp}%
  \BibitemOpen
  \bibfield  {author} {\bibinfo {author} {\bibfnamefont {Masayuki}\
  \bibnamefont {Nakagawa}}, \bibinfo {author} {\bibfnamefont {Tomoo}\
  \bibnamefont {Adachi}}, \bibinfo {author} {\bibfnamefont {Yasuharu}\
  \bibnamefont {Kohyama}}, \ and\ \bibinfo {author} {\bibfnamefont {Naoki}\
  \bibnamefont {Itoh}},\ }\bibfield  {title} {\enquote {\bibinfo {title}
  {{Axion bremsstrahlung in dense stars. II - Phonon contributions}},}\ }\href
  {\doibase 10.1086/166085} {\bibfield  {journal} {\bibinfo  {journal}
  {Astrophys. J.}\ }\textbf {\bibinfo {volume} {326}},\ \bibinfo {pages} {241}
  (\bibinfo {year} {1988})}\BibitemShut {NoStop}%
\bibitem [{\citenamefont {Burleigh}\ \emph {et~al.}(1999)\citenamefont
  {Burleigh}, \citenamefont {Jordan},\ and\ \citenamefont
  {Schweizer}}]{Burleigh:1998pqa}%
  \BibitemOpen
  \bibfield  {author} {\bibinfo {author} {\bibfnamefont {M.~R.}\ \bibnamefont
  {Burleigh}}, \bibinfo {author} {\bibfnamefont {S.}~\bibnamefont {Jordan}}, \
  and\ \bibinfo {author} {\bibfnamefont {W.}~\bibnamefont {Schweizer}},\
  }\bibfield  {title} {\enquote {\bibinfo {title} {{Phase-resolved
  far-ultraviolet hst spectroscopy of the peculiar magnetic white dwarf re
  j0317-853}},}\ }\href {\doibase 10.1086/311794} {\bibfield  {journal}
  {\bibinfo  {journal} {Astrophys. J. Lett.}\ }\textbf {\bibinfo {volume}
  {510}},\ \bibinfo {pages} {L37} (\bibinfo {year} {1999})},\ \Eprint
  {http://arxiv.org/abs/astro-ph/9810109} {arXiv:astro-ph/9810109} \BibitemShut
  {NoStop}%
\bibitem [{\citenamefont {Miller~Bertolami}\ \emph {et~al.}(2014)\citenamefont
  {Miller~Bertolami}, \citenamefont {Melendez}, \citenamefont {Althaus},\ and\
  \citenamefont {Isern}}]{Bertolami:2014wua}%
  \BibitemOpen
  \bibfield  {author} {\bibinfo {author} {\bibfnamefont {Marcelo~M.}\
  \bibnamefont {Miller~Bertolami}}, \bibinfo {author} {\bibfnamefont
  {Brenda~E.}\ \bibnamefont {Melendez}}, \bibinfo {author} {\bibfnamefont
  {Leandro~G.}\ \bibnamefont {Althaus}}, \ and\ \bibinfo {author}
  {\bibfnamefont {Jordi}\ \bibnamefont {Isern}},\ }\bibfield  {title} {\enquote
  {\bibinfo {title} {{Revisiting the axion bounds from the Galactic white dwarf
  luminosity function}},}\ }\href {\doibase 10.1088/1475-7516/2014/10/069}
  {\bibfield  {journal} {\bibinfo  {journal} {JCAP}\ }\textbf {\bibinfo
  {volume} {1410}},\ \bibinfo {pages} {069} (\bibinfo {year} {2014})},\ \Eprint
  {http://arxiv.org/abs/1406.7712} {arXiv:1406.7712 [hep-ph]} \BibitemShut
  {NoStop}%
%%CITATION = ARXIV:1406.7712;%%
\bibitem [{\citenamefont {Srednicki}(1985)}]{Srednicki:1985xd}%
  \BibitemOpen
  \bibfield  {author} {\bibinfo {author} {\bibfnamefont {Mark}\ \bibnamefont
  {Srednicki}},\ }\bibfield  {title} {\enquote {\bibinfo {title} {{Axion
  Couplings to Matter. 1. CP Conserving Parts}},}\ }\href {\doibase
  10.1016/0550-3213(85)90054-9} {\bibfield  {journal} {\bibinfo  {journal}
  {Nucl. Phys. B}\ }\textbf {\bibinfo {volume} {260}},\ \bibinfo {pages}
  {689--700} (\bibinfo {year} {1985})}\BibitemShut {NoStop}%
\bibitem [{\citenamefont {Chang}\ and\ \citenamefont
  {Choi}(1993)}]{Chang:1993gm}%
  \BibitemOpen
  \bibfield  {author} {\bibinfo {author} {\bibfnamefont {Sanghyeon}\
  \bibnamefont {Chang}}\ and\ \bibinfo {author} {\bibfnamefont {Kiwoon}\
  \bibnamefont {Choi}},\ }\bibfield  {title} {\enquote {\bibinfo {title}
  {{Hadronic axion window and the big bang nucleosynthesis}},}\ }\href
  {\doibase 10.1016/0370-2693(93)90656-3} {\bibfield  {journal} {\bibinfo
  {journal} {Phys. Lett. B}\ }\textbf {\bibinfo {volume} {316}},\ \bibinfo
  {pages} {51--56} (\bibinfo {year} {1993})},\ \Eprint
  {http://arxiv.org/abs/hep-ph/9306216} {arXiv:hep-ph/9306216} \BibitemShut
  {NoStop}%
\bibitem [{\citenamefont {Meyer}\ \emph {et~al.}(2013)\citenamefont {Meyer},
  \citenamefont {Horns},\ and\ \citenamefont {Raue}}]{Meyer:2013pny}%
  \BibitemOpen
  \bibfield  {author} {\bibinfo {author} {\bibfnamefont {Manuel}\ \bibnamefont
  {Meyer}}, \bibinfo {author} {\bibfnamefont {Dieter}\ \bibnamefont {Horns}}, \
  and\ \bibinfo {author} {\bibfnamefont {Martin}\ \bibnamefont {Raue}},\
  }\bibfield  {title} {\enquote {\bibinfo {title} {{First lower limits on the
  photon-axion-like particle coupling from very high energy gamma-ray
  observations}},}\ }\href {\doibase 10.1103/PhysRevD.87.035027} {\bibfield
  {journal} {\bibinfo  {journal} {Phys. Rev. D}\ }\textbf {\bibinfo {volume}
  {87}},\ \bibinfo {pages} {035027} (\bibinfo {year} {2013})},\ \Eprint
  {http://arxiv.org/abs/1302.1208} {arXiv:1302.1208 [astro-ph.HE]} \BibitemShut
  {NoStop}%
\bibitem [{\citenamefont {Dessert}\ \emph
  {et~al.}(2020{\natexlab{b}})\citenamefont {Dessert}, \citenamefont {Foster},\
  and\ \citenamefont {Safdi}}]{Dessert:2019dos}%
  \BibitemOpen
  \bibfield  {author} {\bibinfo {author} {\bibfnamefont {Christopher}\
  \bibnamefont {Dessert}}, \bibinfo {author} {\bibfnamefont {Joshua~W.}\
  \bibnamefont {Foster}}, \ and\ \bibinfo {author} {\bibfnamefont
  {Benjamin~R.}\ \bibnamefont {Safdi}},\ }\bibfield  {title} {\enquote
  {\bibinfo {title} {{Hard X-ray Excess from the Magnificent Seven Neutron
  Stars}},}\ }\href {\doibase 10.3847/1538-4357/abb4ea} {\bibfield  {journal}
  {\bibinfo  {journal} {Astrophys. J.}\ }\textbf {\bibinfo {volume} {904}},\
  \bibinfo {pages} {42} (\bibinfo {year} {2020}{\natexlab{b}})},\ \Eprint
  {http://arxiv.org/abs/1910.02956} {arXiv:1910.02956 [astro-ph.HE]}
  \BibitemShut {NoStop}%
\bibitem [{\citenamefont {B\"ahre}\ \emph {et~al.}(2013)\citenamefont {B\"ahre}
  \emph {et~al.}}]{Bahre:2013ywa}%
  \BibitemOpen
  \bibfield  {author} {\bibinfo {author} {\bibfnamefont {Robin}\ \bibnamefont
  {B\"ahre}} \emph {et~al.},\ }\bibfield  {title} {\enquote {\bibinfo {title}
  {{Any light particle search II \textemdash{}Technical Design Report}},}\
  }\href {\doibase 10.1088/1748-0221/8/09/T09001} {\bibfield  {journal}
  {\bibinfo  {journal} {JINST}\ }\textbf {\bibinfo {volume} {8}},\ \bibinfo
  {pages} {T09001} (\bibinfo {year} {2013})},\ \Eprint
  {http://arxiv.org/abs/1302.5647} {arXiv:1302.5647 [physics.ins-det]}
  \BibitemShut {NoStop}%
\bibitem [{Lyn(2018)}]{LynxTeam:2018usc}%
  \BibitemOpen
  \bibfield  {title} {\enquote {\bibinfo {title} {{The Lynx Mission Concept
  Study Interim Report}},}\ }\href@noop {} {\  (\bibinfo {year} {2018})},\
  \Eprint {http://arxiv.org/abs/1809.09642} {arXiv:1809.09642 [astro-ph.IM]}
  \BibitemShut {NoStop}%
\bibitem [{\citenamefont {{Fruscione}}\ \emph {et~al.}(2006)\citenamefont
  {{Fruscione}}, \citenamefont {{McDowell}}, \citenamefont {{Allen}},
  \citenamefont {{Brickhouse}}, \citenamefont {{Burke}}, \citenamefont
  {{Davis}}, \citenamefont {{Durham}}, \citenamefont {{Elvis}}, \citenamefont
  {{Galle}}, \citenamefont {{Harris}}, \citenamefont {{Huenemoerder}},
  \citenamefont {{Houck}}, \citenamefont {{Ishibashi}}, \citenamefont
  {{Karovska}}, \citenamefont {{Nicastro}}, \citenamefont {{Noble}},
  \citenamefont {{Nowak}}, \citenamefont {{Primini}}, \citenamefont
  {{Siemiginowska}}, \citenamefont {{Smith}},\ and\ \citenamefont
  {{Wise}}}]{2006SPIE.6270E..1VF}%
  \BibitemOpen
  \bibfield  {author} {\bibinfo {author} {\bibfnamefont {Antonella}\
  \bibnamefont {{Fruscione}}}, \bibinfo {author} {\bibfnamefont {Jonathan~C.}\
  \bibnamefont {{McDowell}}}, \bibinfo {author} {\bibfnamefont {Glenn~E.}\
  \bibnamefont {{Allen}}}, \bibinfo {author} {\bibfnamefont {Nancy~S.}\
  \bibnamefont {{Brickhouse}}}, \bibinfo {author} {\bibfnamefont {Douglas~J.}\
  \bibnamefont {{Burke}}}, \bibinfo {author} {\bibfnamefont {John~E.}\
  \bibnamefont {{Davis}}}, \bibinfo {author} {\bibfnamefont {Nick}\
  \bibnamefont {{Durham}}}, \bibinfo {author} {\bibfnamefont {Martin}\
  \bibnamefont {{Elvis}}}, \bibinfo {author} {\bibfnamefont {Elizabeth~C.}\
  \bibnamefont {{Galle}}}, \bibinfo {author} {\bibfnamefont {Daniel~E.}\
  \bibnamefont {{Harris}}}, \bibinfo {author} {\bibfnamefont {David~P.}\
  \bibnamefont {{Huenemoerder}}}, \bibinfo {author} {\bibfnamefont {John~C.}\
  \bibnamefont {{Houck}}}, \bibinfo {author} {\bibfnamefont {Bish}\
  \bibnamefont {{Ishibashi}}}, \bibinfo {author} {\bibfnamefont {Margarita}\
  \bibnamefont {{Karovska}}}, \bibinfo {author} {\bibfnamefont {Fabrizio}\
  \bibnamefont {{Nicastro}}}, \bibinfo {author} {\bibfnamefont {Michael~S.}\
  \bibnamefont {{Noble}}}, \bibinfo {author} {\bibfnamefont {Michael~A.}\
  \bibnamefont {{Nowak}}}, \bibinfo {author} {\bibfnamefont {Frank~A.}\
  \bibnamefont {{Primini}}}, \bibinfo {author} {\bibfnamefont {Aneta}\
  \bibnamefont {{Siemiginowska}}}, \bibinfo {author} {\bibfnamefont
  {Randall~K.}\ \bibnamefont {{Smith}}}, \ and\ \bibinfo {author}
  {\bibfnamefont {Michael}\ \bibnamefont {{Wise}}},\ }\bibfield  {title}
  {\enquote {\bibinfo {title} {{CIAO: Chandra's data analysis system}},}\ }in\
  \href {\doibase 10.1117/12.671760} {\emph {\bibinfo {booktitle} {Society of
  Photo-Optical Instrumentation Engineers (SPIE) Conference Series}}},\
  \bibinfo {series} {Society of Photo-Optical Instrumentation Engineers (SPIE)
  Conference Series}, Vol.\ \bibinfo {volume} {6270},\ \bibinfo {editor}
  {edited by\ \bibinfo {editor} {\bibfnamefont {David~R.}\ \bibnamefont
  {{Silva}}}\ and\ \bibinfo {editor} {\bibfnamefont {Rodger~E.}\ \bibnamefont
  {{Doxsey}}}}\ (\bibinfo {year} {2006})\ p.\ \bibinfo {pages}
  {62701V}\BibitemShut {NoStop}%
\bibitem [{\citenamefont {Weisskopf}\ \emph {et~al.}(2003)\citenamefont
  {Weisskopf}, \citenamefont {Aldcroft}, \citenamefont {Bautz}, \citenamefont
  {Cameron}, \citenamefont {Dewey}, \citenamefont {Drake}, \citenamefont
  {Grant}, \citenamefont {Marshall},\ and\ \citenamefont
  {Murray}}]{Weisskopf:2005jy}%
  \BibitemOpen
  \bibfield  {author} {\bibinfo {author} {\bibfnamefont {M.~C.}\ \bibnamefont
  {Weisskopf}}, \bibinfo {author} {\bibfnamefont {T.~L.}\ \bibnamefont
  {Aldcroft}}, \bibinfo {author} {\bibfnamefont {M.}~\bibnamefont {Bautz}},
  \bibinfo {author} {\bibfnamefont {R.~A.}\ \bibnamefont {Cameron}}, \bibinfo
  {author} {\bibfnamefont {D.}~\bibnamefont {Dewey}}, \bibinfo {author}
  {\bibfnamefont {J.~J.}\ \bibnamefont {Drake}}, \bibinfo {author}
  {\bibfnamefont {C.~E.}\ \bibnamefont {Grant}}, \bibinfo {author}
  {\bibfnamefont {H.~L.}\ \bibnamefont {Marshall}}, \ and\ \bibinfo {author}
  {\bibfnamefont {S.~S.}\ \bibnamefont {Murray}},\ }\bibfield  {title}
  {\enquote {\bibinfo {title} {{An Overview of the performance of the Chandra
  X-Ray Observatory}},}\ }\href {\doibase 10.1023/B:EXPA.0000038953.49421.54}
  {\bibfield  {journal} {\bibinfo  {journal} {Exper. Astron.}\ }\textbf
  {\bibinfo {volume} {16}},\ \bibinfo {pages} {1--68} (\bibinfo {year}
  {2003})},\ \Eprint {http://arxiv.org/abs/astro-ph/0503319}
  {arXiv:astro-ph/0503319} \BibitemShut {NoStop}%
\bibitem [{\citenamefont {{Secrest}}\ \emph {et~al.}(2015)\citenamefont
  {{Secrest}}, \citenamefont {{Dudik}}, \citenamefont {{Dorland}},
  \citenamefont {{Zacharias}}, \citenamefont {{Makarov}}, \citenamefont
  {{Fey}}, \citenamefont {{Frouard}},\ and\ \citenamefont
  {{Finch}}}]{2015ApJS..221...12S}%
  \BibitemOpen
  \bibfield  {author} {\bibinfo {author} {\bibfnamefont {N.~J.}\ \bibnamefont
  {{Secrest}}}, \bibinfo {author} {\bibfnamefont {R.~P.}\ \bibnamefont
  {{Dudik}}}, \bibinfo {author} {\bibfnamefont {B.~N.}\ \bibnamefont
  {{Dorland}}}, \bibinfo {author} {\bibfnamefont {N.}~\bibnamefont
  {{Zacharias}}}, \bibinfo {author} {\bibfnamefont {V.}~\bibnamefont
  {{Makarov}}}, \bibinfo {author} {\bibfnamefont {A.}~\bibnamefont {{Fey}}},
  \bibinfo {author} {\bibfnamefont {J.}~\bibnamefont {{Frouard}}}, \ and\
  \bibinfo {author} {\bibfnamefont {C.}~\bibnamefont {{Finch}}},\ }\bibfield
  {title} {\enquote {\bibinfo {title} {{Identification of 1.4 Million Active
  Galactic Nuclei in the Mid-Infrared using WISE Data}},}\ }\href {\doibase
  10.1088/0067-0049/221/1/12} {\bibfield  {journal} {\bibinfo  {journal}
  {\apjs}\ }\textbf {\bibinfo {volume} {221}},\ \bibinfo {eid} {12} (\bibinfo
  {year} {2015})},\ \Eprint {http://arxiv.org/abs/1509.07289} {arXiv:1509.07289
  [astro-ph.GA]} \BibitemShut {NoStop}%
\bibitem [{\citenamefont {{B{\'e}dard}}\ \emph {et~al.}(2020)\citenamefont
  {{B{\'e}dard}}, \citenamefont {{Bergeron}}, \citenamefont {{Brassard}},\ and\
  \citenamefont {{Fontaine}}}]{2020ApJ...901...93B}%
  \BibitemOpen
  \bibfield  {author} {\bibinfo {author} {\bibfnamefont {A.}~\bibnamefont
  {{B{\'e}dard}}}, \bibinfo {author} {\bibfnamefont {P.}~\bibnamefont
  {{Bergeron}}}, \bibinfo {author} {\bibfnamefont {P.}~\bibnamefont
  {{Brassard}}}, \ and\ \bibinfo {author} {\bibfnamefont {G.}~\bibnamefont
  {{Fontaine}}},\ }\bibfield  {title} {\enquote {\bibinfo {title} {{On the
  Spectral Evolution of Hot White Dwarf Stars. I. A Detailed Model Atmosphere
  Analysis of Hot White Dwarfs from SDSS DR12}},}\ }\href {\doibase
  10.3847/1538-4357/abafbe} {\bibfield  {journal} {\bibinfo  {journal} {\apj}\
  }\textbf {\bibinfo {volume} {901}},\ \bibinfo {eid} {93} (\bibinfo {year}
  {2020})},\ \Eprint {http://arxiv.org/abs/2008.07469} {arXiv:2008.07469
  [astro-ph.SR]} \BibitemShut {NoStop}%
\end{thebibliography}%

\clearpage

%\clearpage

\onecolumngrid

\onecolumngrid
\begin{center}
  \textbf{\large Supplementary Material for: No evidence for axions from {\it Chandra} observation of magnetic white dwarf}\\[.2cm]
  \vspace{0.05in}
  { Christopher Dessert, Andrew J. Long, Benjamin R. Safdi}
\end{center}

\twocolumngrid
%%%%%%%%%% Merge with supplemental materials %%%%%%%%%%
\setcounter{equation}{0}
\setcounter{figure}{0}
\setcounter{table}{0}
\setcounter{section}{0}
\setcounter{page}{1}
\makeatletter
\renewcommand{\theequation}{S\arabic{equation}}
\renewcommand{\thefigure}{S\arabic{figure}}
\renewcommand{\thetable}{S\arabic{table}}

\onecolumngrid

This Supplementary Material (SM) is organized as follows.  Sec.~\ref{app:figures} provides Supplementary Figures that are referenced in the main Letter. Sec.~\ref{app:data} gives further information on our data reduction and calibration procedure.  In Sec.~\ref{app:loop} we review the renormalization group evolution of the axion-electron coupling to justify the values taken in the main text.  In Sec.~\ref{app:model} we describe our modeling procedure for the MWD in more detail.  Sec.~\ref{app:Electro-Primakoff} presents our calculation of the Electro-Primakoff axion production rate.  

\section{Supplementary Figures}
\label{app:figures}

\begin{figure}[!htb]
\begin{center}
\includegraphics[width=0.5\textwidth]{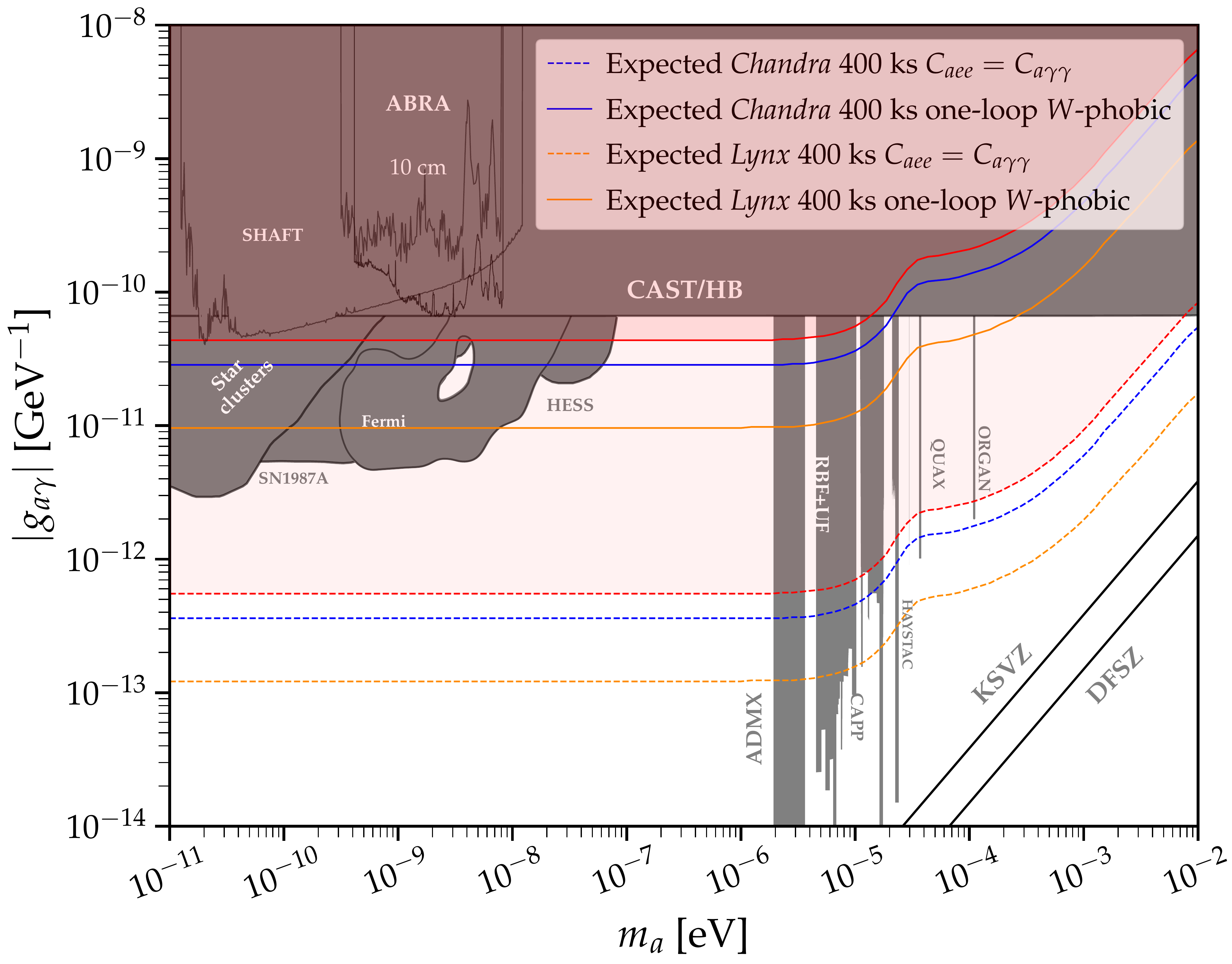} 
\caption{
As in Fig.~\ref{fig:gagg_limits} but projecting future sensitivity from deeper observations of RE J0317-853.  A factor of 10 increase in {\it Chandra} exposure time would lead to the projected expected 95\% upper limits indicated, while in the future the {\it Lynx} $X$-ray observatory will allow for a significant increase in sensitivity. To generate the {\it Lynx} projections, we use the package SOXS to generate expected counts maps, exposure maps, and the {\it Lynx} PSF.  We then run our {\it Chandra} pipeline with the {\it Lynx} files.
}
\label{fig:gagg_limits_prog}
\end{center}
\end{figure}

\begin{figure}[!htb]
\begin{center}
\includegraphics[width=0.5\textwidth]{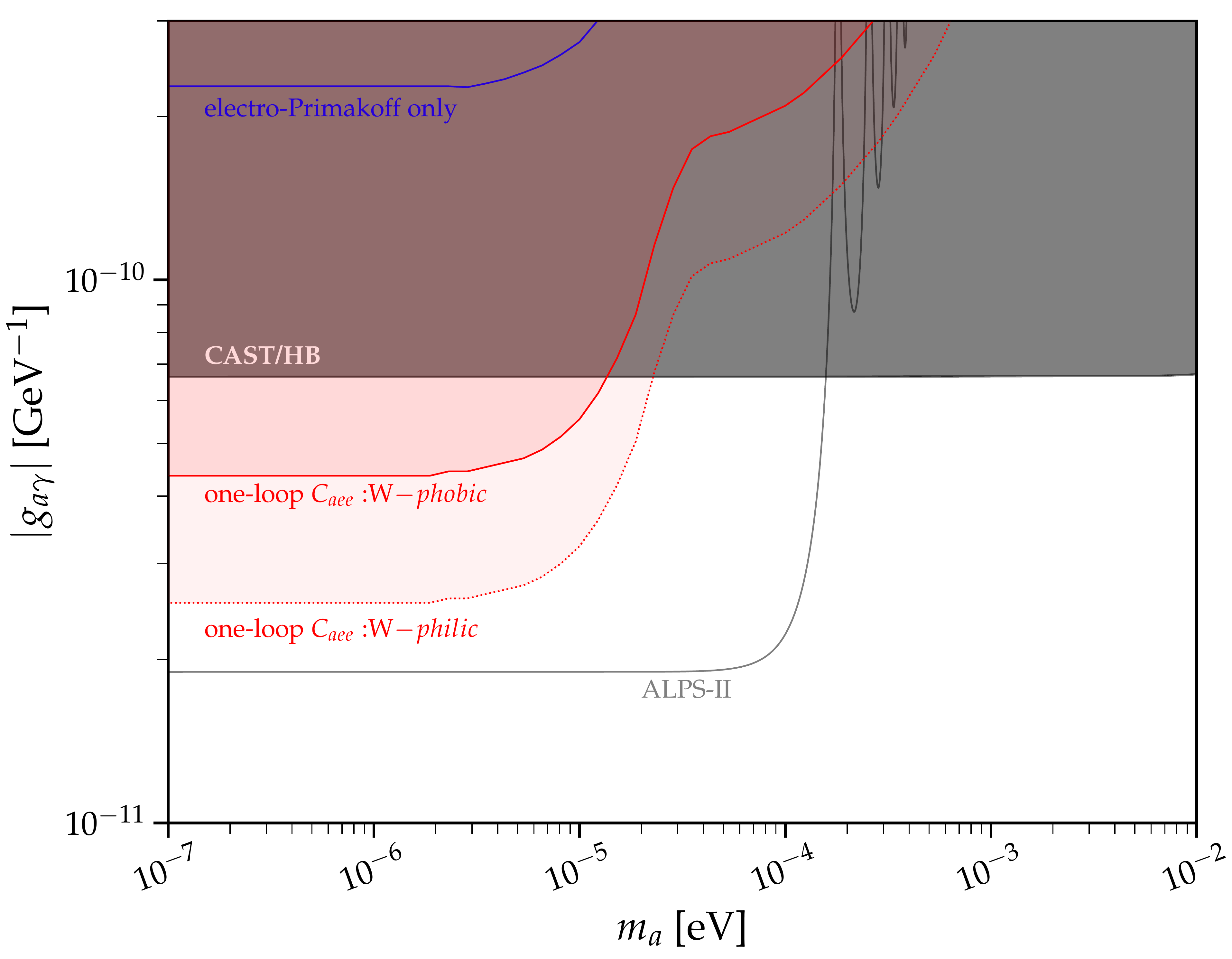} 
\caption{
As in Fig.~\ref{fig:gagg_limits} but showing the 95\% upper limits from this work interpreted in the context of limits on $g_{a\gamma\gamma}$ assuming loop-induced couplings to $C_{aee}$ for the $W$-phobic ($C_{aee} = 1.6 \times 10^{-4} C_{a\gamma\gamma}$) and $W$-philic ($C_{aee} = 4.8 \times 10^{-4} C_{a\gamma\gamma}$) UV completions. Models that couple to both $SU(2)_L$ and $U(1)_Y$ will generically have loop-induced couplings between these two extremes, assuming no fine-tuned cancellations (for example, models that couple in a way that preserve the Grand Unification group symmetry may have $C_{aee} \approx 2.7 \times 10^{-4} C_{a\gamma\gamma}$).  Note that UV contributions to $C_{aee}$ may also exist.  We compare these limits to the projected sensitivity from the ALPS-II experiment.  We also show our limits only accounting for the electro-Primakoff process, which does not involve $C_{aee}$ -- this process is seen to be subdominant compared to the bremsstrahlung process.   
}
\label{fig:gagg_limits_zoom}
\end{center}
\end{figure}

\begin{figure}[!htb]
\begin{center}
\includegraphics[width=0.6\textwidth]{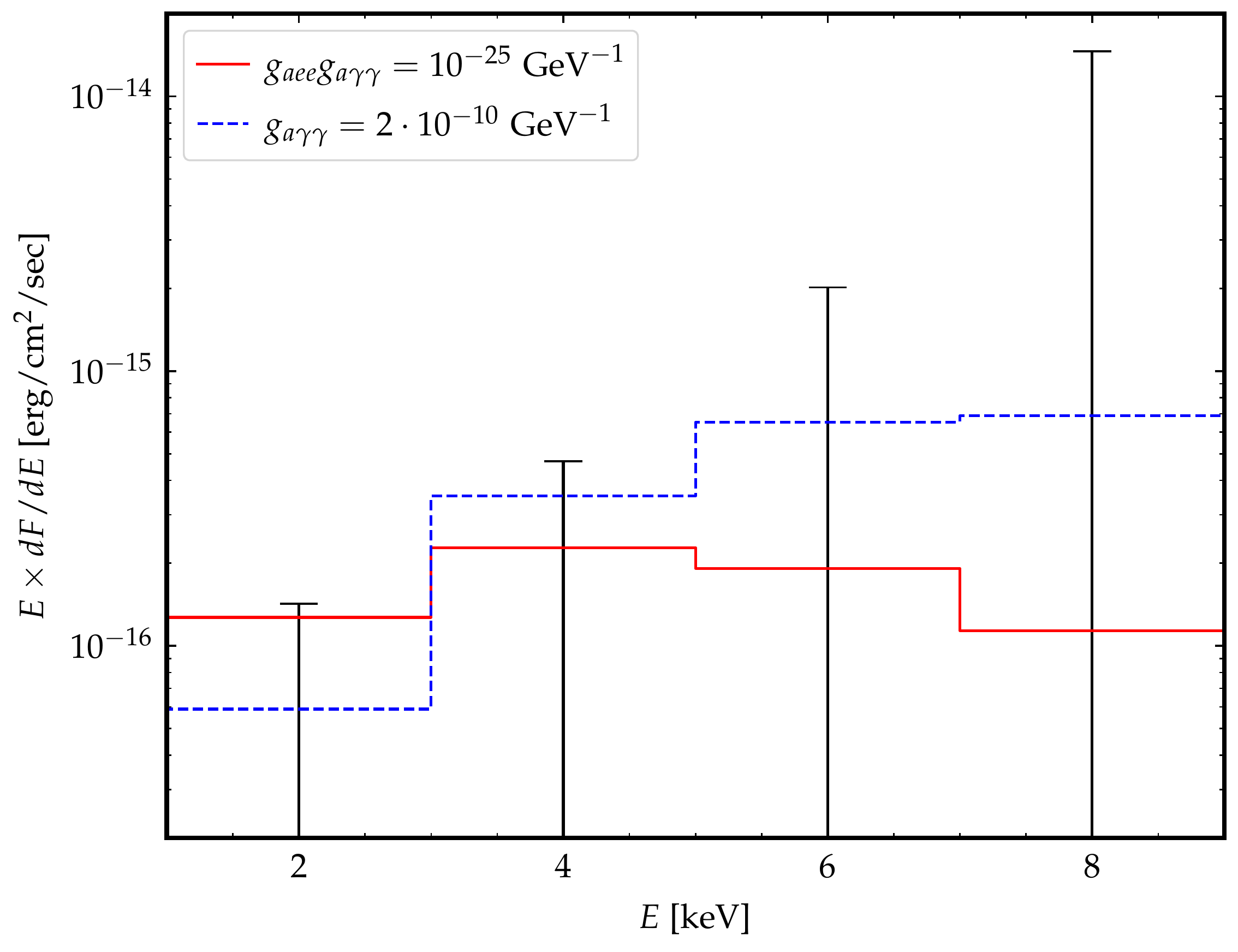} 
\caption{
As in Fig.~\ref{fig:spectrum} but comparing the bremsstrahlung (red) and electro-Primakoff (dashed blue) production rates, for the indicated couplings.   
}
\label{fig:spectrum_ep}
\end{center}
\end{figure}

\begin{figure}[!htb]
\begin{center}
\includegraphics[width=0.6\textwidth]{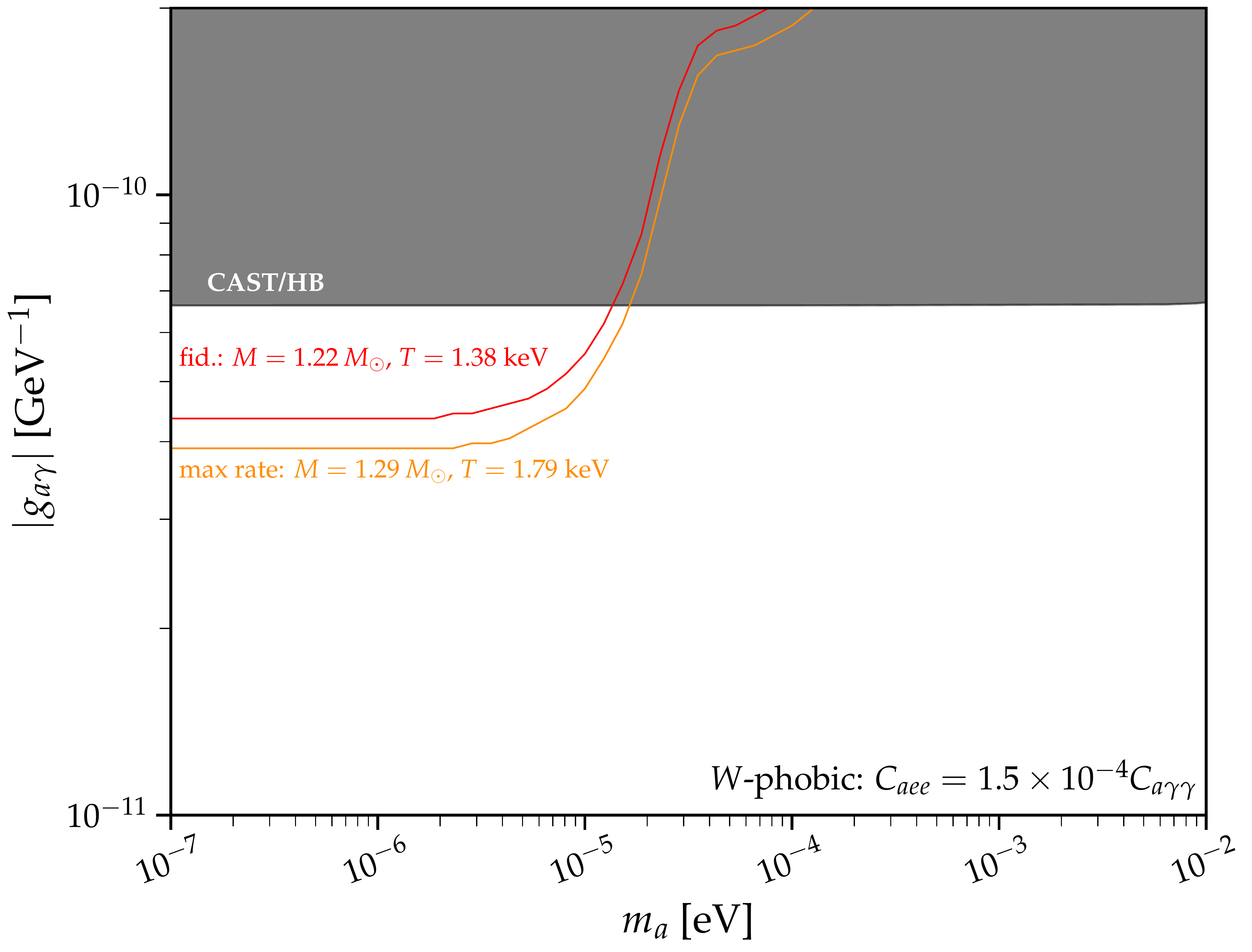} 
\caption{
As in Fig.~\ref{fig:gagg_limits} but comparing the $W$-phobic loop-induced upper limit (red) for our fiducial stellar model to that for the alternate stellar model that differs in two ways: (i) the MWD mass in assumed to be higher at 1.29 $M_\odot$, and (ii) the temperature is taken at the upper value of the 1$\sigma$ containment interval from fitting the stellar model to the {\it Gaia} luminosity data.  The difference between these two limits gives an estimate for the magnitude of the astrophysical uncertainties, which are around 10\%.    
}
\label{fig:gagg_limits_stellar}
\end{center}
\end{figure}

In this section we illustrate Figs.~\ref{fig:gagg_limits_prog},~\ref{fig:gagg_limits_zoom},~\ref{fig:spectrum_ep}, and~\ref{fig:gagg_limits_stellar}, which are cited and described in the main Letter.

%\clearpage

\section{Data reduction and calibration}
\label{app:data}

The data from the 37.42 ks {\it Chandra} ACIS-I \texttt{Timed Exposure} observation of RE J0317--853 (PI Safdi, observation ID \texttt{22326}) is reduced as follows. For the data reduction process, we use the Chandra Interactive Analysis of Observations (CIAO)~\cite{2006SPIE.6270E..1VF} version 4.11. We reprocess the observation with the CIAO task \texttt{chandra\_repro}, which produces an events file filtered for flares and updated for the most recent calibration. We create counts and exposure images (units [cm$^2$s]) with pixel sizes of $0\farcs492$ with \texttt{flux\_image}.

We account for the astrometric uncertainty of {\it Chandra}, which is expected to be on the order of $0\farcs5$~\cite{Weisskopf:2005jy}, 
through the following procedure:  we (i) run the point source (PS) finding algorithm \texttt{celldetect} on the full {\it Chandra} image to find high-significance PSs ($\gtrsim$10$\sigma$ significance), and then (ii) cross-correlate these sources with the {\it Gaia} early data release 3 (EDR3) catalog~\cite{2020arXiv201201533G} evolved to the Dec. 2020 epoch. (Note that there are no already-known $X$-ray sources within the field of view to use as references.)  Two of the high-significance sources have nearby matches with {\it Gaia} sources ({\it Gaia} source IDs 4613614905421384320 and 4613614974140862464). Although we were not able to verify the identity of these two sources from our observation, the Gaia sources both appear in the WISE catalog on active galactic nuclei~\cite{2015ApJS..221...12S}, as J031629.01-852836.0 and J031821.59-852751.5 respectively. 
Both sources are localized by \texttt{celldetect} to within $\sim$$0\farcs2$.  However, both {\it Chandra} sources are displaced from their {\it Gaia} matches by $\sim$$0\farcs6$ in approximately the same direction (the offset is $(0\farcs53,0\farcs25)$ for one source and $(0\farcs57,-0\farcs05)$ for the other, in $({\rm RA} \cos({\rm DEC}), {\rm DEC})$).  We average these two offsets to determine our overall calibration and shift all RA, DEC values accordingly.  The uncalibrated location is shown in Fig.~\ref{fig:cts_map}.  Note that we cannot exclude the possibility that the {\it Chandra} PSs are falsely matched with the {\it Gaia} sources, though this appears less likely given that the two position offsets are nearly the same.  Additionally, using the uncalibrated source location produces nearly identical results to using the calibrated location, since the calibration error is relatively minor and there are no photons in the vicinity of either location.

In addition to the calibration, we also account for the proper motion of the WD.  In particular, RE J0317-853 was observed by {\it Gaia} in the EDR3 with location ${\rm RA} \approx 49^\circ \, 18' \, 42\farcs51$, ${\rm DEC} \approx -85^\circ \, 32' 25\farcs75$ at the reference epoch of J2016.0~\cite{2020arXiv201201533G}.  We use the proper motion measurements from {\it Gaia} to infer the position in December 2020, which accounts for the small shift between {\it Gaia} 2016 and Dec. 2020 shown in Fig.~\ref{fig:cts_map}.

\section{Loop-induced axion-electron coupling}
\label{app:loop}

In this section we review the loop-induced axion-electron coupling in order to justify the fiducial values taken in the main text for the $W$-phobic and $W$-philic axion with no ultraviolet (UV) axion-electron coupling.  Recall that under the renormalization group and at energy scales $\mu > M_Z$, with $M_Z$ the mass of the $Z$-boson,
\es{eq:rge}{
{\mu \, d C_e^\mu \over d \mu} = -{3 \over 64 \pi^4} \left( {3 \over 8} g^4 C_W^\Lambda + {5 \over 8} g'^4 C_B^\Lambda \right) \,,
}
where $C_e^\mu$ is the dimensionless axion-electron coupling at energy scale $\mu < \Lambda$, with $\Lambda$ the UV cutoff~\cite{Srednicki:1985xd,Chang:1993gm,Dessert:2019sgw}.  The dimensionless axion couplings to weak isospin and hypercharge are denoted by $C_W^\Lambda$ and $C_B^\Lambda$, respectively.  Note that these couplings are topologically protected and do not evolve under the renormalization group.  The weak isospin and hypercharge couplings constants are denoted by $g$ and $g'$, respectively. 

It is common to integrate~\eqref{eq:rge} down to $M_Z$ and yet take $g$ and $g'$ to be their low-energy values, at scales well below $M_Z$.  Below $M_Z$ the axion-electron coupling continues to evolve under the renormalization group equation
\es{eq:rge-lowE}{
{\mu \, d C_e^\mu \over d \mu} = -{3 \over 4 \pi^2} \alpha_{\rm EM}^2 (C_W^{\Lambda} + C_B^{\Lambda})  \,,
}
and this contribution to $C_e$ at the scale $\mu = m_e$ is also typically found by integrating~\eqref{eq:rge-lowE} and taking $\alpha_{\rm EM}$ to be the value at the scale $m_e$.  Here, we do not complete a full two-loop computation of $C_e$ but we try to be slightly more precise by accounting for the running of $\alpha_{\rm EM}$, $g$, and $g'$.  To one-loop and within the Standard Model these couplings evolve as 
\es{eq:rge-gauge}{
{\mu \, d g_i \over d \mu} = \frac{b_i}{(4\pi)^2} g_i^3 \,,
}
with $g_1 = \sqrt{5/3} g'$, $g_2 = g$, $b_1 = 41/10$, and $b_2 = -19/6$.  Integrating~\eqref{eq:rge} in conjunction with~\eqref{eq:rge-gauge} from the UV scale $\Lambda$  down to the electroweak scale $M_Z$ leads to the result 
\es{}{
C_e^{M_Z} = C_e^\Lambda + {3 \over 128 \pi^4} \log{\Lambda^2 \over M_Z^2} \left( {3 \over 8} C_W^\Lambda \left[ g(M_Z) g(\Lambda) \right]^2 + {5 \over 8} C_B^\Lambda \left[ g'(M_Z) g'(\Lambda) \right]^2  \right)
\,,
}
where $g(M_Z)$ denotes the coupling at energy scale $M_Z$, while $g(\Lambda)$ is the coupling at the UV scale and similarly for $g'$.  At the $Z$-pole $\alpha_{\rm EM}(M_Z) \approx 1/127$ and $\sin^2 \theta_W \approx 0.231$, with $\theta_W$ the Weinberg angle.  Taking a benchmark value $\Lambda = 10^9$ GeV we then find 
\es{}{
C_e^{M_Z} \approx C_e^\Lambda +  4.2 \times 10^{-4} C_W^\Lambda + 9.8 \times 10^{-5} C_B^\Lambda 
\,.
}
Accounting for the running of $\alpha_{\rm EM}$ from $M_Z$ down to the electron mass we then find 

\es{eq:final_Ce}{
C_e^{M_e} \approx C_e^\Lambda +  4.8 \times 10^{-4} C_W^\Lambda + 1.6 \times 10^{-4} C_B^\Lambda 
\,.
}
Note that the axion-photon coupling is defined by $C_{a\gamma\gamma} = C_W^\Lambda + C_B^\Lambda$.  To be conservative, in our fiducial loop-induced model we consider a ``W-phobic'' axion and take $C_W^\Lambda = 0$ such that $C_{aee} \approx 1.6\times 10^{-4} C_{a\gamma\gamma}$.   
We do note, though, with some amount of fine tuning the loop-induced contribution could be made smaller.  For example, if $C_W^\Lambda \approx -0.33 C_B^\Lambda$  
then the two contributions to $C_e^{M_e}$ would roughly cancel each other.  We do not consider this possibility further because it would require a conspiracy between the UV and IR contributions to the running.  Note, also, that the relations in~\eqref{eq:final_Ce} could be modified by the existence of beyond the Standard Model physics below the UV cutoff $\sim$$10^9$ GeV.

\section{Modeling RE J0317--853}
\label{app:model}

In this section we detail our modeling of the interior of RE J0317-853. To compute the axion luminosity, we need to know the core temperature, the density profile, and the composition profiles. Note that we assume the core temperature is uniform throughout the interior due to the high thermal conductivity of the degenerate matter, while the density and composition can change throughout the interior.

We analyze WD cooling sequences~\cite{2019AA...625A..87C} to infer the core temperature of RE J0317-853. These cooling sequences are improved over older ones in that they take ionic correlations into account, which are expected to be important for RE J0317-853 due to its high mass and low surface temperature. Included with the sequences are corresponding {\it Gaia} DR2 $G$, $G_{\rm BP}$, and $G_{\rm RP}$ band absolute magnitudes as a function of cooling age. The sequences are available for WD masses of $1.10$, $1.16$, $1.22$, and $1.29 M_\odot$.

\begin{figure}[t]
\begin{center}
\includegraphics[width=0.48\textwidth]{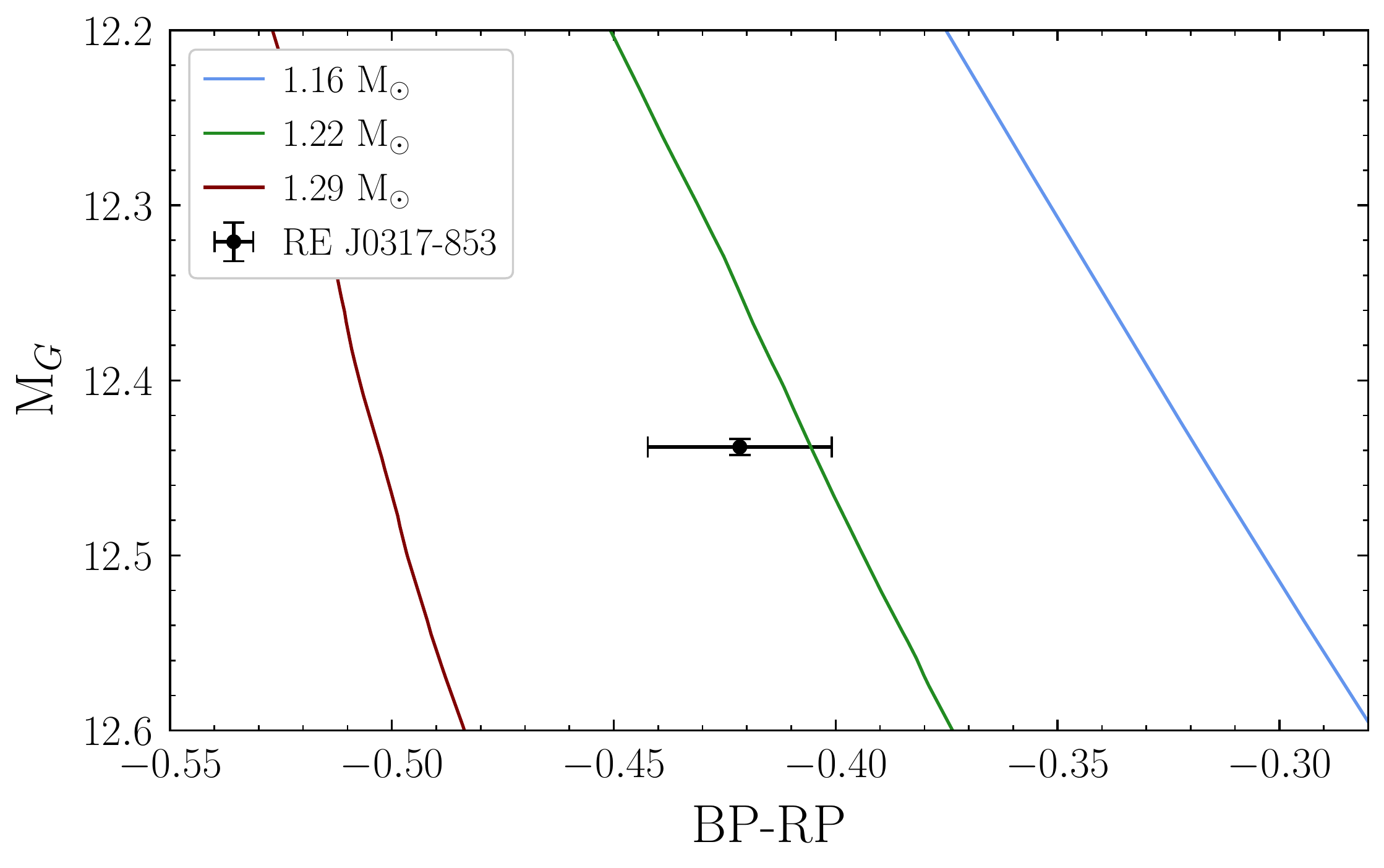} 
\includegraphics[width=0.48\textwidth]{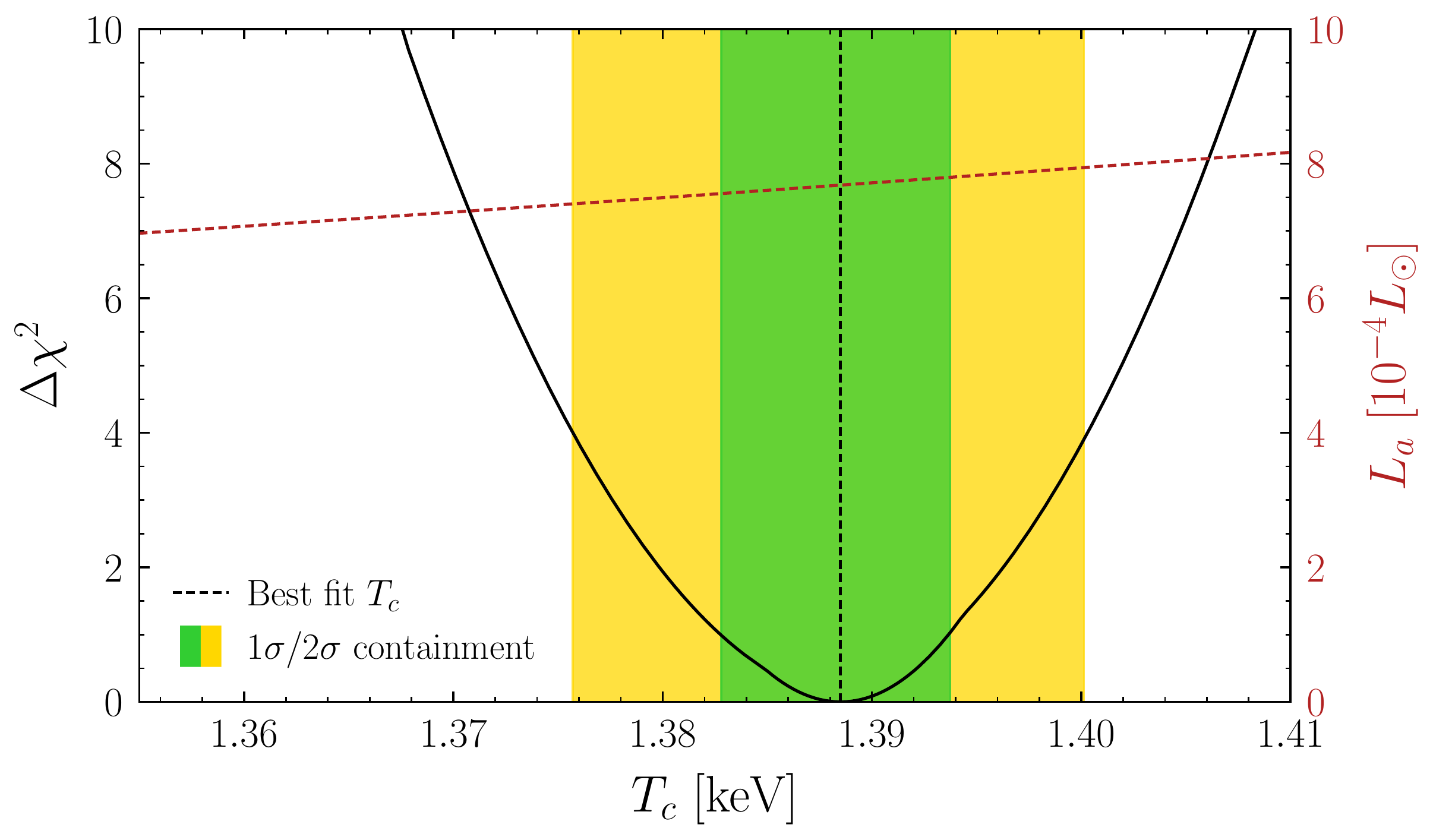} 
\caption{(Left) A color-magnitude diagram with RE J0317-853's {\it Gaia} DR2 data shown with the black error bars. We show the curves predicted by the cooling simulation for three masses: $1.16$, $1.22$, and $1.29 M_\odot$. Note that $M_G$ refers to the absolute $G$-band magnitude, while the color ${\rm BP}-{\rm RP} = G_{\rm BP}-G_{\rm RP}$.  (Right) The likelihood profile for the $1.22 M_\odot$ model as a function of $T_c$. The best fit $T_c$ is shown as the dashed vertical line, while the $1$ and $2\sigma$ containment regions on $T_c$ are shown as green and yellow bands, respectively. We also show, on the right $y$-axis, the axion luminosity (dashed red) as a function of $T_c$ for $g_{aee} = 10^{-13}$.}
\label{fig:GaiaModel}
\end{center}
\end{figure}

RE J0317-853's measured apparent magnitudes in the DR2 {\it Gaia} dataset~\cite{GaiaDR2} are

\begin{align}
    \begin{split}
        G &= 14.779 \pm 0.005 \\
        G_{\rm BP} &= 14.565 \pm 0.017 \\
        G_{\rm RP} &= 14.987 \pm 0.012 
    \end{split}
\end{align}

\noindent where we have converted linear errors on flux to linear errors on magnitude. For reference, the $G$-band covers wavelengths between $\sim$$300$ and $\sim$$1100$ nm, $G_{\rm BP}$ between $\sim$$300$ and $\sim$$700$ nm, and $G_{\rm RP}$ between $\sim$$600$ and $\sim$$1100$ nm, although with wavelength-dependent efficiencies.  Note we use EDR3 astrometric and distance data elsewhere in this work, but there do not yet exist cooling sequences incorporating EDR3 bands. We infer the core temperature $T_c$ of RE J0317-853 with a joint Gaussian likelihood over the three bands as a function of cooling age $t$ for each WD mass available. We find that the $1.22 M_\odot$ model provides the best fit to the data, as shown in the left panel of Fig.~\ref{fig:GaiaModel}. Note that this is a lower mass for RE J0317-853 than previously inferred, but it is a conservative choice with respect to the $1.29 M_\odot$ model, which is closer to previous mass estimates~\cite{Kulebi:2010pd}. 

In the right panel of Fig.~\ref{fig:GaiaModel}, we show the resulting likelihood profile as a function of $T_c$ for the best-fit $1.22 M_\odot$ model. The $\pm 1\sigma$ ages are extracted by solving for the age where $\Delta\chi^2$ increases by 1 on each side of the best-fit point. We find $t = 0.369 \pm 0.003$ Gyr, corresponding to a core temperature $T_c = 1.388 \pm 0.005$ keV. We adopt the lower $1\sigma$ value of $T_c = 1.383$ keV in our fiducial analysis to be conservative. We also show the axion luminosity, for which changes are minor over the range considered. 

The $1.29$ $M_\odot$ model is disfavored in our analysis relative to the $1.22$ $M_\odot$ model at a level $\sim$$5\sigma$ (the measured $G_{\rm BP}$ and $G_{\rm RP}$ are in tension with the model expectations). Therefore, when we determine the properties of RE J0317-853 in the context of the $1.29$ $M_\odot$ model, we broaden the likelihood profile so that at the best-fit point, $\Delta\chi^2/$dof$ = 1$. We find a lower cooling age of $0.301 \pm 0.008$ Gyr and a higher $T_c = 1.77 \pm 0.02$ keV by following the same procedure.  SM Fig.~\ref{fig:gagg_limits_stellar} compares our limits computed using the fiducial model and the 1.29 $M_\odot$ model, with $T_c$ at the upper end of the 1$\sigma$ band; the differences are seen to be minor, indicating that our results are likely not significantly affected by astrophysical mismodeling.

We run simulations with MESA from which we determine the density and composition profiles for RE J0317-853. MESA is a 1-dimensional modular stellar modeling code that outputs these profiles, along with others, as a function of time since stellar birth. We use the default parameters from the test suite inlist \texttt{make\_o\_ne\_wd}, but change the initial stellar mass to $11.1$ ($11.9$) $M_\odot$, which produces a $1.22$ ($1.29$) $M_\odot$ WD. We evolve the star through the pre-WD stages and allow it to cool until its luminosity reaches $10^{-3} L_\odot$. 

\begin{figure}[t]
\begin{center}
\includegraphics[width=0.48\textwidth]{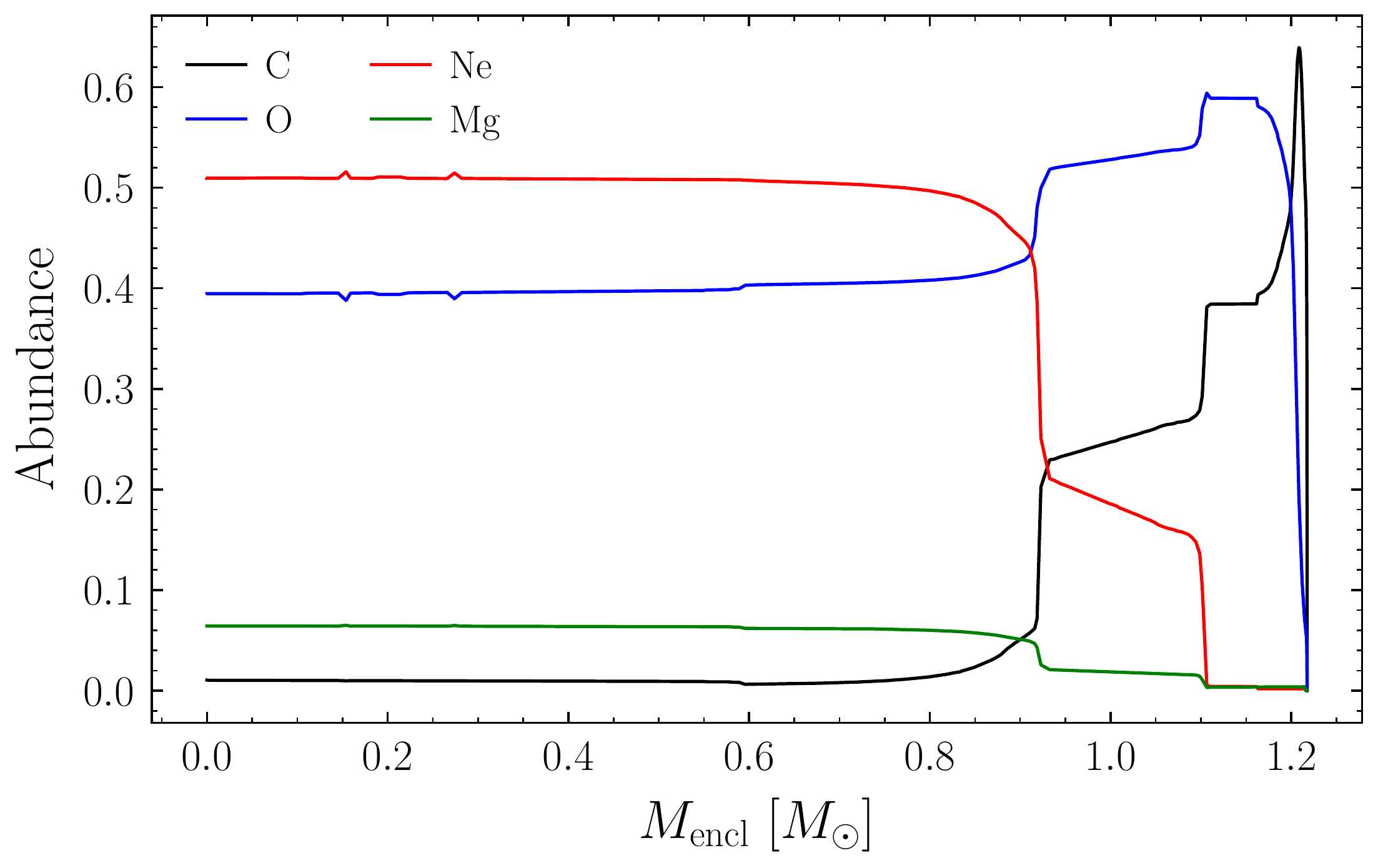} 
\includegraphics[width=0.48\textwidth]{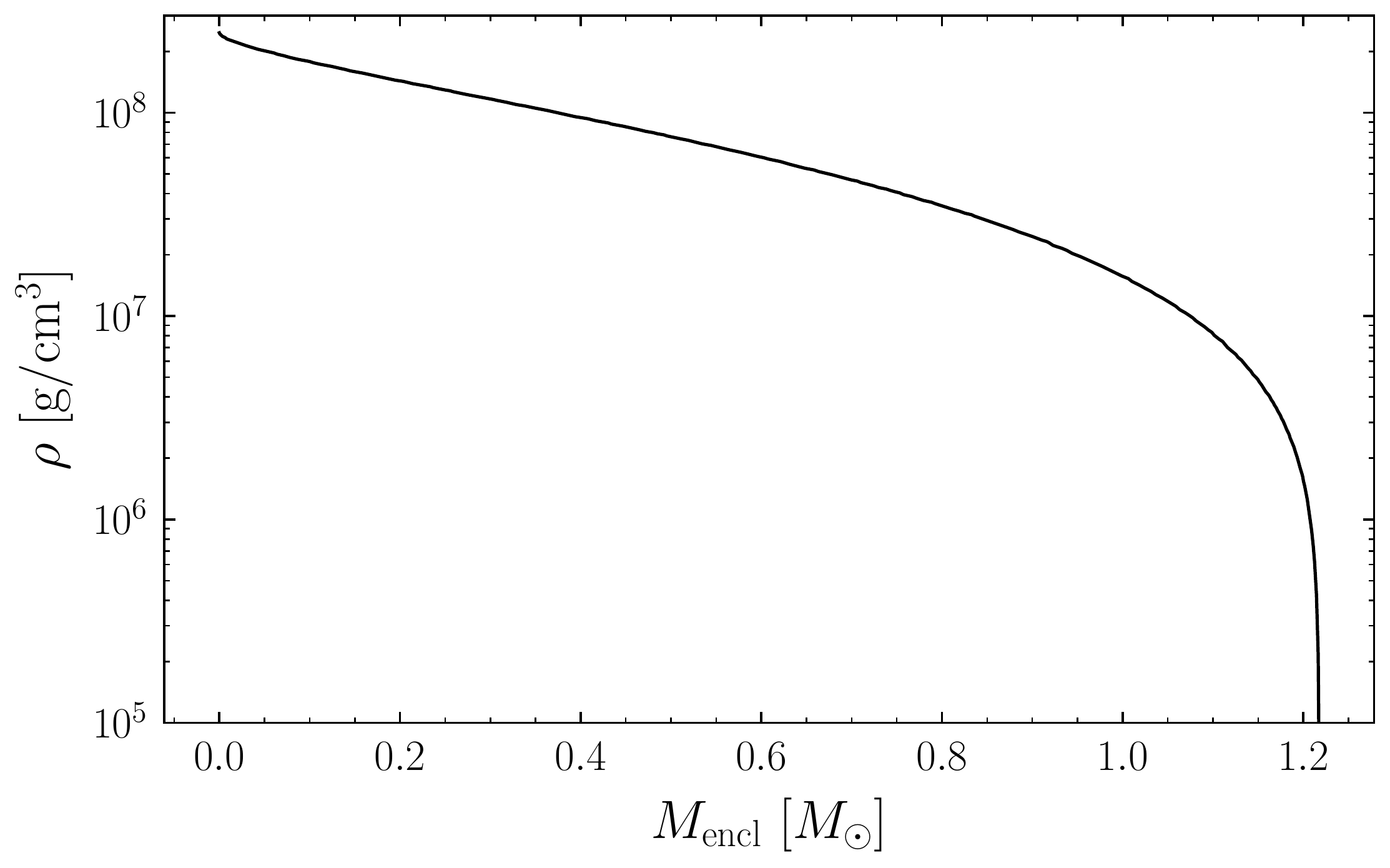} 
\caption{(Left) The carbon, oxygen, neon, and magnesium mass abundances in the MESA simulation for the model most closely matching the observed luminosity of RE J0317-853. The x-axis is the mass coordinate {\it i.e.}, enclosed mass. (Right) The density profile in [g/cm$^3$] for the same model as a function of mass coordinate.}
\label{fig:MESAprofiles}
\end{center}
\end{figure}

\begin{figure}[t]
\begin{center}
\includegraphics[width=0.48\textwidth]{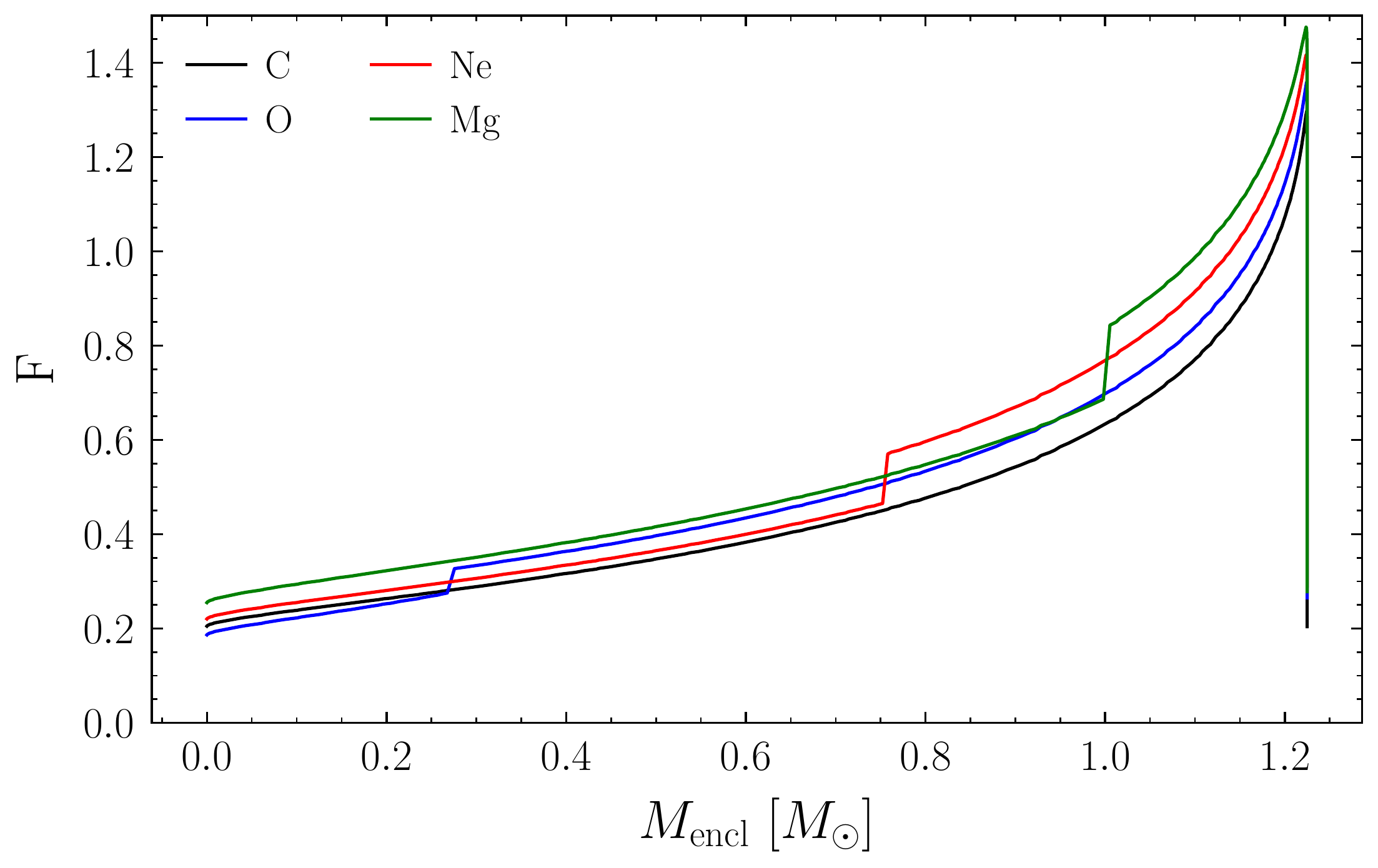} 
\includegraphics[width=0.48\textwidth]{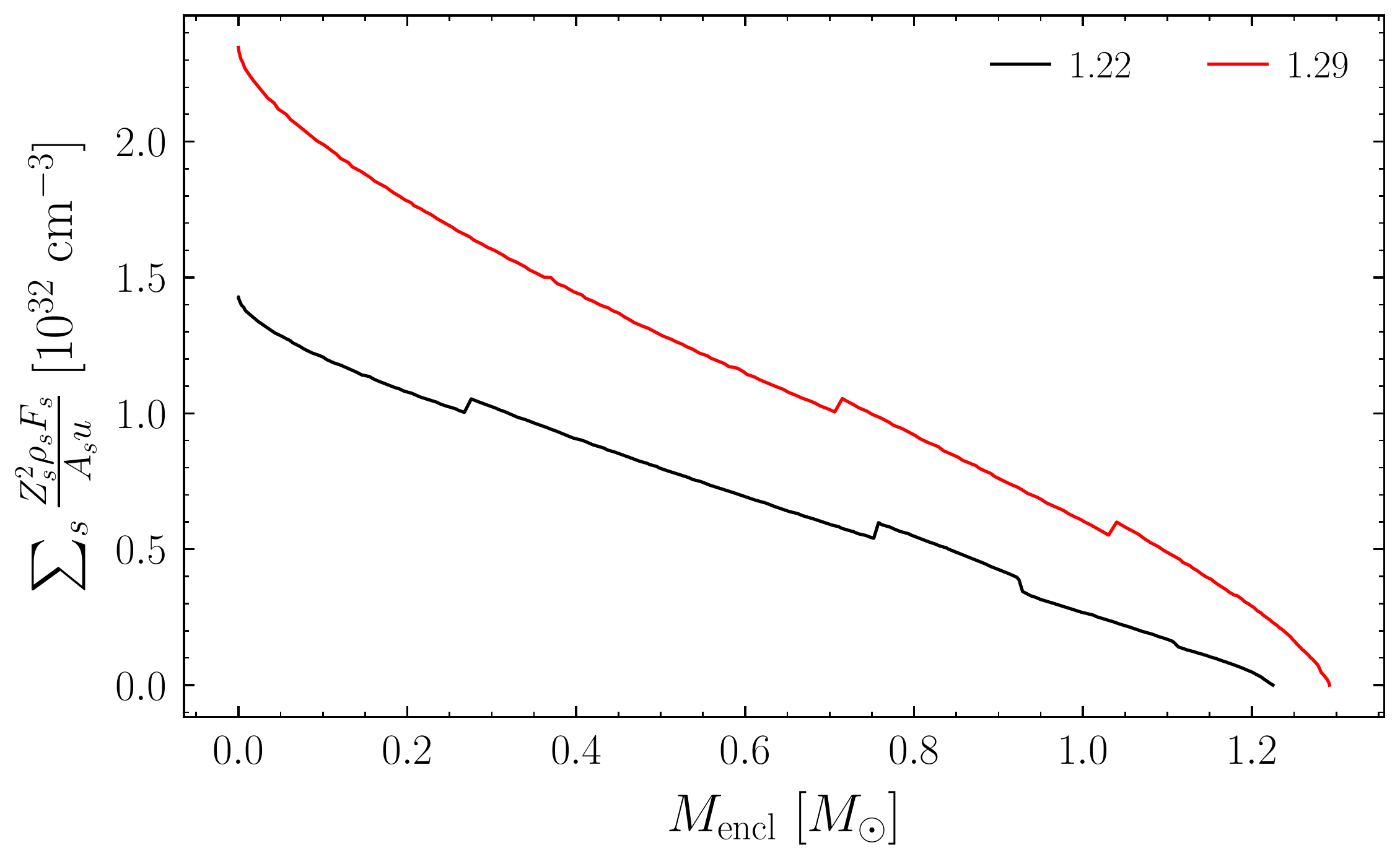} 
\caption{(Left) The $F$-profile evaluated for the 1.22 $M_\odot$ star, evaluated using the parametrization provided by~\cite{Nakagawa:1988rhp}, considered in our emissivity calculation. (Right) The sum in \eqref{eq:epsa_ebrem} evaluated for both mass models (1.22 $M_\odot$ and 1.29 $M_\odot$).}
\label{fig:Fprofiles}
\end{center}
\end{figure}

We then select the model for which the stellar luminosity matches the observed value and choose the profiles corresponding to this model, shown in Fig.~\ref{fig:MESAprofiles}, to be our fiducial density and composition profiles. We find that the core is predominantly oxygen and neon as expected for an isolated WD of its mass, and reaches densities $\rho > 10^6$ g/cm$^3$, which means that the electron gas is strongly correlated. For $\rho \gtrsim 10^7$ g/cm$^3$, the interior transitions to the lattice phase, which tends to reduce the axion emissivity. In the left panel of Fig.~\ref{fig:Fprofiles}, we show the value of $F$ as defined in~\eqref{eq:epsa_ebrem} across the profile of the star for the four dominant ions in our WD model. The discontinuities in the profiles (except carbon) are due to the transition from the liquid phase to the lattice ion structure in the inner core of the WD. In general, $F$ decreases with increasing density, although because the axion emissivity $\varepsilon_a\sim\rho F$, the center of the star is still the most emissive. 

Note that our choice of test suite is not the driving force behind why our WD is modeled as having an oxygen-neon core--this is simply because, under the assumption of single-star evolution, the initial stellar mass of the WD progenitor is high enough so that the star depletes its core carbon on the asymptotic giant branch (this is the case for WDs with masses $\gtrsim 1.1 M_\odot$~\cite{2019AA...625A..87C,2020ApJ...901...93B}). If the star has evolved from a binary channel, then it may host a carbon-oxygen core instead. However, we consider this to be unlikely, as~\cite{Kulebi:2010pd} finds that if RE J0317-853 has an effective temperature $\lesssim$40000 K, the single-star evolution is more likely. Indeed, our {\it Gaia} analysis prefers an effective temperature $25570 \pm 50$ K. Note that although RE J0317-853 has a binary companion, they are too far apart to have interacted~\cite{Kulebi:2010pd}.

Given the core temperature, the density profile, and composition profiles, we have the tools to compute the axion luminosity of RE J0317-853 due to both axion bremsstrahlung and electro-Primakoff. We compute the axion emissivity at each radial slice in the MESA-generated profiles and integrate over the star to obtain the axion luminosity spectrum $dL_a/d\omega$ (in, {\it e.g.}, ergs/s/keV) as

\begin{align}
\label{eq:IntegrateProfile}
    \dfrac{dL_a}{d\omega}(\omega) = 4\pi \int_0^R r^2 dr\frac{d\varepsilon_a}{d\omega}(r)
\end{align}

\noindent for a stellar radius $R$. For axion bremsstrahlung, $d\varepsilon_a/d\omega$ is computed using~\eqref{eq:epsa_ebrem}; for electro-Primakoff,~\eqref{eq:epsa_eP}. Because of the geometric factors in the integrand in~\eqref{eq:IntegrateProfile} that suppress the contribution from the stellar core, the axion luminosity profile $dL_a/dr$ peaks around half the WD radius.

For our fiducial analysis, we model the magnetic field as a dipole field of strength $200$ MG at the pole. To compute the axion-photon conversion probability $p_{a\to\gamma}(\omega)$, we follow the formalism developed in~\cite{Dessert:2019sgw}.
The axion-induced photon flux $dF_{\gamma_a}/d\omega$ at Earth is then

\begin{align}
    \dfrac{dF_{\gamma_a}}{d\omega}(\omega) = \dfrac{dL_a}{d\omega}(\omega) \times p_{a\to\gamma}(\omega) \times \dfrac{1}{4\pi d_{\rm WD}^2}.
\end{align}

%========================
%  Electro-Primakoff Axion Production
%========================
\section{Electro-Primakoff Axion Production}
\label{app:Electro-Primakoff}

%=======
This section provides a derivation of the axion emissivity from the core of a WD from the electro-Primakoff production mechanism. Note that while the bremsstrahlung process dominates for our MWD, the electro-Primakoff process may be important for WDs with higher core temperatures, and this computation has not appeared elsewhere.
%-------------------------
%  Cross section
%-------------------------
\subsection{Cross section} 

\begin{figure}[htb]
\begin{center}
\includegraphics[width=0.48\textwidth]{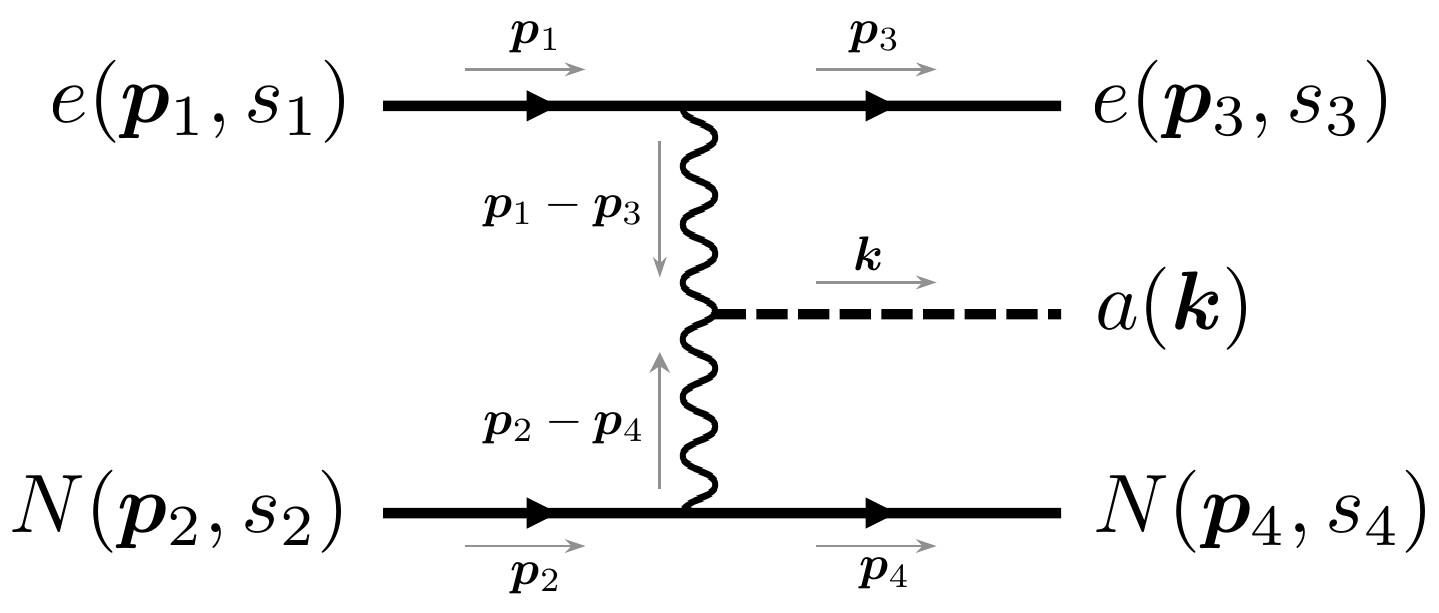} 
\caption{The Feynman graph for axion production via the electro-Primakoff channel. }
\label{fig:eP_graph}
\end{center}
\end{figure}

%=======
Consider the scattering of an electron $e$ and a nucleus $N = (A,Z)$ that results in the emission of an axion $a$: 
\begin{align}\label{eq:reaction}
	e(\pvec_1, s_1) + N(\pvec_2, s_2) \rightarrow e(\pvec_3,s_3) + N(\pvec_4, s_4) + a(\kvec) 
\;.
\end{align}
If the axion-photon coupling is dominant, then axion production is dominated by the electro-Primakoff channel.  
The leading-order Feynman graph is shown in Fig.~\ref{fig:eP_graph}, and the corresponding matrix element is
\bes{
	\Mcal & 
	= \bigl[ \bar{u}_e(p_3,s_3) (+i e) \gamma^\mu u_e(p_1,s_1) \bigr] \Bigl[ - \frac{i g_{\mu\rho}}{(p_1-p_3)^2 + i \epsilon} \Bigr] \bigl[ - i g_{a\gamma\gamma} \epsilon^{\rho\sigma\alpha\beta} (p_1-p_3)_\alpha (p_2-p_4)_\beta \bigr] \\
	& \qquad \times \Bigl[ - \frac{i g_{\sigma\nu}}{(p_2-p_4)^2 + i \epsilon} \Bigr] \bigl[ \bar{u}_N(p_4,s_4) (-i Z e) \gamma^\nu u_N(p_2,s_2) \bigr] 
	\;.
}
Note that the amplitude vanishes as $\omega = k^0 \to 0$, since 4-momentum conservation implies $\epsilon^{\rho\sigma\alpha\beta} (p_1-p_3)_\alpha (p_2-p_4)_\beta = \epsilon^{\rho\sigma\alpha\beta} (p_1-p_3)_\alpha k_\beta$.  
The spin-averaged, squared matrix element is given by $\overline{|\Mcal|^2} = (g_e g_N)^{-1} \sum_s |\Mcal|^2$ where $g_e = g_N = 2$ counts the two spin states of the electron and the nucleus.  

%=======
The differential cross section for axion emission is calculated from the squared matrix element as 
\bes{
	\mathrm{d} \sigma & = \frac{1}{4F_{aN}(p_1,p_2)} \, \mathrm{d}\Pi_e(\pvec_3) \, \mathrm{d}\Pi_N(\pvec_4) \, \mathrm{d}\Pi_a(\kvec) \ 
	(2\pi)^4 \, \delta(p_1 + p_2 - p_3 - p_4 - k) \ 
	|\Mcal|^2 
}
where the Lorentz-invariant flux factor is $F_{aN}(p_1,p_2) = [(p_1 \cdot p_2)^2 - m_e^2 m_N^2]^{1/2}$, and where the Lorentz-invariant phase space volume element is $\mathrm{d} \Pi_s(\pvec) = \mathrm{d}^3 \pvec/(2\pi)^3/2 E_s(\pvec)$ for $s=e,N,a$.  
All 4-momenta are evaluated on shell with $p^0 = E_s(\pvec) = [\pvec^2 + m_s^2]^{1/2}$.  

%-------------------------
%  Thermal-averaging
%-------------------------
\subsection{Thermal-averaging} 

%=======
The thermal environment leads to Pauli-blocking and Bose-enhancement of the final-state particles.  
We take this into account by defining the thermally-suppressed/enhanced differential cross section 

\ba{\label{eq:dsigma_tilde_def}
	\mathrm{d} \tilde{\sigma} = \mathrm{d} \sigma \ \bigl( 1 - f_e(\pvec_3) \bigr) \, \bigl( 1 - f_N(\pvec_4) \bigr) \, \bigl( 1 + f_a(\kvec) \bigr) 
}

where $f_e$, $f_N$, and $f_a$ are the phase space distribution functions for electrons, nuclei, and axions, respectively.  
The electrons are in equilibrium and their distribution function (in the rest frame of the plasma) is given by the Fermi-Dirac distribution 

\begin{subequations}\label{eq:f_def}
\ba{\label{eq:fe_def}
	f_e(\pvec) = \left( e^{[E_e(\pvec) - \mu_e]/T_e} + 1 \right)^{-1}
	\;, 
}
where $T_e$ and $\mu_e$ are the electrons' temperature and chemical potential.  
The nuclei are also in thermal equilibrium, and we could also write their distribution function as a Fermi-Dirac distribution.  
However, since their temperature is so low, $T_N \ll m_N$, it turns out that the nuclei are effectively at rest $v_N \sim \sqrt{T / m_N} \ll 1$.  
To a good approximation we can write the nuclei phase space distribution function (in the rest frame of the plasma) as 

\ba{\label{eq:fN_def}
	f_N(\pvec) = \frac{n_N}{g_N} \ (2\pi)^3 \delta(\pvec) 
	\;,
}
where $n_N$ is the total number density of nuclei and $g_N = 2$ counts the two spin states.  
This also lets us approximate $1 - f_N \approx 1$ in \eqref{eq:dsigma_tilde_def}.  
Finally the axions are out of thermal equilibrium, and their distribution function satisfies

\ba{\label{eq:fa_def}
	f_a(\pvec) \ll 1 
	\;,
}
and we can approximate $1 + f_a \approx 1$ in \eqref{eq:dsigma_tilde_def}.  

\end{subequations}

%-------------------------
%  Axion emissivity
%-------------------------
\subsection{Axion emissivity} 

%=======
Using the differential cross section from \eqref{eq:dsigma_tilde_def}, we construct the thermally-suppressed/enhanced differential scattering rate density, which is 

\ba{\label{eq:gamma_tilde_def}
	\mathrm{d} \tilde{\gamma} & = \frac{\mathrm{d} \tilde{\sigma}}{g_e \, g_N} \, v_\Mol \, \mathrm{d} n_e(\pvec_1) \, \mathrm{d} n_N(\pvec_2) 
}
where the M{\o}ller velocity is $v_\Mol(\pvec_1,\pvec_2) = F_{aN}(p_1,p_2) / E_a(\pvec_1) E_N(\pvec_2)$, where the thermally-weighted differential number density of incident particles is $\mathrm{d} n_s(\pvec) = g_s \mathrm{d}^3 \pvec \, f_s(\pvec) / (2\pi)^3$ for  $s=e,N$, and where $g_e = g_N = 2$ counts the redundant internal degrees of freedom (spin).
The differential axion emissivity (in the rest frame of the plasma) is 

\ba{\label{eq:depsilon_a_def}
	\mathrm{d} \varepsilon_a 
	& = \sum_\mathrm{spins} \mathrm{d} \tilde{\gamma} \ E_a(\kvec) 
	\;,
}
where we multiply by the axion energy and sum over the spins of all the particles.  
Using the expression for $\mathrm{d} \tilde{\gamma}$ gives 

\bes{\label{eq:depsilon_a_integrals}
	\mathrm{d} \varepsilon_a 
	& = 
	\frac{g_e g_N}{32} \, 
	\frac{\mathrm{d}^3 \pvec_1}{(2\pi)^3} \, 
	\frac{\mathrm{d}^3 \pvec_2}{(2\pi)^3} \, 
	\frac{\mathrm{d}^3 \pvec_3}{(2\pi)^3} \, 
	\frac{\mathrm{d}^3 \pvec_4}{(2\pi)^3} \, 
	\frac{\mathrm{d}^3 \kvec}{(2\pi)^3} \, 
	\\ & \hspace{1cm} 
	\times (2\pi) \, \delta\bigl( E_e(\pvec_1) + E_N(\pvec_2) - E_e(\pvec_3) - E_N(\pvec_4) - E_a(\kvec) \bigr) \, 
	\\ & \hspace{1cm} 
	\times (2\pi)^3 \, \delta(\pvec_1 + \pvec_2 - \pvec_3 - \pvec_4 - \kvec) \, 
	\\ & \hspace{1cm} 
	\times f_e(\pvec_1) \, f_N(\pvec_2) \, \bigl( 1 - f_e(\pvec_3) \bigr) \, \bigl( 1 - f_N(\pvec_4) \bigr) \, \bigl( 1 + f_a(\kvec) \bigr) 
	\\ & \hspace{1cm} 
	\times \frac{\overline{|\Mcal|^2}}{E_e(\pvec_1) \ E_N(\pvec_2) \ E_e(\pvec_3) \ E_N(\pvec_4)} 
}
where the factors of $E_a$ have cancelled, and all 4-momenta are on-shell.  

%-------------------------
%  Evaluating phase space integrals
%-------------------------
\subsection{Evaluating phase space integrals}

%=======
To calculate the emissivity, we evaluate the phase space integrals as follows.  
First, we use the momentum-conserving Dirac delta function to evaluate the integral over the recoiling nucleus's momentum, which sets $\pvec_4 = \pvec_1 + \pvec_2 - \pvec_3 - \kvec$.  
Next we write $\pvec_1$, $\pvec_3$, and $\kvec$ in polar coordinates, 
\bes{
	\mathrm{d}^3 \pvec_1 & = p_1^2 \mathrm{d} p_1 \ \mathrm{d} \Omega_1 = p_i \, E_i \, \mathrm{d} E_i \ \mathrm{d} \Omega_i \\ 
	\mathrm{d}^3 \pvec_3 & = p_3^2 \mathrm{d} p_3 \ \mathrm{d} \Omega_3 = p_f \, E_f \, \mathrm{d} E_f \ \mathrm{d} \Omega_f \\ 
	\mathrm{d}^3 \kvec & = k^2 \mathrm{d} k \ \mathrm{d} \Omega_a = k \, \omega \, \mathrm{d} \omega \ \mathrm{d} \Omega_a 
}
where $i$ denotes the initial-state electron, $f$ denotes the final-state electron, and $\omega = E_a(\kvec)$.  
We use the remaining Dirac delta function to evaluate the integral over $E_f$, which gives
\bes{
	\mathrm{d}\varepsilon_a
	& = 
	\frac{g_e g_N}{128 \pi^5} \, 
	\mathrm{d} E_i \, 
	\frac{\mathrm{d} \Omega_i}{4\pi} \, 
	\frac{\mathrm{d}^3 \pvec_2}{(2\pi)^3} \, 
	\frac{\mathrm{d} \Omega_f}{4\pi} \, 
	\mathrm{d} \omega \, 
	\frac{\mathrm{d} \Omega_a}{4\pi} \, 
	\\ & \hspace{1cm} 
	\times f_e(E_i) \, f_N(\pvec_2) \, \bigl( 1 - f_e(E_f) \bigr) \, \bigl( 1 - f_N(\pvec_4) \bigr) \, \bigl( 1 + f_a(\omega) \bigr) 
	\\ & \hspace{1cm} 
	\times \frac{p_i \,  p_f \, k \, \omega}{E_N(\pvec_2) \, E_N(\pvec_4)} \ 
	\overline{|\Mcal|^2} 
    \;.
}
Next we make use of the distribution functions in \eqref{eq:f_def}.  
These let us approximate $1 - f_N \approx 1$ and $1 + f_a \approx 1$.  
Additionally, $f_N \propto \delta(\pvec)$ and the $\pvec_2$ integral sets $\pvec_2 = 0$.  
Finally we note that the scattering is statistically isotropic, since the distributions of incident particles have no preferred direction.   
It suffices to suppose that $\Omega_f$ and $\Omega_a$ are measured with respect to $\Omega_i$, which is then treated as the orientation of the polar axis.  
Then the integral over $\Omega_i$ reduces to the trivial integral over the polar axis (net rotation of the whole system), which just gives $\int \! \mathrm{d} \Omega_i = 4 \pi$, and 
\bes{\label{eq:depsa_dEa_master}
	\mathrm{d}\varepsilon_a
	& = 
	\frac{g_e n_N}{128 \pi^5} \, 
	\mathrm{d} E_i \, 
	\frac{\mathrm{d} \Omega_f}{4\pi} \, 
	\mathrm{d} \omega \, 
	\frac{\mathrm{d} \Omega_a}{4\pi} \, 
	f_e(E_i) \, \bigl( 1 - f_e(E_f) \bigr) \ 
	\frac{p_i \,  p_f \, k \, \omega}{m_N \, E_N(\pvec_4)} \ 
	\overline{|\Mcal|^2} 
	\;.
}

%=======
To evaluate the squared matrix element, we approximate $m_a \approx 0$ implying $\omega \approx |\kvec|$.  
We can also approximate the recoiling nucleus as non-relativistic, implying $E_4 \approx m_N + \pvec_4^2 / (2m_N)$, and here it is important to keep the sub-leading term in the energy expansion, since the would-be leading order contribution to the squared matrix element cancels.  
Then the squared matrix element reduces to 
\bes{\label{eq:Msq_eP}
	& 
	\overline{|\Mcal|^2} \approx 
	\frac{(Z \gagg e^2)^2}{g_e \, g_N} \ 
	\frac{32 m_N^2 \omega^2}{q_{13}^4 \, q_{24}^4} \ 
%	\\ & \hspace{2cm} 
	\biggl[ 
	\bigl( 2 c_{if} c_{ia} c_{fa} p_i^2 p_f^2 - 2 c_{fa}^2 p_i^2 p_f^2 - c_{if} p_i^3 p_f s_{ia}^2 + 2 p_i^2 p_f^2 s_{ia}^2 - c_{if} p_i p_f^3 s_{fa}^2 \bigr) 
	\\ & \hspace{5cm} 
	+ m_e^2 \bigl( 2 c_{if} p_i p_f - 2 c_{ia} c_{fa} p_i p_f - p_i^2 s_{ia}^2 - 
    p_f^2 s_{fa}^2 \bigr) 
   	\\ & \hspace{5cm} 
	+ E_i E_f \bigl( -2 c_{if} p_i p_f + 2 c_{ia} c_{fa} p_i p_f + p_i^2 s_{ia}^2 + p_f^2 s_{fa}^2 \bigr) 
	\biggr] 
}
where we have dropped terms that are $O(m_N^1)$.  
Here we have also written $\pvec_1 \cdot \pvec_3 = p_i p_f c_{if}$ and $\pvec_1 \cdot \kvec = p_i \omega c_{ia}$ and $\pvec_3 \cdot \kvec = p_f \omega c_{fa}$.  
The momentum transfers are
\bes{
	q_{13}^2 
	& = (p_1 - p_3)^2 
	= (E_1 - E_3)^2 - |\pvec_1 - \pvec_3|^2 
	= (E_i - E_f)^2 - p_i^2 - p_f^2 + 2 p_i p_f c_{if} \\ 
	q_{24}^2 
	& = (p_2 - p_4)^2 
	= (E_2 - E_4)^2 - |\pvec_2 - \pvec_4|^2 
	\approx - p_f^2 - \omega^2 - p_i^2 - 2 p_f \omega c_{fa} + 2 p_i p_f c_{if} + 2 p_i \omega c_{ia} 
	\;. 
}
Putting the squared matrix element into~\eqref{eq:depsa_dEa_master} yields the axion emissivity
\bes{
	\mathrm{d}\varepsilon_a
	& = 
	n_N \ 
	\frac{8 Z^2 \alpha_{\rm EM}^2 \aagg}{\pi^2} \ 
	\mathrm{d} E_i \, 
	\frac{\mathrm{d} \Omega_f}{4\pi} \, 
	\mathrm{d} \omega \, 
	\frac{\mathrm{d} \Omega_a}{4\pi} \, 
%	\\ & \quad 
	f_e(E_i) \, \bigl( 1 - f_e(E_f) \bigr) 
	\\ & \quad 
	\times \frac{1}{q_{13}^4 \, q_{24}^4} 
	\biggl[ 
	\bigl( 2 c_{if} c_{ia} c_{fa} p_i^2 p_f^2 - 2 c_{fa}^2 p_i^2 p_f^2 - c_{if} p_i^3 p_f s_{ia}^2 + 2 p_i^2 p_f^2 s_{ia}^2 - c_{if} p_i p_f^3 s_{fa}^2 \bigr) 
	\\ & \hspace{2.5cm} 
	+ m_e^2 \bigl( 2 c_{if} p_i p_f - 2 c_{ia} c_{fa} p_i p_f - p_i^2 s_{ia}^2 - 
    p_f^2 s_{fa}^2 \bigr) 
   	\\ & \hspace{2.5cm} 
	+ E_i E_f \bigl( -2 c_{if} p_i p_f + 2 c_{ia} c_{fa} p_i p_f + p_i^2 s_{ia}^2 + p_f^2 s_{fa}^2 \bigr) 
	\biggr] 
	\;.
}
We have also used $e^2 = 4\pi \alpha_{\rm EM}$ and $\gaee^2 = 4 \pi \aaee$ and set $g_e = g_N = 2$.  
Note that our assumption $E_N(\pvec_4) \approx m_N$ implies the simple relation $E_f \approx E_i - \omega$.  

%==========
If the plasma is degenerate, $T \ll p_F = \sqrt{E_F^2 - m_e^2}$, then the thermal factor can be approximated as 
\bes{
	f_e(E_i) \ \bigl( 1 - f_e(E_f) \bigr) 
	& \approx \frac{1}{e^{\omega/T} -1} \ \Theta(E_F - E_i) \, \Theta(E_i - E_F - \omega)
	\;.
}
Then the integral over $E_i$ sets $E_i \approx E_F$ and $E_f \approx E_F - \omega$ and gives $\mathrm{d} E_i \approx \omega$.  
This lets us write 

\ba{
	\mathrm{d}\varepsilon_a
	= 
	n_N \ 
	\frac{Z^2 \alpha_{\rm EM}^2 \aagg}{2 \pi^2} \ 
	\frac{(m_e^2 + p_F^2) \, \omega^5 \, \mathrm{d} \omega}{m_e^2 \, p_F^2} \, 
	\frac{1}{e^{\omega/T} -1} \ 
	F
    \;,
}
where we have defined 

\ba{
	F \equiv 
	\frac{m_e^2}{m_e^2 + p_F^2} 
	\int \! \frac{\mathrm{d} \Omega_f}{4\pi} 
	\int \! \frac{\mathrm{d} \Omega_a}{4\pi} \, 
	\Bigl[ 
	4 \bigl( 1 - c_{if} - c_{ia}^2 - c_{fa}^2 \bigr) + \bigl( c_{ia} + c_{fa} \bigr)^2 \bigl( 1 + c_{if} \bigr) 
	\Bigr] 
	\frac{16 p_F^8}{\qvec^8} 
	\;,
}
which contains the angular integrals.  
The momentum transfer factors have become 
\bes{
	q_{13}^2 & \approx q_{24}^2 \approx -\qvec^2 + O(\omega p_F) 
	\qquad \text{where} \qquad 
	\qvec^2 \equiv -2 p_F^2 \bigl( 1 - c_{if} \bigr) 
	\;,
}
and we neglect the $\omega$-suppressed terms.  

%-------------------------
%  Emissivity and luminosity
%-------------------------
\subsection{Emissivity and luminosity}

%==========
Now generalizing to a plasma with multiple species of ions, labeled by $s$, the emissivity spectrum is written as 

\ba{\label{eq:depsa_eP}
	\frac{d\varepsilon_a}{d\omega}
	= 
	\frac{\alpha_{\rm EM}^2 \aagg}{2 \pi^2} \ 
	\frac{\omega^5}{e^{\omega/T} -1} \ 
	\sum_s \frac{Z_s^2 \rho_s F_s}{A_s u} \biggl( \frac{1}{m_e^2} + \frac{1}{p_{F,s}^2} \biggr) 
    \;,
}
where we have used $n_s = \rho_s / u$ and $u \approx 931.5 \ \mathrm{MeV}$ is the atomic mass unit, and we have assumed that all species have a common temperature $T_s = T$.  
Note that the emissivity spectrum, $d\varepsilon_a / d\omega_a$, is almost a thermal spectrum, except that there's an additional factor of $\omega^2$, which follows from the momentum-dependent axion-photon coupling.  
The integral over $\omega$ evaluates to $8 \pi^6 T^6 / 63$, and the total emissivity is found to be

\ba{\label{eq:epsa_eP}
	\varepsilon_a
	= 
	\frac{4 \pi^4}{63} \, 
	\alpha_{\rm EM}^2 \aagg T^6 
	\sum_s \frac{Z_s^2 \rho_s F_s}{A_s u} \biggl( \frac{1}{m_e^2} + \frac{1}{p_{F,s}^2} \biggr) 
	\;.
}
Note that these relations hold for either relativistic or non-relativistic electrons; i.e., $p_F \approx E_F \gg m_e$ or $p_F \ll E_F \approx m_e$.  

%==========
In the derivation above, we have neglected medium effects, which are now taken into account following Ref.~\cite{Raffelt:1990yz}.  
Free electrons in the medium will screen the photon propagator, introducing an effective photon mass $k_\mathrm{TF}^2 = 4 \alpha_{\rm EM} p_F E_F / \pi$, which is the Thomas-Fermi screening scale. 
Additionally interference and correlation effects are captured by the static structure factor $S_\mathrm{ions}(|\qvec|)$.  
For a strongly-coupled plasma, such as the one in a WD core, the static structure factor has been calculated in Refs.~\cite{Nakagawa:1987pga,Nakagawa:1988rhp}, and the factor $F$ is also evaluated for axion emission via electron-bremsstrahlung scattering.  
As a rough estimate, we simply carry over that estimate of $F$ here, though future work using this result should calculate $F$ more precisely.  

%==========
The axion luminosity is evaluated by integrating $L_a = \int \! \mathrm{d}V \, \varepsilon_a$ over the volume of the WD star.  
To a good approximation, the core temperature $T \approx T_c$ is approximately uniform throughout the star, due to the degenerate matter's high thermal conductivity.  
On the other hand, the Fermi momenta $p_{F,s}$, medium factors $F_s$, and mass fractions $R_s = \rho_s / \rho_\mathrm{tot}$ have radial-dependent profiles.  
To provide a rough estimate, we neglect these effects and the volume integral gives $\int \! \mathrm{d}V \, \rho_\mathrm{tot} = M$, which is the mass of the star.  
Then the axion luminosity is 
\bes{\label{eq:La_eP}
	L_a
	& \approx  
	\frac{4 \pi^4}{63} \, 
	\alpha_{\rm EM}^2 \aagg \frac{T_c^6 M}{m_e^2 u} 
	\sum_s \frac{Z_s^2 R_s F_s}{A_s} \biggl( 1 + \frac{m_e^2}{p_{F,s}^2} \biggr) 
	\\ 
	& \simeq \bigl( 7.6 \times 10^{-12} \, L_\odot \bigr) \left( \frac{g_{a\gamma\gamma}}{10^{-11} \ \mathrm{GeV}} \right)^2 \left( \frac{T_c}{1 \ \mathrm{keV}} \right)^6 \left( \frac{M}{1 \ M_\odot} \right) \, \sum_s \frac{Z_s^2 R_s F_s}{A_s} \biggl( 1 + \frac{m_e^2}{p_{F,s}^2} \biggr)
	\;,
}
Compared with axion bremsstrahlung emission, the luminosity here is suppressed by a factor of $\alpha_{a\gamma\gamma} T_c^2 / \alpha_{aee}$.  The electro-Primakoff emission spectrum and resulting limits are illustrated in Figs.~\ref{fig:gagg_limits_zoom} and~\ref{fig:spectrum_ep}.

\end{document}